\documentclass[12pt,letterpaper]{article}
\usepackage{putex}
\usepackage{graphicx}
\usepackage{latexsym,amsmath,amsfonts,amssymb,amsthm}
\usepackage{empheq}
\usepackage{bbm}
\usepackage{cite}
\usepackage{indentfirst}
\usepackage{booktabs,array}
\usepackage{placeins}
\usepackage{mathtools}

\DeclarePairedDelimiter\floor{\lfloor}{\rfloor}
\usepackage{colonequals}
\usepackage{cancel}
\usepackage[mode=image|tex]{standalone}
\usepackage[dvipsnames]{xcolor}
\usepackage{tikz}
\usepackage{tikz-cd}
\usepackage{adjustbox}
\usepackage{enumitem}
\usetikzlibrary{decorations.markings}
\usetikzlibrary{decorations.text}
\usepackage{graphicx}
\usepackage[margin=10pt,font=small,labelfont=bf]{caption}
\usepackage{subcaption}
\usepackage{xcolor}
\usepackage{float}
\theoremstyle{definition}

\usepackage{microtype}
\usepackage{hyperref}
\hypersetup{unicode}
\hypersetup{linktoc = all}
\hypersetup{pdfborderstyle={/S/U/W 0.5}}
\hypersetup{linkbordercolor = gray}
\hypersetup{bookmarksnumbered}
\usepackage[T1]{fontenc}
\usepackage[utf8]{inputenc}
\usepackage{lmodern}

\setlist{itemsep=2pt plus 1pt minus 1pt, topsep=2pt plus 1pt minus 1pt}

\renewenvironment{thebibliography}[1]{%
\begin{oldthebibliography}{#1}%
\setlength\itemsep{-2mm}%
}%
{%
\end{oldthebibliography}%
}
\numberwithin{equation}{section}

\begin{document}

%%%%%%% title page %%%%%%%%%

\title{\begin{LARGE}
$\mathcal{N} = 2$ Schur index and line operators
\end{LARGE}}

\authors{Zhaoting Guo, Yutong Li, Yiwen Pan, Yufan Wang
\medskip\medskip\medskip\medskip
 }

\institution{UU}{${}^1$
School of Physics, Sun Yat-Sen University, \cr
$\;\,$ Guangzhou, Guangdong, China}

\abstract{\begin{onehalfspace}{
4d $\mathcal{N} = 2$ SCFTs and their invariants can be often enriched by non-local Bogomol'nyi-Prasad-Sommerfield (BPS) operators. In this paper we study the flavored Schur index of several types of $\mathcal{N} = 2$ SCFTs with and without line operators, using a series of new integration formula of elliptic functions and Eisenstein series. We demonstrate how to evaluate analytically the Schur index for a series of $A_2$ class-$\mathcal{S}$ theories and the $\mathcal{N} = 4$ $SO(7)$ theory. For all $A_1$ class-$\mathcal{S}$ theories we obtain closed-form expressions for $SU(2)$ Wilson line index, and 't Hooft line index in some simple cases. We also observe the relation between the line operator index with the characters of the associated chiral algebras. Wilson line index for some other low rank gauge theories are also studied.
}\end{onehalfspace}}

\preprint{}
\setcounter{page}{0}
\maketitle

%%%%%% end of title page %%%%%%

%%%%%% table of contents %%%%%%
{%\small
\setcounter{tocdepth}{2}
\setlength\parskip{-0.7mm}
\tableofcontents
}
%%%%%%%%%%%%%%%%%%%%%

\section{Introduction}

Any 4d $\mathcal{N} = 2$ superconformal field theory contains a nontrivial protected subsector of Schur operators that form an associated two-dimensional chiral algebra \cite{Beem:2013sza}, providing an important invariants of $\mathcal{N} = 2$ SCFTs. These operators are defined as the cohomology class of some well-chosen supercharge of the $\mathcal{N} = 2$ superconformal algebra. The index that counts these Bogomol'nyi-Prasad-Sommerfield (BPS) operators is called the Schur index $\mathcal{I}(q)$, which happens to be a special limit $t\to q$ of the full $\mathcal{N} = 2$ superconformal index $\mathcal{I}(p,q,t)$ \cite{Gadde:2011uv}. The Schur index plays a central role in the superconformal field theory (SCFT)/vertex operator algebra (VOA) correspondence, as it coincides with the vacuum character of the associated chiral algebra.

Similar to the $S^4$ supersymmetric partition functions \cite{Pestun:2007rz}, the superconformal index \cite{Kinney:2005ej,Romelsberger:2005eg}, and in particular, the Schur index \cite{Gadde:2011uv} is an exactly computable quantity. For theories with a Lagrangian, the Schur index can be computed as an $S^3 \times S^1$ partition function and localizes to a multivariate contour integral along the the unit circles \cite{Nawata:2011un,Pan:2019bor,Jeong:2019pzg,Dedushenko:2019yiw,Oh:2019mcg}. Alternatively for theories of class-$\mathcal{S}$ \cite{Gaiotto:2009we}, the index can be identified as the partition function the 2d $q$-deformed Yang-Mills theory \cite{Gadde:2011ik,Mekareeya:2012tn,Lemos:2012ph,Lemos:2014lua,Buican:2015ina,Xie:2016evu}. There are also instances where the associated chiral algebras are known from other methods, whose module characters have already existed in the literature \cite{2016arXiv161207423K}. There are also other methods to compute the Schur index in different scenarios \cite{Zafrir:2020epd,Gadde:2010te,Agarwal:2018ejn,Kang:2021lic}. Many of these results, although exact, are not given in closed-form in terms of finite combinations of special functions with well-controlled periodic and modular properties. This problem is tackled in several recent works. The unflavored and later the flavored Schur index for the $\mathcal{N} = 4$ $SU(N)$ theories are computed in closed-form using the Fermi-gas formalism in \cite{Bourdier:2015wda,Bourdier:2015sga,Hatsuda:2022xdv} and modular anomaly equation \cite{Huang:2022bry}. In \cite{Beemetal}, the unflavored Schur index for many class-$\mathcal{S}$ theories are computed in terms of quasi-modular forms. In \cite{Pan:2021mrw}, several integration formula are proposed to compute analytically the index for a wide range of Lagrangian theories (and some non-Lagrangian ones) in terms of finite combination of twisted Eisenstein series and Jacobi theta functions.

Line operators can be introduced into 4d $\mathcal{N} = 2$ SCFTs that preserve some amount of supersymmetry \cite{Lee:1996vz,Gomis:2009ir,Gang:2012yr,Cordova:2016uwk}, and the corresponding $S^4$ partition function and superconformal index in their presence have been computed exactly \cite{Pestun:2007rz,Giombi:2009ek,Gomis:2011pf,Dimofte:2011py,Drukker:2015spa,Neitzke:2017cxz}. In the context of the AGT-correspondence, the line operators correspond to the Verlinde network operators in the Liouville/Toda CFT \cite{Drukker:2009id,Drukker:2009tz,Gomis:2010kv,Drukker:2010jp,Bullimore:2013xsa}. In \cite{Gang:2012yr}, the Schur index of supersymmetric Wilson lines and the S-dual 't Hooft lines in different gauge theories are studied, incorporating the monopole bubbling effects. In \cite{Cordova:2016uwk}, the authors propose an infrared computation method using the infrared Seiberg-Witten description and compute the line operator index for Argyres-Douglas theories and $SU(2)$ SQCD. The papers \cite{Watanabe:2016bwr,Watanabe:2017bmi} further studied the Schur index of Wilson-'t Hooft line operators in terms the punctured networks.

In this work, we focus on computing analytically the Schur index with or without line defects in 4d $\mathcal{N} = 2$ SCFTs, generalizing the work in \cite{Pan:2021mrw}. See also \cite{Hatsuda:2023iwi} for extensive analytic results on Wilson line index for the $\mathcal{N} = 2^*$ $U(N)$ theories. The key tools for our purpose are a new set of integration formula that can be applied to a wide range of 4d $\mathcal{N} = 2$ Lagrangian SCFTs, expressing multivariate integrals of elliptic functions in terms of Eisenstein series and rational functions of flavor fugacities. Concretely, we propose integration formula for integrals with the following types (and some variants) of integrand,
\begin{align}
  f(\mathfrak{z}) E_1 \begin{bmatrix}
    \pm 1 \\ za  
  \end{bmatrix}
  E_k \begin{bmatrix}
    \pm 1 \\ zb  
  \end{bmatrix} , \quad
  z^m f(\mathfrak{z}), \quad
  z^m E_k \begin{bmatrix}
    \pm 1 \\ za  
  \end{bmatrix}E_\ell \begin{bmatrix}
    \pm 1 \\ zb  
  \end{bmatrix} \ ,
  \quad
  z^m f(\mathfrak{z}) E_k \begin{bmatrix}
    \mp 1 \\ z a  
  \end{bmatrix}
\end{align}
where $f(\mathfrak{z})$ denotes an elliptic function in $\mathfrak{z}$, and $z = e^{2\pi i \mathfrak{z}}$. These formula can be used to compute the standard Schur index of some $A_2$-type theories of class-$\mathcal{S}$, Wilson line index in $A_1$-type theories of class-$\mathcal{S}$, $SU(N)$ SQCD and $\mathcal{N} = 4$ theories of gauge group $SO(N)$. We also compute the 't Hooft line index in some simplest cases. 

Let us explain the $A_1$ case in a bit more detail. Recall that an $SU(2)$ Wilson line operator\footnote{We consider the simplest Wilson lines charged under one $SU(2)$ gauge group, and leave more general Wilson lines and their correlators to future work.} in a $A_1$ class-$\mathcal{S}$ theories $\mathcal{T}[\Sigma_{g, n}]$ is dual to a line operator on the $\Sigma_{g, n}$, and in particular, such a line operator resides at some long tube in the pants-decomposition of $\Sigma_{g,n}$ which provides a gauge theory description for $\mathcal{T}[\Sigma_{g,n}]$. We find that for $A_1$-type theories $\mathcal{T}[\Sigma_{g, n}]$, there are two major types of Wilson line operators: if the relevant tube separates $\Sigma_{g,n}$ into two disconnected parts, the index is called type-2, and otherwise type-1. Note that type-1 line operators exist only when genus $g \ge 1$. It turns out that the type-1 Wilson index is easy to compute and we are able to obtain an elegant compact closed-form, 
\begin{align}
  \langle W_{j \in \mathbb{Z}}\rangle^{(1)}_{g \ge 1, n}
  = & \ \mathcal{I}_{g,n}
  - \left[\prod_{i = 1}^{n} \frac{i \eta(\tau)}{\vartheta_1(2 \mathfrak{b}_i)}
  \right]
        \sum_{\substack{m = - j\\m \ne 0}}^{+ j}
    \left[\frac{\eta(\tau)}{q^{m/2} - q^{-m /2}}\right]^{2g - 2}
    \prod_{i = 1}^{n} \frac{b_i^m - b_i^{-m}}{q^{m/2} - q^{- m /2}} \ ,\\
  \langle W_{j \in \mathbb{Z} + \frac{1}{2}}\rangle^{(1)}_{g \ge 1, n} = & \ 0 \ .
\end{align}
Note that given a $\Sigma_{g, n}$, as long as the Wilson operator is of type-1, the index is independent of the specific tube it resides. Generalizing the observation in \cite{Cordova:2016uwk}, the type-1 index can be viewed as a linear combination of the vacuum character $\mathcal{I}_{g,n}$ of the associated chiral algebra of $\mathcal{T}[\Sigma_{g,n}]$ and a non-vacuum module character $\eta(\tau)^{2 g - 2}\prod_{i = 1}^{n}\frac{i \eta(\tau)}{\vartheta_i(2 \mathfrak{b}_i)}$, where the coefficient is a rational function
\begin{align}
  \sum_{\substack{m = -j \\ m \ne 0}}^{+j}\frac{1}{(q^{m/2} - q^{-m/2})^{2 g - 2}} \prod_{i = 1}^n \frac{b_i^m - b_i^{-m}}{q^{m/2} - q^{-m/2}} \ .
\end{align}
For the type-2 Wilson line, the closed-form index takes a less elegant form. Still, we are able to identify a similar structure of finite linear combination of characters for the $SU(2)$ SQCD, $\mathcal{T}[\Sigma_{2,1}]$ and all $\mathcal{T}[\Sigma_{g, 0}]$. In particular, the type-2 Wilson index in $\mathcal{T}[\Sigma_{2,1}]$ provides two new linearly independent (combinations of) characters of the associated chiral algebra, which were previously not visible from analyzing surface defects in $\mathcal{T}[\Sigma_{1,2}]$.

This paper is organized as follows. In section 2, we demonstrate that the generalized integration formula can be used to compute analytically the Schur index of a series of $A_2$-type class-$\mathcal{S}$ theories with or without Lagrangian, and of $SO(7)$ $\mathcal{N} = 4$ SYM. In section 3, we compute both type-1 and type-2 line operator index for $A_1$-type class-$\mathcal{S}$ theories. In section 4, we further compute line operator index for some other higher rank gauge theories. The appendix \ref{app:special-functions} contains a quick review of the relevant special functions, and appendix \ref{app:integration-formula} collects a series of new integration formula that help compute Schur index or and without line operators.

\section{More on Schur index}

Several integration formula were proposed in \cite{Pan:2021mrw}, which can be used to analytically compute some multivariate contour integral of elliptic functions. Those formula were enough to compute exactly the Schur index of $A_1$ class-$\mathcal{S}$ theories and some low rank $\mathcal{N} = 4$ theories. However, they were insufficient for more general $A_N$ class-$\mathcal{S}$ theories. In this section, with the help from some new integration formula, we explore the exact computation the Schur index of a series of $A_2$ theories and the $\mathcal{N} = 4$ $SO(7)$ theory, generalizing the results in \cite{Pan:2021mrw}. The computation in this section is relatively technical, and uninterested readers may skip to section \ref{section:Wilson-index-A1-theories} for the computation of line operator index.

\subsection{\texorpdfstring{$A_2$ theories of class-$\mathcal{S}$}{}}

First we recall the Schur index of the $SU(3)$ SQCD. It can be computed as a contour integral
\begin{equation}
  \mathcal{I}_{\text{SQCD}} = - \frac{1}{3!} \eta(\tau)^{16} \oint \prod_{A = 1}^2 \frac{da_A}{2\pi i a_A}
  \frac{\prod_{A \ne B} \vartheta_1(\mathfrak{a}_A - \mathfrak{a}_B)}{\prod_{A = 1}^3 \prod_{i = 1}^{6} \vartheta_4(\mathfrak{a}_A - \mathfrak{m}_i)}
  \equiv \oint \prod_{A = 1}^2 \frac{da_A}{2\pi i a_A}\mathcal{Z}(\mathfrak{a})\ ,
\end{equation}
where $\mathfrak{a}_3 = -\mathfrak{a}_1 - \mathfrak{a}_2$, $a_3 = (a_1 a_2)^{-1}$, and $a_i = e^{2\pi i \mathfrak{a}_i}$, $m_i = e^{2\pi i \mathfrak{m}_i}$. See also Appendix \ref{app:special-functions} for the definitions and properties of the Eisenstein series $E_k\big[\substack{\phi\\\theta}\big]$ and the Jacobi theta functions. The integral can be performed by applying the integration formula in Appendix \ref{app:special-functions}, which yields the exact (albeit slightly complicated) result\footnote{Here the prefactors $R_{j_1j_2}$ are given by \begin{align}
      R_{0 j_2} \coloneqq & \ \frac{
        i \eta(\tau)^{13}\vartheta_1(2 \mathfrak{m}_{j_2})
        \vartheta_4(\mathfrak{m}_{j_2})^3
      }{
        6 \prod_{i \ne j_2}\vartheta_1(\mathfrak{m}_{j_2} - \mathfrak{m}_i)
        \vartheta_1(\mathfrak{m}_{j_2} + \mathfrak{m}_i)
        \prod_{i \ne j_2}\vartheta_4(\mathfrak{m}_i)
      }\nonumber \\
      R_{j_1 j_2} \coloneqq & \ \frac{
        \eta(\tau)^{10}
        \vartheta_4(2 \mathfrak{m}_{j_1} + \mathfrak{m}_{j_2})
        \vartheta_4(\mathfrak{m}_{j_1} + 2\mathfrak{m}_{j_2})}{
        6
        \prod_{i\ne j_1, j_2}\vartheta_1(\mathfrak{m}_{j_1} - \mathfrak{m}_i)\vartheta_1(\mathfrak{m}_{j_2} - \mathfrak{m}_i)
        \prod_{i \ne j_1, j_2} \vartheta_4(\mathfrak{m}_{j_1}+ \mathfrak{m}_{j_2} + \mathfrak{m}_i)
      } \nonumber\\
      R_{j_1} = R_{j_1 0} \coloneqq & \ \frac{i\eta(\tau)^{13}\vartheta_1(2 \mathfrak{m}_{j_1}) \vartheta_4(\mathfrak{m}_{j_1})^3}{
        \prod_{i \ne j}\vartheta_1(\mathfrak{m}_j + \mathfrak{m}_i)
        \vartheta_1(\mathfrak{m}_j - \mathfrak{m}_i)
        \vartheta_4(\mathfrak{m}_i)
      } = R_{0 j_1}\nonumber\ .
\end{align}},
\begin{align}\label{index-SQCD}
  \mathcal{I}_{\text{SQCD}} = & \ \sum_{j_2 = 1}^{6}2R_{0j_2}E_1\left[\begin{matrix}
    -1 \\ m_{j_2}
  \end{matrix}\right] \nonumber \\
  & \ + \sum_{j_1}^{6}
    \left(
    R_{j_1}(\mathfrak{a}_2 = 0) + \sum_{j_2 = 1}^6
    \Big(  
    R_{j_1 j_2}E_1\left[\begin{matrix}
      -1 \\ m_{j_2}
    \end{matrix}\right]
    + R_{j_1 j_2} E_1\left[\begin{matrix}
      -1 \\ m_{j_1}m_{j_2}q^{ - \frac{1}{2}}
    \end{matrix}\right]  \Big)
  \right)E_1\left[\begin{matrix}
    -1 \\ m_{j_1}
  \end{matrix}\right]\nonumber\\
  & \ + \sum_{j_1, j_2 = 1}^{6} R_{j_1 j_2}\left(
      E_2\left[\begin{matrix}
        1 \\ m_{j_1} m_{j_2}
      \end{matrix}\right]
      - E_2\left[
      \begin{matrix}
        1 \\ m_{j_2}q^{ - \frac{1}{2}}
      \end{matrix}\right]
    \right)  \ .
\end{align}
From the computation of $SU(3)$ SQCD index, we already see that the complexity is far above the $A_1$ type. Therefore, we shall focus on arguing that the index can be computed using the existing integration formula. The complexity could decrease once more optimized integration formula is found, which we leave to future work.

As a class-$\mathcal{S}$ theory, the SQCD has manifest flavor symmetry $SU(3)^{(1)} \times SU(3)^{(2)} \times U(1)^{(1)} \times U(1)^{(2)}$. We shall denote the fugacities of $SU(3)^{(\alpha)}$ as $c^{(\alpha)}$, and those of $U(1)^{(\alpha)}$ as $d^{(\alpha)}$. They are related to $m_j$ by
\begin{align}
  c^{(1)}_1 = & \ m_1/d^{(1)}, \qquad c^{(1)}_2 = m_2 / d^{(1)}, \qquad
  d^{(1)} = (m_1 m_2 m_3)^{1/3} \ ,\\
  c^{(2)}_1 = & \ m_4/d^{(2)}, \qquad c^{(2)}_2 = m_5/d^{(2)}, \qquad
  d^{(2)} =  (m_4 m_5 m_6)^{1/3} \ .
\end{align}

Starting from $SU(3)$ SQCD, one can build $SU(3)$ linear quiver theories by successively gauging in $9$ hypermultiplets one after another. Let us perform one such computation. The gauging procedure multiplies to $\mathcal{I}_\text{SQCD}$ factors
\begin{align}
  \mathcal{I}_\text{VM} \sim \prod_{\substack{A,B = 1 \\ A\ne B}}^{3} \vartheta_1 (\mathfrak{a}_A - \mathfrak{a}_B) \ ,\qquad
  \mathcal{I}_\text{HM} = \prod_{A, B = 1}^{3}\frac{\eta(\tau)}{\vartheta_4(- \mathfrak{a}_A + \mathfrak{c}^{(3)}_B + \mathfrak{d}^{(3)})} \ ,
\end{align}
where again $\mathfrak{a}_3 = - \mathfrak{a}_1 - \mathfrak{a}_2$, $\mathfrak{c}^{(3)}_3 = - \mathfrak{c}^{(3)}_ 1 - \mathfrak{c}^{(3)}_2$. The gauging also identifies $\mathfrak{c}^{(2)}_A$ with $\mathfrak{a}_A$, and a contour integral of $a_1, a_2$ should be performed, 
\begin{align}
  \mathcal{I} = \oint \frac{da_1}{2\pi i a_1}
  \frac{da_2}{2\pi i a_2} \mathcal{I}_\text{SQCD}(c^{(1)}, a, d^{(1)}, d^{(2)}) \mathcal{I}_\text{VM}(a) \mathcal{I}_\text{HM}(a, c^{(3)}, d^{(3)}) \ .
\end{align}

Let us look at the various terms in this integral. First of all, we have an integral of
\begin{align}
  \sum_{j_2 = 1}^6
  R_{0 j_2} \mathcal{I}_\text{VM} \mathcal{I}_\text{HM} E_1 \begin{bmatrix}
    -1 \\ m_{j_2}
  \end{bmatrix} \ .
\end{align}
It is straightforward to verify that, as a function of $\mathfrak{a}_{1,2}$, the factor $R_{0j_2}\mathcal{I}_\text{VM}\mathcal{I}_\text{HM}$ is elliptic with respect to both $\mathfrak{a}_{1,2}$. Moreover, after the replacing $m$ with the $c, d$ fugacities and $a$,
\begin{align}
  (m_1, \ldots, m_6) = (
  c^{(1)}_1 d^{(1)},
  c^{(1)}_2 d^{(1)},
  \frac{d^{(1)}}{c^{(1)}_1 c^{(1)}_2},
  a_1d^{(2)},
  a_2 d^{(2)},
  \frac{d^{(2)}}{a_1 a_2}
  ) \ ,
\end{align}
and similarly $m_{j_1} m_{j_2 \ne j_1} \sim ((\ldots),  a_1^{\pm 1} (\ldots), a_2^{\pm 1}(\ldots), (a_1 a_2)^{\pm 1} (\ldots) )$ where $(\ldots)$ denotes combinations of $c^{(1)}, d^{(1)}, d^{(2)}$. Therefore, one can perform the $a_1$ integral using (\ref{integration-formula-fE-1}) or (\ref{integration-formula-fE-2}). For all $j_2$, there are several types of poles from $R_{0j_2}\mathcal{I}_\text{VM}\mathcal{I}_\text{HM}$,
\begin{align}
  \mathfrak{a}_1 = & \ [1,2], \quad  [3],
  \quad - \mathfrak{a}_2 + [1,2], \quad
   - \mathfrak{a}_2 + [3] \ .
\end{align}
Here $[1,2]$ and $[3]$ denote respectively linear combinations of $\mathfrak{c}^{(1)}, d^{(1,2)}$ and $\mathfrak{c}^{(3)}, \mathfrak{d}^{(3)}$. The Eisenstein series $E_1 \big[\substack{-1\\m_{j_2 = 1,2,3}}\big]$ are independent of $a_1, a_2$, and will never participate in subsequent integrations or gauging. The variables $a_1, a_2$ in $E_1 \big[ \substack{-1\\ m_{j_2 = 4,5,6}} \big]$ appear in the form $a_1$, or $a_2$, or a product $a_1 a_2$. The $a_1$ integral using the integration formula will produce $E_1 \big[ \substack{\pm 1\\ [1,2]} \big], E_1 \big[ \substack{\pm 1\\ [3]} \big]$, $E_2\big[ \substack{\pm1\\ a_2[1,2]} \big]$ or $E_2\big[ \substack{\pm1\\ a_2[3]} \big]$, where $[1,2]$ and $[3]$ denote respectively combinations of the flavor fugacities $c^{(1)}, d^{(1)}, d^{(2)}$, and of $c^{(3)}, d^{(3)}$. The $a_2$-integration of these terms can be further carried out, and we have Eisenstein structure, 
\begin{equation}
  E_{1} \begin{bmatrix}
    \pm 1\\ [1,2]
\end{bmatrix}E_{1} \begin{bmatrix}
    \pm 1\\ [1,2]
\end{bmatrix}, \quad
  E_{1} \begin{bmatrix}
    \pm 1\\ [1,2]
\end{bmatrix}E_{1} \begin{bmatrix}
    \pm 1\\ [3]
\end{bmatrix},  \quad
  E_{1,2} \begin{bmatrix}
    \pm 1\\ [1,2,3]
\end{bmatrix} \ ,
\end{equation}
where $[1,2,3]$ denotes products of $c^{(1)}, c^{(3)}, d^{(1,2,3)}$.

Next we have $R_{j_1}(\mathfrak{a}_2 = 0)\mathcal{I}_\text{VM}\mathcal{I}_\text{HM}E_1 \big[\substack{-1 \\ m_{j_1}}\big]$ integral. Again, the prefactor $R_{j_1}(\mathfrak{a}_2 = 0)\mathcal{I}_\text{VM}\mathcal{I}_\text{HM}$ is separately elliptic with respect to both $\mathfrak{a}_{1,2}$. This factor again has $\mathfrak{a}_1$-poles of the form
\begin{align}
  \mathfrak{a}_1 = & \ [1,2], \quad  [3],
  \quad - \mathfrak{a}_2 + [1,2], \quad
   - \mathfrak{a}_2 + [3] \ .
\end{align}
Therefore, the $a_1$, $a_2$ integral can also be straightforwardly performed with a reference point $\mathfrak{a}_1 = 0$. The Eisenstein structure is the same as that of the previous term.

Let us also look at the last two terms in (\ref{index-SQCD}),
\begin{align}
  R_{j_1 j_2}\left(
  E_2 \begin{bmatrix}
    1 \\ m_{j_1}m_{j_2}
  \end{bmatrix}
  - E_2 \begin{bmatrix}
    1 \\ m_{j_2}
  \end{bmatrix}
  \right) \ .
\end{align}
We note that $R_{j_1 j_2} = 0$ when $j_1 = j_2$. One can also directly verify that $R_{j_1j_2} \mathcal{I}_\text{VM} \mathcal{I}_\text{HM}$ is elliptic, with $\mathfrak{a}_1$ poles of the same simple form as the above. Hence, one can also proceed with both $a_1, a_2$ integral using the integration formula (\ref{integration-formula-fE-1}), (\ref{integration-formula-fE-2}). The Eisenstein structure of the result involves
\begin{align}
  E_1 \begin{bmatrix}
    \pm 1 \\ [1,2]  
  \end{bmatrix}
  E_2 \begin{bmatrix}
    \pm 1 \\ [1,2]  
  \end{bmatrix}
  E_1 \begin{bmatrix}
    \pm 1 \\ [1,2] \text{ or } [3]
  \end{bmatrix} \ ,\quad
  E_1 \begin{bmatrix}
    \pm 1 \\ [3]  
  \end{bmatrix}
  E_2 \begin{bmatrix}
    \pm 1 \\ [1,2]  
  \end{bmatrix}
  E_1 \begin{bmatrix}
    \pm 1 \\ [1,2] \text{ or } [3]
  \end{bmatrix} \ ,
\end{align}

We are now ready to deal with the middle two terms in (\ref{index-SQCD}). Again, the factor in front of the Eisenstein series is suitably elliptic. But now this elliptic function is multiplying with
\begin{align}
  E_1 \begin{bmatrix}
    -1 \\ m_{j_2}
  \end{bmatrix}E_1 \begin{bmatrix}
    -1 \\ m_{j_1}  
  \end{bmatrix}, \qquad
  E_1 \begin{bmatrix}
    -1 \\ m_{j_1} m_{j_2}q^{-1/2}
  \end{bmatrix}E_1 \begin{bmatrix}
    -1 \\ m_{j_1}  
  \end{bmatrix} \ .
\end{align}
When substituting in the $a, c, d$ fugacities we will need to integrate
\begin{align}
  f(a_1, a_2)E_1 \begin{bmatrix}
    -1 \\ a_A[1,2]
  \end{bmatrix}
  E_1 \begin{bmatrix}
    -1 \\ a_B [1,2]
  \end{bmatrix}, \quad
  f(a_1, a_2)
  E_1 \begin{bmatrix}
    -1 \\ a_A [1,2]
  \end{bmatrix}
  E_1 \begin{bmatrix}
    -1 \\ a_1 a_2 [1,2]
  \end{bmatrix} \ .
\end{align}
We can carry out the $a_1$ integral which involves poles of the same form as the above, $\mathfrak{a}_1 = \text{expressions of } \mathfrak{c}, \mathfrak{d}$ and $\mathfrak{a}_1 = - \mathfrak{a}_2 + \text{expressions of } \mathfrak{c}, \mathfrak{d}$. The integral can be performed using the integration formula (\ref{integration-formula-fE-1}). After the $a_1$ integral, we will have the following type of integrand left to integrate (factors independent of $a_2$ are omitted),
\begin{align}
  & \ f(a_2)E_{k = 1,2} \begin{bmatrix}
      \pm 1 \\ a_2(\ldots)
    \end{bmatrix}, \qquad \text{ or },\qquad
  f(a_2)E_{k = 1,2} \begin{bmatrix}
      \pm 1 \\ a_2(\ldots)
    \end{bmatrix}
    E_1 \begin{bmatrix}
        \pm 1 \\ (\ldots)
      \end{bmatrix}\ .
\end{align}
To illustrate this, we can look at a term in the sum, for example,
\begin{align}
  f(a_1, a_2) E_1 \begin{bmatrix}
    -1 \\ a_1 a_2 (\ldots)
  \end{bmatrix}
  E_1 \begin{bmatrix}
    -1 \\ a_1 a_2 (\ldots)'
  \end{bmatrix} \ .
\end{align}
Since $f(a_1, a_2)$ has poles only of the form $a_1 = (\ldots)$ and $a_1 = a_2^{-1} (\ldots)$, the integral of the above will produce Eisenstein series with arguments
\begin{align}
  \frac{a_2(\ldots)}{a_2(\ldots)'}, \quad e^{2\pi i 0} a_2^{-1}(\ldots), \quad e^{2\pi i 0}(\ldots), \quad a_2(\ldots) a_2^{-1}(\ldots), \quad
  a_2(\ldots) (\ldots) \ , \quad \text{etc.}
\end{align}
Here we have chosen the reference point as $\mathfrak{a}_1 = 0$. Therefore, although tedious, the leftover $a_2$ integral can be dealt with, and it produces the exact Schur index for the $SU(3) \times SU(3)$ linear quiver theory. In the end, the exact index contains Eisenstein structures
\begin{align}
  E_3 \begin{bmatrix}
    \pm 1\\ (\ldots)  
  \end{bmatrix}
  , \quad
  E_1 \begin{bmatrix}
    \pm 1\\ (\ldots)  
  \end{bmatrix}E_2 \begin{bmatrix}
    \pm 1\\ (\ldots)  
  \end{bmatrix} \ .
\end{align}

The above analysis can be repeated for longer linear $SU(3)$ quiver theories, where we will encounter integrals in the presence of
\begin{align}
  E_n \begin{bmatrix}
    \pm 1 \\ z(\ldots)
  \end{bmatrix}, \qquad
  E_1 \begin{bmatrix}
    \pm 1 \\ z(\ldots)  
  \end{bmatrix} E_n \begin{bmatrix}
    \pm 1 \\ z(\ldots)  
  \end{bmatrix} \ .
\end{align}
These integrals can be treated using the integration formula in the appendix \ref{app:integration-formula}, and therefore Schur index of all linear $SU(3)$-quiver are computable, though rather tedious, with the current method.

\begin{figure}
  \centering
  \tikzset{every picture/.style={line width=0.75pt}} %set default line width to 0.75pt        

  \begin{tikzpicture}[x=0.75pt,y=0.75pt,yscale=-1,xscale=1, scale=0.8]
  %uncomment if require: \path (0,300); %set diagram left start at 0, and has height of 300

  %Shape: Circle [id:dp4320163713313452] 
  \draw   (49,159.88) .. controls (49,116.87) and (83.87,82) .. (126.88,82) .. controls (169.89,82) and (204.76,116.87) .. (204.76,159.88) .. controls (204.76,202.89) and (169.89,237.76) .. (126.88,237.76) .. controls (83.87,237.76) and (49,202.89) .. (49,159.88) -- cycle ;
  %Shape: Circle [id:dp9005557880817314] 
  \draw   (85,122.88) .. controls (85,117.98) and (88.98,114) .. (93.88,114) .. controls (98.78,114) and (102.76,117.98) .. (102.76,122.88) .. controls (102.76,127.78) and (98.78,131.76) .. (93.88,131.76) .. controls (88.98,131.76) and (85,127.78) .. (85,122.88) -- cycle ;
  %Shape: Circle [id:dp8128693514347864] 
  \draw   (84,193.88) .. controls (84,188.98) and (87.98,185) .. (92.88,185) .. controls (97.78,185) and (101.76,188.98) .. (101.76,193.88) .. controls (101.76,198.78) and (97.78,202.76) .. (92.88,202.76) .. controls (87.98,202.76) and (84,198.78) .. (84,193.88) -- cycle ;
  %Shape: Circle [id:dp013466054384840831] 
  \draw   (159,157.88) .. controls (159,152.98) and (162.98,149) .. (167.88,149) .. controls (172.78,149) and (176.76,152.98) .. (176.76,157.88) .. controls (176.76,162.78) and (172.78,166.76) .. (167.88,166.76) .. controls (162.98,166.76) and (159,162.78) .. (159,157.88) -- cycle ;
  %Straight Lines [id:da7199354398184901] 
  \draw    (167.88,149) -- (277.76,149) ;
  %Straight Lines [id:da8638636540047111] 
  \draw    (167.88,166.76) -- (277.76,166.76) ;
  %Shape: Circle [id:dp10915438442528558] 
  \draw   (268.88,157.88) .. controls (268.88,152.98) and (272.86,149) .. (277.76,149) .. controls (282.67,149) and (286.64,152.98) .. (286.64,157.88) .. controls (286.64,162.78) and (282.67,166.76) .. (277.76,166.76) .. controls (272.86,166.76) and (268.88,162.78) .. (268.88,157.88) -- cycle ;
  %Shape: Circle [id:dp38443858243949536] 
  \draw   (242,157.88) .. controls (242,114.87) and (276.87,80) .. (319.88,80) .. controls (362.89,80) and (397.76,114.87) .. (397.76,157.88) .. controls (397.76,200.89) and (362.89,235.76) .. (319.88,235.76) .. controls (276.87,235.76) and (242,200.89) .. (242,157.88) -- cycle ;
  %Shape: Circle [id:dp47858825931637883] 
  \draw   (334.88,120.88) .. controls (334.88,115.98) and (338.86,112) .. (343.76,112) .. controls (348.67,112) and (352.64,115.98) .. (352.64,120.88) .. controls (352.64,125.78) and (348.67,129.76) .. (343.76,129.76) .. controls (338.86,129.76) and (334.88,125.78) .. (334.88,120.88) -- cycle ;
  %Shape: Circle [id:dp9614118989595697] 
  \draw  [fill={rgb, 255:red, 0; green, 0; blue, 0 }  ,fill opacity=1 ] (345.88,192.5) .. controls (345.88,191.12) and (347,190) .. (348.38,190) .. controls (349.76,190) and (350.88,191.12) .. (350.88,192.5) .. controls (350.88,193.88) and (349.76,195) .. (348.38,195) .. controls (347,195) and (345.88,193.88) .. (345.88,192.5) -- cycle ;

  \end{tikzpicture}
  \caption{\label{fig:T3-HM} An $SU(3)$-type class-$\mathcal{S}$ theory at genus-zero with three maximal (circle) and a minimal puncture (dot).}
\end{figure}
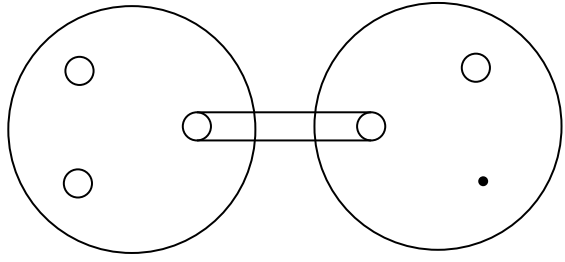

Now that gauging a $SU(3)$ symmetry with fugacities $c^{(2)}$ can be carried out using the integration formula, we are able to also compute Schur index of some non-Lagrangian theories. Consider the $E_6$ superconformal field theory of Minahan and Nemeschansky \cite{Minahan:1996fg}, whose index can be computed by exploiting the Argyres-Seiberg duality \cite{Argyres:2007cn} and an inversion formula \cite{Gadde:2010te,Razamat:2012uv},
\begin{align}
  & \ \mathcal{I}_{E_6}(\mathbf{c}^{(1)}, \mathbf{c}^{(2)}, (wr, w^{-1}r, r^{-2})) \nonumber\\
  = & \ \frac{\mathcal{I}_{\text{SQCD}}(\mathbf{c}^{(1)}, \mathbf{c}^{(2)}, \frac{w^{\frac{1}{3}}}{r}, \frac{w^{- \frac{1}{3}}}{r})_{w \to q^{\frac{1}{2}}w}}{\theta(w^2)} + \frac{\mathcal{I}_{\text{SQCD}}(\mathbf{c}^{(1)}, \mathbf{c}^{(2)}, \frac{w^{\frac{1}{3}}}{r}, \frac{w^{- \frac{1}{3}}}{r})_{w \to q^{ - \frac{1}{2}}w}}{\theta(w^{ - 2})} \ ,
\end{align}
where the denominator is related to $\vartheta_1$ by
\begin{equation}
  \theta(z) \equiv \frac{\vartheta_1( \mathfrak{z})}{i z^{\frac{1}{2}} q^{\frac{1}{8}} (q;q)} \ .
\end{equation}
With the known closed-form expression of $\mathcal{I}_\text{SQCD}$, the above formula also provides a closed-form $\mathcal{I}_{E_6}$\footnote{Unfortunately, the current closed-form $\mathcal{I}_{E_6}$ only makes manifest the $SU(3)_a \times SU(3)_b\subset E_6$ symmetry. It would be interesting to further explore a better formula with explicit $E_6$ Weyl-invariance.}. Note that two of the $SU(3)$ flavor symmetries of the $E_6$ theory share identical fugacities $\mathbf{c}^{(1)}, \mathbf{c}^{(2)}$ with those of the $SU(3)$ SQCD. The above formula allows one to directly compute the Schur index of, for instance, a theory of class-$\mathcal{S}$ with three maximal and one minimal punctures (see Figure \ref{fig:T3-HM}),
\begin{align}
  \mathcal{I}
  = \oint \frac{d\mathbf{a}}{2\pi i \mathbf{a}} \sum_{\pm} 
  \frac{\mathcal{I}_{\text{SQCD}}(\mathbf{c}^{(1)}, \mathbf{a}^{-1}, \frac{w^{\frac{1}{3}}}{r}, \frac{w^{- \frac{1}{3}}}{r})_{w \to q^{\pm \frac{1}{2}}w}}{\theta(w^{\pm 2})} \mathcal{I}_\text{VM}(\mathbf{a}) \mathcal{I}_\text{HM}(\mathbf{a}, c^{(3)}, d^{(3)}) \ .
\end{align}
As we have argued, this can be computed exactly with the currently available formula in the appendix \ref{app:integration-formula}.

\subsection{\texorpdfstring{$\mathcal{N}=4$ $SO(7)$ SYM}{}}

The Schur index of the $\mathcal{N} = 4$ $SO(7)$ theory is a contour integral of the following integrand,
\begin{align}
\mathcal{Z}\left(a_1,a_2,a_3\right)=\left(\frac{\vartheta_1^{\prime}(0)}{\vartheta_4(0)}\right)^3\prod_{\alpha,\beta}\prod_{i<j}\frac{\vartheta_1(\alpha\mathfrak{a}_i+\beta\mathfrak{a}_j,q)}{\vartheta_4(\alpha\mathfrak{a}_i+\beta\mathfrak{a}_j+\mathfrak{b},q)}\prod_{\alpha}\prod_{i=1}^3 \frac{\vartheta_1(\alpha\mathfrak{a}_i,q)}{\vartheta_4(\alpha\mathfrak{a}_i+\mathfrak{b},q)} \ ,
\end{align}
which is separately elliptic with respect to all three variables $\mathfrak{a}_{1,2,3}$.

The integral can be performed analytically by integrating $a_1, a_2, a_3$ one after another using the integration formula collected in the appendix \ref{app:integration-formula}. The $a_1$ integration involves the following simple poles which are all imaginary,
\begin{align}
\alpha\mathfrak{b}+\frac{\tau}{2},
\quad \alpha\mathfrak{a}_2+\beta\mathfrak{b}+\frac{\tau}{2},
\quad \alpha\mathfrak{a}_3+\beta\mathfrak{b}+\frac{\tau}{2} \ .
\end{align}
The residues of these simple poles are denoted by $\mathcal{P}_{\alpha}$, $\mathcal{Q}_{\alpha\beta}$ and $\tilde{\mathcal{Q}}_{\alpha\beta}$.
Using the integration formula (\ref{integration-formula-f}), the $a_1$ integration leaves an integrand
\begin{align}
\mathcal{Z}_1(a_2,a_3)
= & \ \oint_{|a_1|=1}\frac{da_1}{2\pi i a_1}\mathcal{Z}(a_1,a_2,a_3) \nonumber\\
= & \ \sum_{\alpha}\mathcal{P}_\alpha E_1\begin{bmatrix}
-1\\
b^\alpha
\end{bmatrix}+\sum_{\alpha,\beta}\mathcal{Q}_{\alpha\beta}E_1\begin{bmatrix}
-1\\
a_2^{\alpha}b^{\beta}
\end{bmatrix}+\sum_{\alpha,\beta}\tilde{\mathcal{Q}}_{\alpha\beta}E_1\begin{bmatrix}
-1\\
a_3^{\alpha}b^\beta \
\end{bmatrix}.
\end{align}
The poles and residues of $\mathcal{P}, \mathcal{Q}, \tilde {\mathcal{Q}}$ are listed in Table \ref{poles-residues-SO(7)}, which are used in the $a_2$-integration.
{
  \renewcommand{\arraystretch}{1.8}
  \begin{table}[h!]
    \centering
    \begin{tabular}{c|c|c}
      & poles  & residues  \\
      \hline
      $\mathcal{P}_{\alpha}$ & $2\gamma\mathfrak{b}$ & $\mathcal{P}_{\alpha\gamma}$\\
      &  $\frac{1}{2}\left(2\gamma\mathfrak{a}_3+2\beta\mathfrak{b}+\tau\right)$    &  $\mathcal{P}_{\alpha\beta\gamma}$   \\
      \hline
       $\mathcal{Q}_{\alpha\beta}$  & $\frac{k}{2}+\frac{\ell}{2}\tau$,$\quad$  $(k,\ell) = \{(0,1),(1,0),(1,1)\}$ & $ \mathcal{Q}_{\alpha\beta}^{(k,\ell)} $\\
      & $-\alpha\beta\mathfrak{b}+\frac{k}{2}+\frac{\ell}{2}\tau$, $\quad $ $\{(k,\ell)\}=\{(0,0),(1,0),(1,1)\}$  & $-\mathcal{Q}_{\alpha\beta}^{(k,\ell+1)}$ \\
      & $\alpha\beta\mathfrak{b}+\frac{\tau}{2}$  & $\gamma\mathcal{P}_{\alpha\gamma}$ \\
      & $\gamma\mathfrak{a}_3+\alpha\beta\mathfrak{b}+\frac{\tau}{2}$ &  $\mathcal{Q}_{\alpha\beta\gamma}$   \\
      &   $\gamma\mathfrak{a}_3-2\alpha\beta\mathfrak{b}$           &  $\mathcal{Q}_{\alpha\beta -\gamma}$  \\
      \hline
      $\tilde{\mathcal{Q}}_{\alpha\beta}$ &    $\gamma\mathfrak{b}+\frac{\tau}{2}$    &    $-\mathcal{P}_{\gamma\beta\alpha}$   \\
      &  $-\gamma\mathfrak{a}_3+\alpha\beta\gamma\mathfrak{b}+\frac{\tau}{2}$  &   $\tilde{\mathcal{Q}}_{\alpha\beta\gamma}$   \\
      &  $\gamma(-\mathfrak{a}_3-2\alpha\beta\mathfrak{b})$ &  $-\gamma\mathcal{Q}_{\alpha\beta 1}$   
    \end{tabular}
    \caption{Poles and residues of $\mathcal{P}_{\alpha}$, $\mathcal{Q}_{\alpha\beta}$ and $\tilde{\mathcal{Q}}_{\alpha\beta}$ with respect to the variable $\mathfrak{a}_2$. Here $\alpha, \beta, \gamma=\pm 1$.\label{poles-residues-SO(7)}}
  \end{table}
}

Using the integration formula (\ref{integration-formula-fE-1}), the $a_2$ integration leaves a final integrand
\begin{align}
\mathcal{Z}_2(a_3)=\oint \frac{da_2}{2\pi i a_2}\mathcal{Z}_1(a_2,a_3)= I_1+ I_2 +I_3,
\end{align}
where 

\begin{align}
I_1=&\ \sum_{\alpha=\pm 1}\oint\frac{da_2}{2\pi i a_2}\mathcal{P}_{\alpha}E_1\begin{bmatrix}
-1\\
b^\alpha
\end{bmatrix}\nonumber\\
= & \sum_{\alpha, \gamma=\pm 1}\mathcal{P}_{\alpha\gamma}E_1\begin{bmatrix}
-1\\
b^{2\gamma-1}q^{\frac{1}{2}}
\end{bmatrix}E_1\begin{bmatrix}
-1\\
b^\alpha
\end{bmatrix}+\sum_{\alpha, \beta,\gamma=\pm 1}\mathcal{P}_{\alpha\beta\gamma}E_1\begin{bmatrix}
-1\\
a_3^\gamma b^{\beta-1}
\end{bmatrix}E_1\begin{bmatrix}
-1\\
b^\alpha
\end{bmatrix}.
\end{align}
and 
\begin{align}
I_2= & \ \sum_{\alpha=\pm 1}\oint \frac{da_2}{2\pi i a_2}\mathcal{Q}_{\alpha\beta}E_1\begin{bmatrix}
  -1\\
  a_2^{\alpha} b^{\beta}
\end{bmatrix}=\sum_{\alpha=\pm 1}\sum_{k,\ell,m,n=0}^{1}(-1)^n \alpha \mathcal{Q}^{(k,\ell)}_{\alpha\beta}\mathcal{S}_{m}E_1\begin{bmatrix}
(-1)^m\\
(-1)^k b^{(n-m)\alpha\beta}q^{\frac{n+(-1)^n \ell}{2}}
\end{bmatrix}\nonumber\\
& \ -\sum_{\alpha,\beta,\gamma=\pm 1}\sum_{k=0}^1 \alpha\mathcal{Q}_{\alpha\beta\gamma}\mathcal{S}_{k}\left(E_{2-k}\begin{bmatrix}
(-1)^k\\
a_3^{-\gamma}b^{-(k+1)\alpha\beta}q^{\frac{1}{2}}
\end{bmatrix}+E_{2-k}\begin{bmatrix}
(-1)^k\\
a_3^\gamma b^{(2-k)\alpha\beta}
\end{bmatrix}\right)\nonumber\\
& \ -\sum_{\alpha,\beta=\pm 1}\sum_{k=0}^1 \alpha\gamma \mathcal{P}_{\alpha\gamma}\mathcal{S}_k E_{2-k}\begin{bmatrix}
(-1)^k\\
b^{(2-k)\alpha\beta}
\end{bmatrix},
\end{align}
and
\begin{align}
I_3= & \ \sum_{\alpha,\beta=\pm 1}\oint \frac{da_2}{2\pi i a_2}\tilde{Q}_{\alpha\beta}E_1\begin{bmatrix}
-1\\
a_3^{\alpha}b^\beta
\end{bmatrix}\nonumber\\
= & \ \sum_{\alpha,\beta=\pm 1}\left.\tilde{Q}_{\alpha\beta}\right|_{\mathfrak{a}_2=0} E_1\begin{bmatrix}
-1\\
a_3^{\alpha}b^\beta
\end{bmatrix}-\sum_{\alpha,\beta,\gamma=\pm 1}\gamma \mathcal{Q}_{\alpha\beta 1}E_1\begin{bmatrix}
-1\\
a_3^{-\gamma}b^{-2\alpha\beta\gamma}q^{\frac{1}{2}}
\end{bmatrix}E_1\begin{bmatrix}
-1\\
a_3^{\alpha}b^\beta
\end{bmatrix}\notag\\
& \ -\sum_{\alpha,\beta,\gamma=\pm 1}\mathcal{P}_{\gamma\beta\alpha}E_1\begin{bmatrix}
-1\\
b^\gamma
\end{bmatrix}E_1\begin{bmatrix}
-1\\
a_3^{\alpha}b^\beta
\end{bmatrix}+\sum_{\alpha,\beta,\gamma=\pm 1}\tilde{\mathcal{Q}}_{\alpha\beta\gamma}E_1\begin{bmatrix}
-1\\
a_3^{-\gamma}b^{\alpha\beta\gamma}
\end{bmatrix}E_1\begin{bmatrix}
-1\\
a_3^{\alpha}b^\beta
\end{bmatrix}.
\end{align}
The closed-form of $\mathcal{N}=4$ $SO(7)$ Schur index is then given by
\begin{align}
\mathcal{I}=\oint_{|a_3|=1}\frac{da_3}{2\pi i a_3}\mathcal{Z}_2(a_3).
\end{align}
At this stage we encounter the following types of integral
\begin{align}
\oint_{|z|=1}\frac{dz}{2\pi i z}f(z)E_k\begin{bmatrix}
\pm 1\\
z a
\end{bmatrix}, \quad \oint_{|z|=1}\frac{dz}{2\pi i z}f(z)E_1\begin{bmatrix}
-1\\
z a
\end{bmatrix}E_1\begin{bmatrix}
-1\\
z b
\end{bmatrix} \ ,
\end{align}
which can be computed using the integration formula (\ref{integration-formula-fE-1}), (\ref{integration-formula-fE-2}), and (\ref{integration-formula-fEE-1}). Since the computation of integrand is somewhat technical and tedious, we will only present the final result without the details. To do so, we define $\mathbf{I}_{1,2,3}$ as some intricate combinations of Eisenstein series,
\begin{align}
(\mathbf{I}_1)_{\alpha\beta} \coloneqq & \ E_1\begin{bmatrix}
-1\\
b
\end{bmatrix}\left(-\sum_{k=0}^2 (-1)^{\lceil\frac{k}{2}\rceil}(3-k)E_2\begin{bmatrix}
(-1)^{\beta+k+1}\\
(-1)^\alpha b^k
\end{bmatrix}\right. \nonumber\\
&\ \left. +(-1)^\beta\left(\sum_{\pm 1}\mp E_1\begin{bmatrix}
-1\\
(-1)^\alpha b^{\frac{1}{2}} q^{\pm\frac{1}{4}}
\end{bmatrix}+\frac{1}{4}\right)\right),
\end{align}
and
\begin{align}
(\mathbf{I}_2)_{\beta\gamma} \coloneqq & \ 4(-1)^\beta \left(3E_3\begin{bmatrix}
(-1)^{\beta+\gamma+1}\\
(-1)^\beta b^2
\end{bmatrix}-3 E_3\begin{bmatrix}
(-1)^{\beta+\gamma}\\
(-1)^\beta b
\end{bmatrix}+E_3\begin{bmatrix}
(-1)^{\beta+\gamma}\\
(-1)^\beta b^3
\end{bmatrix}\right) \nonumber\\
& \ -2 \sum_{k=0}^1 (-1)^{\beta+k} E_1\begin{bmatrix}
-1\\
b
\end{bmatrix}E_2\begin{bmatrix}
(-1)^{\beta+\gamma}\\
(-1)^\beta b^{2k+1}
\end{bmatrix}-4(-1)^\beta E_2\begin{bmatrix}
1\\
b^2
\end{bmatrix} E_1\begin{bmatrix}
(-1)^{\beta+\gamma+1}\\
(-1)^\beta b^2
\end{bmatrix} \nonumber \\
& \ +2(-1)^\beta E_1\begin{bmatrix}
(-1)^{\beta+\gamma+1}\\
(-1)^\beta b^2
\end{bmatrix}\sum_{k\in\{0,1,3\}}\left(4-k\right)(-1)^{\lceil\frac{k}{2}\rceil}E_2\begin{bmatrix}
(-1)^{\beta+\gamma+k}\\
b^k
\end{bmatrix}\\
& \ +\sum_{\pm}\sum_{k=0}^1 E_1\begin{bmatrix}
  1\\
  (-1)^\beta b^{\frac{1}{2}+k} q^{\pm\frac{(-1)^k}{4}}
\end{bmatrix}\left( 2(-1)^\beta  \left(E_2\begin{bmatrix}
1\\
b^2
\end{bmatrix}-E_2\begin{bmatrix}
-1\\
b
\end{bmatrix}\right)\pm (-1)^\gamma E_1\begin{bmatrix}
-1\\
b
\end{bmatrix}\right) \nonumber\\
& \ -\frac{5+4\gamma -4\beta}{2}E_1\begin{bmatrix}
(-1)^{\beta+\gamma+1}\\
(-1)^\beta b^2
\end{bmatrix}+5\beta\gamma E_1\begin{bmatrix}
-1\\
-b^2
\end{bmatrix} \nonumber\\
& \ +\sum_{\pm}\sum_{k=0}^1 \frac{(-1)^\beta}{4}E_1\begin{bmatrix}
1\\
(-1)^\beta b^{\frac{1}{2}+k} q^{\pm\frac{(-1)^k}{4}}
\end{bmatrix}, \nonumber
\end{align}
and
\begin{align}
\mathbf{I}_3 \coloneqq & \ -4\left(-\sum_{k=0}^1\sum_{\alpha=\pm 1}E_1\begin{bmatrix}
1\\
b^{k+\frac{3}{2}}q^{\frac{\alpha}{4}}
\end{bmatrix}+E_1\begin{bmatrix}
-1\\
b^5
\end{bmatrix}\right)\left(\sum_{i=1}^2 (-1)^{i+1}E_2\begin{bmatrix}
(-1)^i\\
b^i
\end{bmatrix}-\frac{1}{8}\right) \nonumber\\
& \ -2\sum_{k=1}^2 E_1\begin{bmatrix}
-1\\
b^{2k+1}
\end{bmatrix}E_1\begin{bmatrix}
-1\\
b
\end{bmatrix}^2+E_1\begin{bmatrix}
-1\\
b^3
\end{bmatrix}\left(\sum_{k\in\{1,2,4\}}B_k E_2\begin{bmatrix}
(-1)^k\\
b^k
\end{bmatrix}+4\right) \nonumber\\
& \ +E_1\begin{bmatrix}
-1\\
b
\end{bmatrix}\left(\sum_{\alpha=\pm 1}2\alpha \left(\sum_{k=1}^2 k E_1\begin{bmatrix}
(-1)^{k+1}\\
b^{\frac{5}{2}}q^{\frac{\alpha}{4}}
\end{bmatrix}- E_1\begin{bmatrix}
1\\
b^{\frac{3}{2}}q^{\frac{\alpha}{4}}
\end{bmatrix}\right)+\sum_{k=1}^3 C_k E_2\begin{bmatrix}
(-1)^k\\
b^k
\end{bmatrix}+\frac{9}{2} \right) \nonumber\\
& \ -2\prod_{k=0}^2 E_1\begin{bmatrix}
-1\\
b^{2k+1}
\end{bmatrix}-2 E_1\begin{bmatrix}
-1\\
b
\end{bmatrix}^3+\sum_{n=1}^4 A_n E_3\begin{bmatrix}
(-1)^n\\
b^n
\end{bmatrix}.
\end{align}
With the above three definitions, we can express the $\mathcal{N} = 4$ $SO(7)$ Schur index as
\begin{align}
\mathcal{I}_{\mathcal{N} = 4 \ SO(7)}
= \frac{i}{48}\sum_{\alpha,\beta}\left(\mathbf{R}_{\alpha\beta}(\mathbf{I}_1)_{\alpha\beta}+\mathbf{T}_{\alpha\beta}(\mathbf{I}_2)_{\alpha\beta}\right)+\frac{i}{48}\mathbf{W}\mathbf{I}_3.
\end{align}
In this formula, the Greek indices $(\alpha,\beta)$ sum over the set $\{(0,0),(1,0),(1,1)\}$.
Note that
\begin{align}
&\mathbf{R}_{\alpha\beta} \coloneqq b^{-2\beta}\frac{\vartheta_1\left(\mathfrak{b}+\frac{\alpha+\beta\tau}{2}\right)\vartheta_4\left(\frac{\alpha+\beta\tau}{2}\right)\vartheta_4(\mathfrak{b})^3}{\vartheta_1\left(3\mathfrak{b}+\frac{\alpha+\beta\tau}{2}\right)\vartheta_4\left(2\mathfrak{b}+\frac{\alpha+\beta\tau}{2}\right)\vartheta_1(2\mathfrak{b})^3},\\
&\mathbf{T}_{\beta\gamma} \coloneqq  b^{\gamma-\beta}\frac{\vartheta_4\left(\frac{\beta+(\beta-\gamma)\tau}{2}\right)\vartheta_4\left(2\mathfrak{b}+\frac{\beta+(\beta-\gamma)\tau}{2}\right)\vartheta_4(\mathfrak{b})}{\vartheta_1\left(3\mathfrak{b}+\frac{\beta+(\beta-\gamma)\tau}{2}\right)\vartheta_1\left(\mathfrak{b}+\frac{\beta+(\beta-\gamma)\tau}{2}\right)\vartheta_1(4\mathfrak{b})}
\quad
\mathbf{W} \coloneqq  \prod_{k=0}^2\frac{\vartheta_4\left((2k+1)\mathfrak{b}\right)}{\vartheta_1\left((2k+2)\mathfrak{b}\right)}.
\end{align}

Finally, it is straightforward to check that the unflavored limit of $\mathcal{I}_{\mathcal{N} = 4 \ SO(7)}$ satisfies a monic $\Gamma^0(2)$ modular differential equation at order 10,
\begin{footnotesize}
\begin{align}
  &\mathcal{D}^{\mathcal{N}=4}_{\mathfrak{so}(7)}=\mathcal{D}_q^{(10)}+\left(\frac{64169}{45888}\Theta_{1,1}-\frac{116269}{45888}\Theta_{0,2}\right)\mathcal{D}_q^{(8)}+\left(\frac{99455}{68832}\Theta_{0,3}-\frac{51397}{22944}\Theta_{1,2}\right)\mathcal{D}_q^{(7)}+\left(\frac{4531009 \Theta_{0,4}}{13215744}\right.\nonumber\\
  &\left.-\frac{3779273\Theta_{1,3}}{3303936}+\frac{5245697\Theta_{2,2}}{4405248}\right)\mathcal{D}^{(6)}_q+\left(-\frac{1133653\Theta_{0,5}}{4405248}+\frac{2557903\Theta_{1,4}}{4405248}-\frac{55973\Theta_{2,3}}{2202624}\right)\mathcal{D}^{(5)}_q\nonumber\\
  &+\left(\frac{1190885473\Theta_{0,6}}{22836805632}-\frac{924970757\Theta_{1,5}}{3806134272}+\frac{3937715525\Theta_{2,4}}{7612268544}-\frac{2505775369\Theta_{3,3}}{11418402816}\right)\mathcal{D}_q^{(4)}+\left(-\frac{117336059\Theta_{0,7}}{22836805632}\right.\nonumber\\
  &\left.+\frac{2991097351\Theta_{1,6}}{22836805632}-\frac{380366011\Theta_{2,5}}{2537422848}+\frac{930902663\Theta_{3,4}}{22836805632}\right)\mathcal{D}_q^{(3)}+\left(-\frac{274137107749\Theta_{0,8}}{26308000088064}-\frac{3736889371\Theta_{1,7}}{3288500011008}\right.\nonumber\\
  &\left.-\frac{330134662435\Theta_{2,6}}{6577000022016}+\frac{199201642115\Theta_{3,5}}{3288500011008}+\frac{41820786289\Theta_{4,4}}{26308000088064}\right)\mathcal{D}_q^{(2)}+\left(\frac{240693275531\Theta_{0,9}}{39462000132096}\right.\nonumber\\
  &\left.-\frac{88992212869\Theta_{1,8}}{4384666681344}-\frac{8099874757\Theta_{2,7}}{1096166670336}+\frac{330064570085\Theta_{3,6}}{3288500011008}-\frac{174451260571\Theta_{4,5}}{2192333340672}\right)\mathcal{D}_q^{(1)}+\left(-\frac{256921875\Theta_{0,10}}{256624295936}\right.\nonumber\\
  &\left.+\frac{477416835\Theta_{1,9}}{128312147968}-\frac{59821335\Theta_{2,8}}{256624295936}+\frac{559460601\Theta_{3,7}}{32078036992}-\frac{13109319531\Theta_{4,6}}{128312147968}+\frac{10552431897\Theta_{5,5}}{128312147968}\right),
\end{align}
\end{footnotesize}
and non-monic $\Gamma^0(2)$ equation at order 9,
\begin{footnotesize}
\begin{align}
&\mathcal{D}_{\mathfrak{so}(7)}^{\mathcal{N}=4}=\Theta_{0,1}\mathcal{D}_q^{(9)}+\left(\frac{2477\Theta_{1,1}}{1912}-\frac{3407\Theta_{0,2}}{1912}\right)\mathcal{D}_q^{(8)}+\left(\frac{6971\Theta_{1,2}}{11472}-\frac{10377\Theta_{0,3}}{3824}\right)\mathcal{D}_q^{(7)}+\left(\frac{1339781\Theta_{0,4}}{275328} \right.\nonumber\\
&\left.-\frac{15625\Theta_{1,3}}{1434}+\frac{1767811\Theta_{2,2}}{275328}\right)\mathcal{D}_q^{(6)}+\left(\frac{17516635\Theta_{0,5}}{13215744}-\frac{37436353\Theta_{1,4}}{13215744}+\frac{12423443\Theta_{2,3}}{6607872}\right)\mathcal{D}_q^{(5)}+\nonumber\\
&\left(-\frac{44438921\Theta_{0,6}}{317177856}+\frac{24643855\Theta_{1,5}}{52862976}-\frac{177147775\Theta_{2,4}}{35241984}+\frac{767384147\Theta_{3,3}}{158588928}\right)\mathcal{D}_q^{(4)}+\left(\frac{436377635\Theta_{0,7}}{11418402816}\right.\nonumber\\
&\left.-\frac{2631122143\Theta_{1,6}}{11418402816}+\frac{6486893273\Theta_{2,5}}{3806134272}-\frac{17028452303\Theta_{3,4}}{11418402816}\right)\mathcal{D}_q^{(3)}+\left(\frac{287431763\Theta_{0,8}}{30449074176}+\frac{4632486095\Theta_{1,7}}{22836805632}\right.\nonumber\\
&\left.-\frac{3490714387\Theta_{2,6}}{2537422848}+\frac{19973363051\Theta_{3,5}}{7612268544}-\frac{133319985805\Theta_{4,4}}{91347222528}\right)\mathcal{D}_q^{(2)}+\left(-\frac{247722449497\Theta_{0,9}}{26308000088064}\right.\nonumber\\
&\left.+\frac{663592423\Theta_{1,8}}{36087791616}-\frac{105995132111\Theta_{2,7}}{243592593408}+\frac{2910273174797\Theta_{3,6}}{2192333340672}-\frac{439516599949\Theta_{4,5}}{487185186816}\right)\mathcal{D}_q^{(1)}+\left(\frac{49415625\Theta_{0,10}}{32078036992}\right.\nonumber\\
&\left.-\frac{52688223\Theta_{1,9}}{16039018496}-\frac{2541812427\Theta_{2,8}}{32078036992}+\frac{3001266891\Theta_{3,7}}{4009754624}-\frac{36356798079\Theta_{4,6}}{16039018496}+\frac{25650617139\Theta_{5,5}}{16039018496}\right).
\end{align}
\end{footnotesize}
In the above modular linear differential equations, the operator $D_q^{(n)}$ is the so called Serre derivative given by
\begin{equation}
  D_q^{(n)} \coloneqq \partial_q^{(2n - 2)} \circ \cdots \partial_q^{(2)} \circ \partial_q^{(0)} , \qquad \partial_q^{(k)} \coloneqq q\partial_q + k E_2(\tau) \ ,
\end{equation}
and the functions $\Theta_{(m,n)} \coloneqq \vartheta_2^{4m}\vartheta_3^{4n} + \vartheta_2^{4n}\vartheta_3^{4m}$ are weight $2(m+n)$ $\Gamma^0(2)$ modular forms \cite{Beem:2017ooy}.
We also note that each row vector $(\mathbf{R}_{00},\mathbf{R}_{10},\mathbf{R}_{11})$, and $(\mathbf{T}_{00},\mathbf{T}_{10},\mathbf{T}_{11})$ forms the same 3-dimensional representation $\rho$ of $\Gamma^0(2)$, following from the modularity of Jacobi-theta function. In particular, the representation matrix of $STS$ is given by
\begin{align}
\rho(STS)=\left(\begin{array}{ccc}
-1 & 0 & 0\\
0  & 0 & -1\\
0  & -1 & 0
\end{array}\right),
\end{align}
acting to these two row vectors. The factor $\mathbf{W}$ form a one-dimensional representation of $\Gamma^0(2)$. It would be interesting to further investigate the relation between the many ingredients in the above closed-form of $\mathcal{I}_{\mathcal{N} = 4 \ SO(7)}$ and the highest weight characters of the associated chiral algebra $\mathbb{V}_{\mathcal{N} = 4 \ SO(7)}$, which we leave for future work.

\section{\texorpdfstring{Line operator index of $A_1$-theories of class-$\mathcal{S}$}{}\label{section:Wilson-index-A1-theories}}

In this and the following section we discuss the Schur index in the presence of a line operator. For a Lagrangian 4d $\mathcal{N} = 2$ SCFT with gauge group $G$ and flavor group $f$, the Schur index in the absence of operator insertion can be computed by a multivariate contour integral \cite{Gadde:2011uv,Beem:2013sza}
\begin{align}
  \mathcal{I} = \oint \left[\frac{da}{2\pi i a}\right] \mathcal{Z}(a, b) \ ,
\end{align}
where the integrand $\mathcal{Z}(a, b)$ is elliptic with respect to the ``exponent variables'' $\mathfrak{a}_i$ separately, and captures contributions from the vector multiplets and hypermultiplets in a gauge theory description. Variables $b$ denote the flavor fugacities with respect to the flavor symmetry $f$.

One can introduce half line operators in the 4d theory that extend from the origin to infinity while preserving certain amount of supercharges \cite{Cordova:2016uwk}. In particular, there are line operators that preserve the supercharges used to construct the Schur index. In the presence of such a BPS half Wilson line operator in the representation $\mathcal{R}$ of the gauge group, the half Wilson line index can be computed simply by\footnote{For simplicity we omit the normalization factor $\mathcal{I}^{-1}$.} \cite{Gang:2012yr,Cordova:2016uwk}
\begin{align}
  \langle W_{\mathcal{R}}\rangle = \oint \left[\frac{da}{2\pi i a}\right]
  \chi_\mathcal{R}(a) \mathcal{Z}(a) \ ,
\end{align}
where $\chi_\mathcal{R}(a)$ denotes the character of representation $\mathcal{R}$ of $G$. A full Wilson line operator in representation $\mathcal{R}$ can be thought of as a junction at the origin of two half Wilson line operators in complex-conjugating representation $\mathcal{R}, \overline{\mathcal{R}}$, and hence the full Wilson line index can be computed by
\begin{align}
  \langle W_{\mathcal{R}}^\text{full}\rangle = \oint \left[\frac{da}{2\pi i a}\right]
  \chi_\mathcal{R}(a)\chi_{\overline {\mathcal{R}}}(a) \mathcal{Z}(a) \ .
\end{align}
In our notation, we will only add the superscript ``full'' when dealing with a full Wilson line operator.

One can also consider correlators of half Wilson line operators, which take the form
\begin{align}
  \langle W_{\mathcal{R}_1} \cdots W_{\mathcal{R}_n}\rangle
  = \oint \bigg[\frac{da}{2\pi i a}\bigg]
  \bigg[\prod_{i=1}^n\chi_{\mathcal{R}_i}(a)\bigg]
  \mathcal{Z}(a)\ .
\end{align}
One can consider applying the tensor product decomposition $\otimes_{i = 1}^n \mathcal{R}_i = \sum_{j} m_j \mathcal{R}^{(j)}$ and reduce the product of characters on the right to a sum of characters of the irreducible representations $\mathcal{R}^{(j)}$ of the gauge group,
\begin{align}
  \langle W_{\mathcal{R}_1} \cdots W_{\mathcal{R}_n}\rangle = \sum_{j} m_j \langle W_{\mathcal{R}^{(j)}} \rangle \ .
\end{align}
In this sense, half Wilson line indices in irreducible representations are the basic building blocks for correlators of half/full Wilson line, which will be our main focus.

In the following we will study line operator index for $A_1$ theories of class-$\mathcal{S}$. We will start with some simple examples where we are able to compute both the Wilson line index and the $S$-dual 't Hooft line index. Eventually we will analyze in detail the half Wilson line index for general $A_1$ theories of class-$\mathcal{S}$.

In many cases, we are able to expand the Wilson line operator index as a linear combination of chiral algebra module characters. At the computational level, these characters come from residues of the elliptic integrand $\mathcal{Z}$ which are related to Gukov-Witten type surface defects. As already discussed \cite{Cordova:2016uwk,Neitzke:2017cxz}, the appearance of non-vacuum chiral algebra modules is somewhat expected. Recall that the associated chiral algebra of a Lagrangian theory can be constructed using a set of small $bc$ ghost\footnote{Smallness means the zero mode $c_0$ of the $c$ ghost is removed from the algebra.} and symplectic bosons $\beta \gamma$ through a BRST reduction that imposes gauge-invariance. Let us denote the vacuum character of the $bc \beta \gamma$ system as $\mathcal{Z}_{bc \beta \gamma}$. The Wilson line index in an irreducible $G$-representation $\mathcal{R}$ can be written more explicitly as
\begin{equation}
  \oint \left[\frac{da}{2\pi i a}\right]_\text{Haar} \chi_\mathcal{R}(a) \mathcal{Z}_{bc\beta \gamma}(a) \ ,
\end{equation}
Hence the Wilson index account for the local operators formed from the normal ordered product of the $bc \beta \gamma$ that are gauge-variant and can compensate the charge $\mathcal{R}$ at the end of the Wilson line. These operators are acted on by the chiral algebra $\mathbb{V}(\mathcal{T})$ and naturally form a reducible module $\mathcal{R}^* \otimes M(\mathcal{R})$ of $G \times \mathbb{V}(\mathcal{T})$, since the operators in $\mathbb{V}(\mathcal{T})$ are gauge-invariant under $G$. The Wilson index then picks up a trace over the reducible module $M$,
\begin{equation}
  \langle W_\mathcal{R} \rangle = \operatorname{tr}_{M(\mathcal{R})} q^{L_0 - \frac{c}{24}}b^f \ .
\end{equation}
In general, $M(\mathcal{R})$ may be decomposed as an infinite tower of irreducible highest weight modules $M_j$ of $\mathbb{V}(\mathcal{T})$. Therefore it is natural to expect that the trace returns a weighted sum of irreducible characters $\operatorname{ch}M_j$
\begin{equation}
  \langle W_\mathcal{R}\rangle = \sum_j L_j(b, q) \operatorname{ch}M_j \ ,
\end{equation}
where $L_j(b, q)$ are rational functions of the flavor fugacities and $q$. These irreducible modules often arise from different types of surface defects in the 4d $\mathcal{N} = 2$ SCFT \cite{Cordova:2017mhb,Beem:2017ooy,Nishinaka:2018zwq,Bianchi:2019sxz,Pan:2021ulr,Zheng:2022zkm}, and (linear combinations of) the module characters correspond to the defect Schur index. For more general line insertion, it is less obvious how the the chiral algebra modules arise in the line index. One may argue from a pure two dimensional perspective. A temporal line $\mathcal{L}$ insertion into the trace gives a trace over the Hilbert space $\mathcal{H}_\mathcal{L}$ of the line $\mathcal{L}$,
\begin{equation}
  \operatorname{tr}_{\mathcal{H}_\mathcal{L}} q^{L_0 - \frac{c}{24}} \ .
\end{equation}
When the line is topological (or, commute with the chiral algebra at hand), the line Hilbert space $\mathcal{H}_\mathcal{L}$ can be decomposed into chiral algebra modules. This is the case for the Verlinde lines in rational theories, and the corresponding trace has been shown to be expanded in characters of the primaries \cite{Cordova:2016uwk,Chang:2018iay}. In any case, it would be interesting to further investigate the precise origin of the surface defect index's appearance in line operator index, as well as the physical meaning of the rational functions $L_j(b, q)$.

\subsection{\texorpdfstring{$\mathcal{N} = 4 $  $ SU(2)$ theory}{}\label{section:N4SU(2)}}

\subsubsection{Half Wilson line index}

The associated chiral algebra $\mathbb{V}_{\mathcal{N} = 4}$ of the $\mathcal{N} = 4$ theory with an $SU(2)$ gauge group is given by the 2d small $\mathcal{N} = 4$ superconformal algebra. The Schur index, which is identified with the vacuum character of $\mathbb{V}_{\mathcal{N} = 4}$, can be computed by the contour integral
\begin{align}
  \mathcal{I}_{\mathcal{N} = 4} 
  = & \ - \frac{1}{2}\frac{\eta(\tau)^3}{\vartheta_4(\mathfrak{b})}
  \oint_{|a| = 1} \frac{da}{2\pi i a} 
  \frac{
    \vartheta_1(2\mathfrak{a})\vartheta_1(- 2\mathfrak{a})
  }{
    \vartheta_4(2\mathfrak{a} + \mathfrak{b})
    \vartheta_4(-2\mathfrak{a} + \mathfrak{b})
  }
  \coloneqq \oint \frac{da}{2\pi i a} \mathcal{Z}(a)\\
  = & \ \frac{i\vartheta_4(\mathfrak{b})}{\vartheta_1(2 \mathfrak{b})} E_1 \begin{bmatrix}
    -1 \\ b  
  \end{bmatrix} \ . \nonumber
\end{align}
In the following we consider the index in the presence of a half Wilson line operator in the spin-$j$ representation. The index is then given by the integral
\begin{align}
  \langle W_j\rangle =
  \oint_{|a| = 1} \frac{da}{2\pi i a} \left[\sum_{m = - j}^{j} a^{2m}\right]
  \mathcal{Z}(a)\ .
\end{align}
Here the spin-$j$ character is given by $\chi_j(a) = \sum_{m = -j}^j a^{2m}$, and in particular, the adjoint character is $a^2 + 1 + a^{-2}$.

To proceed, we note that there are a collection of poles from the elliptic integrand,
\begin{align}
  \mathfrak{a}_{k\ell}^\pm = \pm \frac{\mathfrak{b}}{2} + \frac{(2k + 1)\tau}{4} + \frac{\ell}{2}, \qquad
  k, \ell = 0, 1 \ .
\end{align}
Due to the presence of $\tau/4$, all these poles are imaginary, with essentially the same residues
\begin{align}
  R^\pm_{k\ell} = \mp \frac{i}{4} \frac{\vartheta_4(\mathfrak{b})}{\vartheta_1(2 \mathfrak{b})} \ .
\end{align}
Applying the integral formula (\ref{integration-formula-monomial}), the index reads
\begin{align}
  \langle W_j\rangle = \mathcal{I}_{\mathcal{N} = 4}\delta_{j \in \mathbb{Z}} - \frac{i}{4} \frac{\vartheta_4(\mathfrak{b})}{\vartheta_1(2 \mathfrak{b})}\sum_{\substack{m = -j \\ m \ne 0}}^{j}\sum_{k, \ell = 0, 1} \frac{(-1)^{2\ell m} (b^m - b^{-m})q^{( - \frac{1}{2} + k )m}}{q^{m} - q^{-m}} \ .
\end{align}
Note that for $j \in \mathbb{Z}$, the character $\chi_j(a)$ contains a constant term $1$, which upon integration leads to the original Schur index $\mathcal{I}_{\mathcal{N} = 4}$. On the other hand, when $j \in \mathbb{Z} + \frac{1}{2}$, the entire expression vanishes identically thanks to the summation over $\ell = 0, 1$. Therefore, we have
\begin{align}
  \langle W_{j \in \mathbb{Z}}\rangle
  = + \mathcal{I}_{\mathcal{N} = 4} - \frac{i}{2} \frac{\vartheta_4(\mathfrak{b})}{\vartheta_1(2 \mathfrak{b})}\sum_{\substack{m = -j \\ m \ne 0}}^{j} \frac{b^m - b^{-m}}{q^{m/2} - q^{-m/2}} \ ,
  \qquad
  \langle W_{j \in \mathbb{Z} + \frac{1}{2}}\rangle = 0 \ .
\end{align}
The first term $\mathcal{I}_{\mathcal{N} = 4} = \operatorname{ch}_0$ is identified with the vacuum character of the associated chiral algebra $\mathbb{V}_{\mathcal{N} = 4}$. The factor $\frac{i\vartheta_4(\mathfrak{b})}{\vartheta_1(2 \mathfrak{b})}$ in the second term is the residue of the integrand $\mathcal{Z}$ which is related to the Schur index of Gukov-Witten type surface defect in the $\mathcal{N} = 4$ theory \cite{Pan:2021ulr}. It can be shown that the residue satisfies $\frac{i\vartheta_4(\mathfrak{b})}{\vartheta_1(2 \mathfrak{b})} = \operatorname{ch}_0 + \operatorname{ch}_M$ where $M$ is another irreducible module $M$ of $\mathbb{V}_{\mathcal{N} = 4}$ \cite{Adamovic:2014lra,Bonetti:2018fqz,Pan:2021ulr}. As module characters of $\mathbb{V}_{\mathcal{N} = 4}$, both $\operatorname{ch}_0$ and $\operatorname{ch}_M$ satisfy the flavored modular differential equations arising from null states in $\mathbb{V}_{\mathcal{N} = 4}$ \cite{Gaberdiel:2008pr,Gaberdiel:2009vs,Beem:2017ooy,Pan:2021ulr,Zheng:2022zkm}. Therefore, the line index can be written as a combination of the two irreducible characters,
\begin{align}
  \langle W_{j \in \mathbb{Z}}\rangle
  = \bigg(1 - \frac{1}{2}\sum_{\substack{m = -j \\ m \ne 0}}^{+j}\frac{b^m - b^{-m}}{q^{m/2} - q^{- m /2}}\bigg)\operatorname{ch}_0
  - \frac{1}{2}\Big(\sum_{\substack{m = -j \\ m \ne 0}}^{+j}\frac{b^m - b^{-m}}{q^{m/2} - q^{- m /2}}\Big) \operatorname{ch}_M \ .
\end{align}
However, the coefficients of the linear combination are rational functions of $b$ and $q$. This is quite different from the Schur index of surface defects, which are expected to be linear combinations of characters with constant coefficients\footnote{Possibly up to some overall factors of $q$ \cite{Bianchi:2019sxz}}. In particular, the line index does not solve the flavored modular differential equations in \cite{Pan:2021ulr}.

\subsubsection{'t Hooft line index}

In the 4d $\mathcal{N} = 4$ SYM (and in general $\mathcal{N} = 2$ superconformal gauge theories), one can define 't Hooft line operators by specifying certain singular profile for the gauge fields and scalars in the path integral. By the Dirac quantization condition, the magnetic charge $B$ of a 't Hooft operator is valued in the cocharacter lattice $\Lambda_\text{cochar}$ inside the Cartan $\mathfrak{h}$ of the gauge group $G$. This lattice $\Lambda_\text{cochar}$ corresponds to the weights of the Langland dual group $G^\vee$, and therefore a dominant integral element $B$ corresponds to a $G^\vee$-representation $\mathcal{R}^\vee_B$. The cocharacters as weights in $\mathcal{R}^\vee_B$ are obtained from $B$ by subtracting suitable coroot element $\alpha^\vee$, and weights related by the Weyl group $W$ of the gauge group $G$ are identified. A weight $v$ in $\mathcal{R}^\vee_B$ that is not Weyl-related to $B$ can screen the 't Hooft operator and signals monopole bubbling effect \cite{Lee:1996vz,Gomis:2009ir,Ito:2011ea,Brennan:2018yuj}.

Under the S-duality, a full Wilson line in a $\mathcal{N} = 4$ SYM is mapped to a 't Hooft line. If the magnetic charge of a 't Hooft operator corresponds to a minuscule representation of $G^\vee$, then its index is safe from monopole bubbling effect, and the index can be computed by a relatively simple contour integral \cite{Gang:2012yr}. In particular, For the $\mathcal{N} = 4$ $U(2)$ theory, the 't Hooft line with minimal magnetic charge $(1,0)$ corresponds to a minuscule representation, and is dual to the  full Wilson operator in the fundamental representation. The 't Hooft index can be written as a contour integral \cite{Gang:2012yr},
\begin{align}\label{U2-t-hooft}
  \langle H_{(1,0)}^\text{full} \rangle
  = - \oint \frac{da}{2\pi i a} \frac{(a - b)(-1 + a b)}{(\sqrt{q} - a)(-1 + \sqrt{q}a)b}
  \frac{\eta(\tau)^6 \vartheta_4(\mathfrak{a})^2}{
    \vartheta_1(\mathfrak{a} - \mathfrak{b})
    \vartheta_1(\mathfrak{a} + \mathfrak{b})
    \vartheta_4(\mathfrak{b})^2
  } \ .
\end{align}
Note that the parameters and integration variables have been renamed and reorganized compared to the double contour integral in \cite{Gang:2012yr}. In series expansion,
\begin{align}
  \langle H_{(1,0)}^\text{full}\rangle = 1 + 2(b + b^{-1})\sqrt{q}
  + (1 + 3b^2 + 3b^{-2}) q
  + 4(b^3 + b^{-3})q^{3/2} + \cdots \ .
\end{align}
The ratio of $\vartheta$ functions in $\langle H^\text{full}\rangle$ are essentially identical to the original integrand that computes $\mathcal{I}_{\mathcal{N} = 4}$, up to a shift from $\vartheta_{1, 4} \to \vartheta_{4,1}$. It is therefore elliptic in $\mathfrak{a}$, with real poles $\mathfrak{a} = \pm \mathfrak{b}$. The rational factor in the integrand can also be expanded in the $SU(2)$ characters,
\begin{align}
  - \frac{(a-b)(-1 + ab)}{(\sqrt{q} - a)(-1 + \sqrt{q}a)}
  = (1 + b^2) \sum_{n = 0}^{+\infty}q^{\frac{n}{2}}\chi_{j = \frac{n}{2}}(a)
  - b \sum_{n = 0}^{+\infty}q^{n/2}
  \chi_{j = \frac{1}{2}}(a)\chi_{j = \frac{n}{2}}(a) \\
  = (1 + b^2) \sum_{n = 0}^{+\infty}q^{n/2}\chi_{j = \frac{n}{2}}(a)
  - b \sum_{n = 0}^{+\infty}q^{n/2}
  \chi_{j = \frac{n}{2} + \frac{1}{2}}(a)
  - b \sum_{n = 0}^{+\infty}q^{n/2}
  \chi_{j = \frac{n}{2} - \frac{1}{2}}(a) \ . 
\end{align}
Hence, the integral $\langle H^\text{full}\rangle$ can be computed directly and exactly using (\ref{integration-formula-χf}). In this case, the residues of two real poles $a = b^{\pm}$ are given by
\begin{align}
  R_\pm = \pm \frac{i \eta(\tau)^3}{\vartheta_1(2\mathfrak{b})} \ .
\end{align}
After some algebra, we have
\begin{align}
  \langle H_{(1,0)}^\text{full}\rangle
  = \frac{i \eta(\tau)^3}{\vartheta_1(2\mathfrak{b})}
  (q^{\frac{1}{2}} & \ + q^{-\frac{1}{2}} - b - b^{-1})
  \sum_{n = 0}^{+\infty}
  \sum_{\substack{m = - n/2 \\ m \ne 0}}^{+ n/2}
  q^{\frac{n}{2}} \frac{b^{2m} - b^{-2m}}{1 - q^{-2m}}
  \nonumber\\
  & \ + \frac{2(b + b^{-1} - 2q^{\frac{1}{2}})}{1-q} \frac{i \eta(\tau)^3}{\vartheta_1(2\mathfrak{b})} E_1 \begin{bmatrix}
    -1 \\ b  
  \end{bmatrix} \ ,
\end{align}
where in the second line we applied
\begin{align}
  \oint \frac{da}{2\pi i a}\frac{\eta(\tau)^6 \vartheta_4(\mathfrak{a})^2}{
    \vartheta_1(\mathfrak{a} - \mathfrak{b})
    \vartheta_1(\mathfrak{a} + \mathfrak{b})
    \vartheta_4(\mathfrak{b})^2
  } = \frac{2i \eta(\tau)^3}{\vartheta_1(2 \mathfrak{b})} E_1 \begin{bmatrix}
    -1 \\ b  
  \end{bmatrix} \ .
\end{align}
After stripping off the free contribution $\eta(\tau)/\vartheta_4(\mathfrak{b})$, here we see explicitly a combination of two characters $\frac{i \vartheta_4(\mathfrak{b})}{\vartheta_1(2 \mathfrak{b})}$ and vacuum character $\frac{i \vartheta_4(\mathfrak{b})}{\vartheta_1(2 \mathfrak{b})} E_1 \big[\substack{-1 \\b}\big]$. However, the physical meaning of the prefactors is unclear to the authors.

The dual Wilson operator index can be computed a lot more easily with (\ref{integration-formula-monomial}),
\begin{align}
  \langle W^\text{full}_{j = 1/2}\rangle
  = & \ - \frac{1}{2}\frac{\eta(\tau)^6}{\vartheta_4(\mathfrak{b})^2}
  \oint_{|a| = 1} \frac{da}{2\pi i a}
  (a + \frac{1}{a})^2 \frac{
    \vartheta_1(2\mathfrak{a})\vartheta_1(- 2\mathfrak{a})
  }{
    \vartheta_4(2\mathfrak{a} + \mathfrak{b})
    \vartheta_4(-2\mathfrak{a} + \mathfrak{b})
  } \\
  = & \ \langle W_{j = 1}\rangle_{U(2)} + \mathcal{I}_{\mathcal{N} = 4 \ U(2)}
  = q^{-\frac{1}{2}}
  \frac{i\eta(\tau)^3}{\vartheta_4(\mathfrak{b})}
  \frac{\vartheta_4(\mathfrak{b})}{\vartheta_1(2\mathfrak{b})}\left(
  2E_1\begin{bmatrix}
    -1 \\b  
  \end{bmatrix} -  \frac{b -b^{-1}}{q^{1/2} - q^{-1/2}}
  \right) \ . \nonumber
\end{align}
As required by S-duality, $\langle W^\text{full}_{j = 1/2}\rangle = \langle H_{(1,0)}^\text{full}\rangle$. This equality indeed follows analytically from the identity (\ref{E1-expansions}).

Let us also consider 't-Hooft operators with non-minimal charge $B = (2,0)$. In this case, the index receives contribution from monopole bubbling with $v = (1,1)$, and is expected to equal the $U(2)$ Wilson index in the tensor product of fundamental representation. The 't Hooft index reads
\begin{align}
  \langle H^\text{full}_{(2,0)}\rangle
  = q^{-1/2} \oint \frac{da}{2\pi i a} \mathcal{Z}(a) \frac{\eta(\tau)^6}{\vartheta_4(\mathfrak{b})^2} \frac{\vartheta_1(\mathfrak{a})^2}{\vartheta_4(\pm \mathfrak{a} + \mathfrak{b})} \ ,
\end{align}
where
\begin{align}
  \mathcal{Z}(a)
  = \frac{(1 - \frac{\sqrt{q}}{ab}) (1 - \frac{a\sqrt{q}}{b})}{(1 - \frac{1}{a})(1 - a)}& \ \frac{(1 - \frac{b \sqrt{q}}{a})(1 - a b \sqrt{q})}{(1 - \frac{q}{a})(1 - aq)} \nonumber \\
  & \ + \frac{1}{2} \left[\frac{(q - 1)^2 + (b + \frac{1}{b})\sqrt{q}(1 + q) - 2q (a + \frac{1}{a}) }{(1 - \frac{q}{a})(1 - a q)}\right]^2 \ .
\end{align}
Note that
\begin{align}
  \frac{1}{(1 - \frac{q}{a})(1 - aq)} = \sum_{j \in \frac{1}{2}\mathbb{N}}q^{2j} \chi_j(a) \ ,\quad
  (1 - \frac{b^\pm \sqrt{q}}{a})(1 - a b^\pm \sqrt{q}) = (1 + b^{\pm2} q) - b^\pm q^{\frac{1}{2}} \chi_{\frac{1}{2}}(a) \ . \nonumber
\end{align}
Inserting these expansions, we have
\begin{align}
  \mathcal{Z}
  = & \ \frac{1}{(1-z)(1-1/z)} \left[A - B\chi_{1/2}(a) + q \chi_1(a)
  \right]\sum_{j\in \frac{1}{2}\mathbb{N}} q^{2j}\chi_{j}(a) \nonumber \\
  & \ + \Big[
  4q^2(1 + \chi_1 (a)) - C^2 - 2C q \chi_{\frac{1}{2}}(a)
  \Big]\sum_{j, j', j'' \in \frac{1}{2}\mathbb{N}} q^{2(j + j')}N_{j j'}^{j''} \chi_{j''}(a) \\
  \coloneqq & \ \frac{1}{(1-a)(1-1/a)} \sum_{j \in \frac{1}{2}\mathbb{N}} \mathcal{Z}_j \chi_j(a) + \sum_{j \in \frac{1}{2}\mathbb{N}}\mathcal{Z}_j'\chi_j(a),
\end{align}
where
\begin{align}
  A & \ \coloneqq (1 + b^2q)(1 + \frac{q}{b^2}) + q, 
  & B \coloneqq & \ (b + b^{-1})\sqrt{q}(1+q)\\
  C & \ \coloneqq (q-1)^2 + (b + \frac{1}{b}) \sqrt{q}(1 + q), 
  & \chi_{J}(a)\chi_{J'}(a) = & \ \sum_{J''}N_{JJ'}^{J''} \chi_{J''}(a) \ ,
\end{align}
and $\mathcal{Z}_j$, $\mathcal{Z}'_j$ are polynomials of $b, q$ from applying the tensor product rule for the $SU(2)$ characters,
\begin{align}
   \sum_{j \in \frac{1}{2}\mathbb{N}}\mathcal{Z}_j \chi_j(a)  = & \ [A - B \chi_{1/2}(a) + q \chi_1(a)] \sum_{j \in \frac{1}{2}\mathbb{N}}q^{2j} \chi_j(a)\\
   \sum_{j \in \frac{1}{2}\mathbb{N}}\mathcal{Z}'_j \chi_j(a) = & \ \Big[
     4q^2(1 + \chi_1 (a)) - C^2 - 2C q \chi_{\frac{1}{2}}(a)
     \Big]\sum_{j, j', j'' \in \frac{1}{2}\mathbb{N}} q^{2(j + j')}N_{j j'}^{j''} \chi_{j''}(a) \ ,
\end{align}
while their explicit expressions will be left implicit. Plugging this expansion into the integral, we have
\begin{align}
  \langle H^\text{full}_{(2,0)} \rangle
  = & \ \frac{i \eta(\tau)^3}{\vartheta_1(2 \mathfrak{b})}\sum_{j \in \frac{1}{2}\mathbb{N}} \mathcal{Z}_j \left(
  - \lfloor (j + \frac{1}{2})^2 \rfloor 2E_1 \begin{bmatrix}
    -1 \\ b  
  \end{bmatrix}
  + \sum_{m = -j}^{+j}\sum^{+\infty}_{\substack{k = 0 \\ k+2m \ne 0}}
  \frac{k(b^{k + 2m} - b^{-k -2m})}{q^{\frac{k}{2} + m} - q^{- \frac{k}{2} - m}}
  \right) \nonumber \\
  & \ + \frac{i \eta(\tau)^3}{\vartheta_1(2\mathfrak{b})}\sum_{j \in \frac{1}{2} \mathbb{N}} \mathcal{Z}'_j \left(
    \delta_{j \in \mathbb{Z}} 2E_1 \begin{bmatrix}
      -1 \\ b  
    \end{bmatrix}
    - \sum_{\substack{m = -j \\ m \ne 0}}^{+j} \frac{b^{2m} - b^{-2m}}{q^m - q^{-m}}
  \right) \ .
\end{align}
Unfortunately, we are unable to recast the expression to a more elegant form, therefore we do not prove $\langle W_{\mathbf{2} \otimes \mathbf{2}}^\text{full}\rangle_{U(2)} = \langle H_{(2,0)}^\text{full}\rangle$ analytically.

\subsection{\texorpdfstring{$SU(2)$ theory with four flavors}{}}

Next we consider the $\mathcal{N} = 2$ $SU(2)$ gauge theory with four fundamental flavors. In terms of the class-$\mathcal{S}$ description, the theory is associated to the four-punctured sphere $\Sigma_{0,4}$ and it admits three weak coupling limits corresponding to three different pants-decompositions. For any such limit, we can insert a half or full Wilson line operator of the $SU(2)$ gauge group in the spin-$j$ representation, which is illustrated in Figure \ref{SQCD-Wilson-line}. The half Wilson index can be computed by the following integral,
\begin{figure}
  \centering

  \tikzset{every picture/.style={line width=0.75pt}} %set default line width to 0.75pt        

  \begin{tikzpicture}[x=0.75pt,y=0.75pt,yscale=-1,xscale=1]
  %uncomment if require: \path (0,300); %set diagram left start at 0, and has height of 300

  %Shape: Circle [id:dp955395337317696] 
  \draw   (100,150) .. controls (100,122.39) and (122.39,100) .. (150,100) .. controls (177.61,100) and (200,122.39) .. (200,150) .. controls (200,177.61) and (177.61,200) .. (150,200) .. controls (122.39,200) and (100,177.61) .. (100,150) -- cycle ;
  %Shape: Circle [id:dp3632616563983295] 
  \draw   (230,150) .. controls (230,122.39) and (252.39,100) .. (280,100) .. controls (307.61,100) and (330,122.39) .. (330,150) .. controls (330,177.61) and (307.61,200) .. (280,200) .. controls (252.39,200) and (230,177.61) .. (230,150) -- cycle ;
  %Shape: Circle [id:dp48461029883265794] 
  \draw  [fill={rgb, 255:red, 0; green, 0; blue, 0 }  ,fill opacity=1 ] (120.5,127.5) .. controls (120.5,126.12) and (121.62,125) .. (123,125) .. controls (124.38,125) and (125.5,126.12) .. (125.5,127.5) .. controls (125.5,128.88) and (124.38,130) .. (123,130) .. controls (121.62,130) and (120.5,128.88) .. (120.5,127.5) -- cycle ;
  %Shape: Circle [id:dp9228260278753102] 
  \draw  [fill={rgb, 255:red, 0; green, 0; blue, 0 }  ,fill opacity=1 ] (284,168) .. controls (284,166.62) and (285.12,165.5) .. (286.5,165.5) .. controls (287.88,165.5) and (289,166.62) .. (289,168) .. controls (289,169.38) and (287.88,170.5) .. (286.5,170.5) .. controls (285.12,170.5) and (284,169.38) .. (284,168) -- cycle ;
  %Curve Lines [id:da9178219413280493] 
  \draw    (167.47,141) .. controls (183.27,151.41) and (253.67,152.21) .. (267.47,141) ;
  %Curve Lines [id:da7411040940511331] 
  \draw    (168.27,161.8) .. controls (188.07,152.21) and (253.27,151.41) .. (268.27,161.8) ;
  %Shape: Arc [id:dp9118343768594901] 
  \draw  [draw opacity=0][line width=1.5]  (217.44,160.33) .. controls (216.72,164.25) and (215.53,166.8) .. (214.18,166.8) .. controls (211.97,166.8) and (210.18,159.9) .. (210.18,151.38) .. controls (210.18,142.87) and (211.97,135.97) .. (214.18,135.97) .. controls (215.53,135.97) and (216.72,138.52) .. (217.44,142.43) -- (214.18,151.38) -- cycle ; \draw  [line width=1.5]  (217.44,160.33) .. controls (216.72,164.25) and (215.53,166.8) .. (214.18,166.8) .. controls (211.97,166.8) and (210.18,159.9) .. (210.18,151.38) .. controls (210.18,142.87) and (211.97,135.97) .. (214.18,135.97) .. controls (215.53,135.97) and (216.72,138.52) .. (217.44,142.43) ;  
  %Shape: Circle [id:dp10699332934939365] 
  \draw  [fill={rgb, 255:red, 0; green, 0; blue, 0 }  ,fill opacity=1 ] (120.5,164.8) .. controls (120.5,163.42) and (121.62,162.3) .. (123,162.3) .. controls (124.38,162.3) and (125.5,163.42) .. (125.5,164.8) .. controls (125.5,166.18) and (124.38,167.3) .. (123,167.3) .. controls (121.62,167.3) and (120.5,166.18) .. (120.5,164.8) -- cycle ;
  %Shape: Circle [id:dp7385326402103962] 
  \draw  [fill={rgb, 255:red, 0; green, 0; blue, 0 }  ,fill opacity=1 ] (284,127.5) .. controls (284,126.12) and (285.12,125) .. (286.5,125) .. controls (287.88,125) and (289,126.12) .. (289,127.5) .. controls (289,128.88) and (287.88,130) .. (286.5,130) .. controls (285.12,130) and (284,128.88) .. (284,127.5) -- cycle ;

  % Text Node
  \draw (125,130.9) node [anchor=north west][inner sep=0.75pt]    {$b_{1}$};
  % Text Node
  \draw (288.5,171.4) node [anchor=north west][inner sep=0.75pt]    {$b_{4}$};
  % Text Node
  \draw (125,168.2) node [anchor=north west][inner sep=0.75pt]    {$b_{2}$};
  % Text Node
  \draw (288.5,130.9) node [anchor=north west][inner sep=0.75pt]    {$b_{3}$};

  \end{tikzpicture}
  \caption{$SU(2)$ SQCD, a weak coupling limit of the $A_1$ theory of genus-zero and four punctures. The black arc denotes the half Wilson operator associated to the $SU(2)$ gauge group.\label{SQCD-Wilson-line}}

\end{figure}
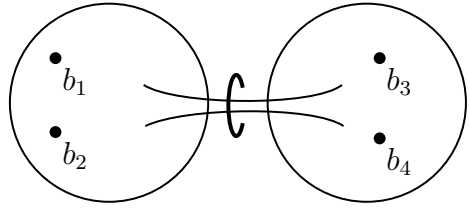
\begin{align}
  \langle W_j\rangle_{0,4} = - \frac{1}{2} \oint \frac{da}{2\pi i z} \left[\sum_{m = -j}^{j} a^{2m}\right] \frac{da}{2\pi i a}
  \vartheta_1(2\mathfrak{a}) \vartheta_1(-2\mathfrak{a})
  \prod_{j = 1}^{4} \frac{\eta(\tau)^2}{\vartheta_1(\mathfrak{a} + \mathfrak{m}_j)
  \vartheta_1(- \mathfrak{a} + \mathfrak{m}_j)} \ .
\end{align}
The poles of the integrand are all imaginary, given by $\mathfrak{a}_i^\pm = \pm \mathfrak{m}_i + \frac{\tau}{2}$ with residues
\begin{align}
  R_{i, \pm} = \pm \frac{i}{2} \frac{\vartheta_1(2 \mathfrak{m}_i)}{\eta(\tau)}
  \prod_{\ell \ne i} \frac{\eta(\tau)}{\vartheta_1(\mathfrak{m}_i + \mathfrak{m}_\ell) \vartheta_1(\mathfrak{m}_i - \mathfrak{m}_\ell)}
  \coloneqq \pm R_i
\end{align}
Applying the integration formula (\ref{integration-formula-monomial}), we have
\begin{align}\label{Wilson-index-SQCD}
  \langle W_j \rangle_{0,4} =  & \ \mathcal{I}_{0,4}\delta_{j \in \mathbb{Z}} - \sum_{\substack{m = - j \\ m \ne 0}}^{+ j} \sum_{\pm} \sum_{i = 1}^4 R_{i, \pm} \frac{1}{q^{2m} - 1} (b_i^\pm q^{\frac{1}{2}})^{2m} \nonumber\\
  = & \ \mathcal{I}_{0,4}\delta_{j \in \mathbb{Z}}
  - \sum_{i = 1}^{4} \left(\sum_{\substack{m = - j \\ m \ne 0}}^{+j} \frac{M_i^{2m} - M_i^{-2m}}{q^{m} - q^{-m}}\right)R_i \ , 
\end{align}
where $M_i \coloneqq e^{2\pi i \mathfrak{m}_i}$. The theory is of class-$\mathcal{S}$ associated to the four-punctured sphere. The $SU(2)^4$ fugacities $b_i$ are related to the $m_i$ by
\begin{align}
  M_1 = b_1 b_2, \quad
  M_2 = b_1/b_2, \quad
  M_3 = b_3 b_4, \quad
  M_4 = b_3/b_4 \ .
\end{align}

In \cite{Cordova:2016uwk}, several Wilson line index in $SU(2)$ SQCD were computed, and the results can be organized as linear combinations of the infinitely many highest weight characters $\chi_{[m, n, 0,0,0]}$ of $\widehat{\mathfrak{so}}(8)_{-2}$ which were obtained from the Kazhdan-Lusztig formula \cite{Lusztig1979}. Our new computation improves the result and relates all $\langle W_j\rangle_{0,4}$ to just five highest weight characters, with respect to finite weights $\lambda = 0, -2 \omega_1, - \omega_2, - 2 \omega_3, -2 \omega_4$, of the simple vertex operator algebra $\widehat{\mathfrak{so}}(8)_{-2}$ \cite{Arakawa:2015jya,Arakawa:2016hkg}. Indeed, the four residues $R_i$ in the above are related to the Schur index of Gukov-Witten type surface defects, and also to the module characters \cite{Peelaers,Pan:2021mrw,2023arXiv230409681L,Pan:2023jjw,Arai:2020qaj},
\begin{align}
  \operatorname{ch}_{-2\widehat \omega_1} = & \ \operatorname{ch}_0 - 2R_1\\
  \operatorname{ch}_{-\widehat \omega_2} = & \ -2 \operatorname{ch}_0 + 2R_1 + 2R_2\\
  \operatorname{ch}_{-2\widehat \omega_3} = & \ \operatorname{ch}_0 - R_1 - R_2 - R_3 - R_4\\
  \operatorname{ch}_{-2\widehat \omega_4} = & \ \operatorname{ch}_0 - R_1 - R_2 - R_3 + R_4 \ ,
\end{align}
where $\operatorname{ch}_0$ is the vacuum character of $\widehat{\mathfrak{so}}(8)_{-2}$, identified with the Schur index $\mathcal{I}_{0,4}$. Therefore, one may write the half Wilson line index as a linear combination of the five module characters,
\begin{align}
  \langle W_j\rangle_{0,4}
  = (\delta_{j \in \mathbb{Z}} - \frac{1}{2}\mathcal{M}_{1j} - \frac{1}{2}\mathcal{M}_{2j})\operatorname{ch}_0
  + & \ \frac{1}{2}(\mathcal{M}_{1j} - \mathcal{M}_{2j})\operatorname{ch}_1
  + \frac{1}{2}(\mathcal{M}_{3j} - \mathcal{M}_{2j})\operatorname{ch}_2 \nonumber\\
  & \ + \frac{1}{2}(\mathcal{M}_{3j} + \mathcal{M}_{4j})\operatorname{ch}_3
  + \frac{1}{2}(\mathcal{M}_{3j} - \mathcal{M}_{4j})\operatorname{ch}_4 \ , \nonumber
\end{align}
where we define the rational functions
\begin{align}
  \mathcal{M}_{ij} \coloneqq \sum_{\substack{m = - j\\m \ne 0}}^{+j}\frac{M_i^{2m} - M_i^{-2m}}{q^m - q^{-m}} \ .
\end{align}

With the half-Wilson index, the index of a full Wilson line operator in the fundamental representation is then given by
\begin{align}
  \langle W_{j = \frac{1}{2}}^\text{full}\rangle_{0,4} = \langle W_{j = \frac{1}{2}} W_{j = \frac{1}{2}}\rangle_{0,4}
  = \mathcal{I}_{0,4} + \langle W_{j = 1}\rangle_{0,4} \ .
\end{align}
By S-duality, this Wilson operator is mapped to the 't Hooft operator with a minimal magnetic charge $B = (-1, 1)$ which receives contribution from monopole bubbling \cite{Gang:2012yr}. The 't Hooft index is given by a slightly more involved contour integral,
\begin{align}
  \langle H_{1,-1}\rangle_{0,4}
  = & \ \oint \frac{da}{2\pi i a} \frac{2q^{\frac{5}{12}}\prod_{i =1}^{4}(a - M_i)(-1 + aM_i)}{(-1 + a^2)^2 (a^2 - q)(-1 + a^2 q) \prod_{i = 1}^{4}M_i}
  \left(- \frac{1}{2}\vartheta_1(\pm 2 \mathfrak{a})\right) \prod_{i = 1}^{4}\frac{\eta(\tau)^2}{\vartheta_1(\pm \mathfrak{a} + M_i)} \nonumber \\
  & \ + q^{-\frac{7}{12}} \oint \frac{da}{2\pi i a} Z_\text{mono}\left(- \frac{1}{2}\vartheta_1(\pm 2 \mathfrak{a})\right) \prod_{i = 1}^{4}\frac{\eta(\tau)^2}{\vartheta_4(\pm \mathfrak{a} + M_i)} \ ,
\end{align}
where
\begin{align}
  Z_\text{mono} = \frac{1}{q \prod_{i =1}^{4}M_i}\left[
  - \left( q + \prod_{i = 1}^{4}M_i\right)
  + \sum_{\pm}\frac{\prod_{i = 1}^{4}(q^{\frac{1}{2}}a^\pm - M_i)}{(1 - a^{\pm 2}) (1 - q a^{\pm 2})}
  \right]^2 \ .
\end{align}
We can rewrite
\begin{align}
  & \ \frac{2q^{\frac{5}{12}}\prod_{i =1}^{4}(a - M_i)(-1 + aM_i)}{(-1 + a^2)^2 (a^2 - q)(-1 + a^2 q) \prod_{i = 1}^{4}M_i} \nonumber \\
  = & \ \frac{2q^{\frac{5}{12}}}{(1-a^2)(1-a^{- 2})}
  \left[\sum_{J \in \mathbb{N}}q^{J}\sum_{j = 0}^{J}(-1)^j\chi_{J - j}(a)\right]
  \prod_{i =1}^{4} \Big(  \chi_{1/2}(a) - \chi_{1/2}(M_i)  \Big) \nonumber\\
  \coloneqq & \ \frac{2q^{\frac{5}{12}}}{(1-a^2)(1-a^{- 2})} \sum_{J \in \frac{1}{2}\mathbb{N}} \mathcal{Z}_J\chi_J(a) \ ,  \\
  Z_\text{mono} = & \ \frac{1}{q \prod_{i = 1}^{4}M_i}
  \left[
  \sum_{j \in \frac{1}{2}\mathbb{N}}(-1)^{2j + 1}\chi_j(a) g_j(M) q^{1 + j}
  \right]^2 \coloneqq \sum_{J \in\frac{1}{2} \mathbb{N}} \mathcal{Z}'_J \chi_J(a) \ .
\end{align}
where
\begin{align}
  g_{J \in \mathbb{N}}(M) \coloneqq 1 + \sum_{\substack{i, j = 1 \\ i < j}}^4M_i M_j + \prod_{i = 1}^{4}M_i \ ,\quad
  g_{J \in \mathbb{N} + \frac{1}{2}}(M) \coloneqq \sum_{i = 1}^{4}M_i + \sum_{\substack{i,j,k = 1\\i < j < k}}^{4}M_i M_j M_k \ ,
\end{align}
and $\mathcal{Z}_J$ and $\mathcal{Z}'_J$ are rational functions of $q$ and fugacities $M$ which simply follow from expanding tensor product of $SU(2)$ irreducible representations; their explicit form will be left implicit. Therefore, we have the exact formula for the 't-Hooft index,
\begin{align}
  \langle H_{1, -1}\rangle_{0,4}
  = & \ \sum_{J \in \frac{1}{2}\mathbb{N}}\mathcal{Z}_J
  \sum_{m = -J}^{+J}\sum_{\substack{k = 1 \\ 2k + 2m \ne 0}}^{+\infty}\left[\sum_{i, \pm}R_{i} \frac{k(M_i^{2k + 2m} - M_i^{ - 2k - 2m})}{q^{\frac{2k + 2m}{2}} - q^{- \frac{2k + 2m}{2}}}
      + \frac{2m}{2}\delta_{\frac{2m}{2} \in \mathbb{Z}_{< 0}} \mathcal{I}_{0,4}\right] \nonumber \\
  & \ + \sum_{J \in \frac{1}{2}  \mathbb{N}} \mathcal{Z}'_J\sum_{m = -J}^{+J} 
  \sum_{i = 1}^4 R_{i} \frac{M_i^{2m} - M_i^{-2m}}{q^m - q^{-m}} \ .
\end{align}
Unfortunately we are unable to reorganize the expression into a more elegant form. Therefore we do not further compare analytically between this 't-Hooft index with the corresponding Wilson index.

\subsection{Genus-one theory with two punctures \label{section:genus-one-two-punctures}}

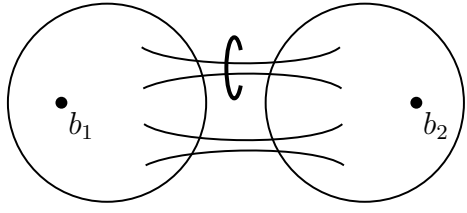
\begin{figure}
  
  \centering
  \tikzset{every picture/.style={line width=0.75pt}} %set default line width to 0.75pt        

  \begin{tikzpicture}[x=0.75pt,y=0.75pt,yscale=-1,xscale=1]
  %uncomment if require: \path (0,300); %set diagram left start at 0, and has height of 300

  %Shape: Circle [id:dp052847036642745815] 
  \draw   (100,150) .. controls (100,122.39) and (122.39,100) .. (150,100) .. controls (177.61,100) and (200,122.39) .. (200,150) .. controls (200,177.61) and (177.61,200) .. (150,200) .. controls (122.39,200) and (100,177.61) .. (100,150) -- cycle ;
  %Shape: Circle [id:dp737996419387422] 
  \draw   (230,150) .. controls (230,122.39) and (252.39,100) .. (280,100) .. controls (307.61,100) and (330,122.39) .. (330,150) .. controls (330,177.61) and (307.61,200) .. (280,200) .. controls (252.39,200) and (230,177.61) .. (230,150) -- cycle ;
  %Shape: Circle [id:dp3979186478695349] 
  \draw  [fill={rgb, 255:red, 0; green, 0; blue, 0 }  ,fill opacity=1 ] (124.5,150) .. controls (124.5,148.62) and (125.62,147.5) .. (127,147.5) .. controls (128.38,147.5) and (129.5,148.62) .. (129.5,150) .. controls (129.5,151.38) and (128.38,152.5) .. (127,152.5) .. controls (125.62,152.5) and (124.5,151.38) .. (124.5,150) -- cycle ;
  %Shape: Circle [id:dp8259854936352573] 
  \draw  [fill={rgb, 255:red, 0; green, 0; blue, 0 }  ,fill opacity=1 ] (303.5,150) .. controls (303.5,148.62) and (304.62,147.5) .. (306,147.5) .. controls (307.38,147.5) and (308.5,148.62) .. (308.5,150) .. controls (308.5,151.38) and (307.38,152.5) .. (306,152.5) .. controls (304.62,152.5) and (303.5,151.38) .. (303.5,150) -- cycle ;
  %Curve Lines [id:da23783623263485554] 
  \draw    (167.47,122) .. controls (183.27,132.41) and (253.67,133.21) .. (267.47,122) ;
  %Curve Lines [id:da7983967355558208] 
  \draw    (168.27,142.8) .. controls (188.07,133.21) and (253.27,132.41) .. (268.27,142.8) ;
  %Curve Lines [id:da4259021517247188] 
  \draw    (168.67,160.8) .. controls (184.47,171.21) and (254.87,172.01) .. (268.67,160.8) ;
  %Curve Lines [id:da1621698308216466] 
  \draw    (169.47,181.6) .. controls (189.27,172.01) and (254.47,171.21) .. (269.47,181.6) ;
  %Shape: Arc [id:dp4098887515399521] 
  \draw  [draw opacity=0][line width=1.5]  (217.44,141.33) .. controls (216.72,145.25) and (215.53,147.8) .. (214.18,147.8) .. controls (211.97,147.8) and (210.18,140.9) .. (210.18,132.38) .. controls (210.18,123.87) and (211.97,116.97) .. (214.18,116.97) .. controls (215.53,116.97) and (216.72,119.52) .. (217.44,123.43) -- (214.18,132.38) -- cycle ; \draw  [line width=1.5]  (217.44,141.33) .. controls (216.72,145.25) and (215.53,147.8) .. (214.18,147.8) .. controls (211.97,147.8) and (210.18,140.9) .. (210.18,132.38) .. controls (210.18,123.87) and (211.97,116.97) .. (214.18,116.97) .. controls (215.53,116.97) and (216.72,119.52) .. (217.44,123.43) ;  

  % Text Node
  \draw (129,153.4) node [anchor=north west][inner sep=0.75pt]    {$b_{1}$};
  % Text Node
  \draw (308,153.4) node [anchor=north west][inner sep=0.75pt]    {$b_{2}$};

  \end{tikzpicture}
  \caption{An $A_1$ class-$\mathcal{S}$ theory of genus one and two punctures, where a half Wilson operator (indicated by the black arc) is inserted at one of the tubes which denotes an $SU(2)$ gauge group in this particular weak coupling limit.\label{fig:genus-one-type-1}}

\end{figure}

Let us consider a higher rank theory with $g = 1$ and $n = 2$, which can be obtained by gauging a diagonal $SU(2) \times SU(2)$ subgroup of the flavor symmetry of two copies of trinion theories $\mathcal{T}_{0,3}$. There are essentially two different weak-coupling frames one can consider, and here we focus on the frame illustrated in Figure \ref{fig:genus-one-type-1}. In this frame, the original Schur index is given as a contour integral
\begin{align}
  \mathcal{I}_{1,2} 
  = \oint \prod_{i = 1}^{2}\frac{da_i}{2\pi i a_i}
  \prod_{j = 1}^{2}\prod_{\pm \pm} \frac{\eta(\tau)}{\vartheta_4(\mathfrak{b}_j \pm \mathfrak{a}_1 \pm \mathfrak{a}_2)}
  \prod_{i = 1}^{2}\left(- \frac{1}{2}\vartheta(\pm 2 \mathfrak{a}_i)\right)
  \coloneqq \oint \left[\frac{da}{2\pi i a}\right]\mathcal{Z}_{1,2}(a) \ .
\end{align}
Let us consider a half Wilson line operator associated to one of the $SU(2)$ gauge groups, whose index is given by the integral
\begin{align}
  \langle W_j\rangle_{1, 2}^{(1)} = \oint \prod_{i = 1}^{2}\frac{da_i}{2\pi i a_i}
  \left(\sum_{m = - j}^{j}a_1^{2m}\right)
  \prod_{j = 1}^{2}\prod_{\pm \pm} \frac{\eta(\tau)}{\vartheta_4(\mathfrak{b}_j \pm \mathfrak{a}_1 \pm \mathfrak{a}_2)}
  \prod_{i = 1}^{2}\left(- \frac{1}{2}\vartheta(\pm 2 \mathfrak{a}_i)\right)\ .
\end{align}

The integral can be evaluated in two different orders: first $a_1$  or first $a_2$. We choose to integrate over $a_1$ first, where the relevant poles are $ \mathfrak{a}_1 = \alpha \mathfrak{b}_j + \beta \mathfrak{a}_2 + \frac{\tau}{2}$ with residues (where $\alpha, \beta = \pm 1$)
\begin{align}
  R_{i \alpha \beta} = \frac{
    i \eta(\tau)^5 \vartheta_1(2 \beta \mathfrak{a}_2) \vartheta_1( 2 \beta \mathfrak{a}_2 + 2 \alpha \mathfrak{b}_i)
  }
  {
  4\vartheta_1(2 \alpha \mathfrak{b}_i)
  \vartheta_1(\alpha \mathfrak{b}_i - \beta \mathfrak{b}_{3-i})
  \vartheta_1(\alpha \mathfrak{b}_i + \beta \mathfrak{b}_{3-i})
  \vartheta_1(2 \mathfrak{a}_2 + \alpha \beta \mathfrak{b}_i - \mathfrak{b}_{3-i})
  \vartheta_1(2 \mathfrak{a}_2 + \alpha \beta \mathfrak{b}_i + \mathfrak{b}_{3-i}) 
  } \ , \nonumber
\end{align}
The $a_1$ integral leaves integrals of the form
\begin{align}
  \oint \frac{da_2}{2\pi i a_2} f(\mathfrak{a}_2) a_2^{n} \ ,
\end{align}
which can be carried out using formula \eqref{integration-formula-monomial}. Finally, the index in the presence of the Wilson line operator gives
\begin{align}
  \langle W_{j \in \mathbb{Z}}\rangle_{1, 2}
  = \mathcal{I}_{1,2}
     + \frac{\eta(\tau)^2}{2 \prod_{i = 1}^2 \vartheta_1(2\mathfrak{b}_i)}
     \sum_{\substack{m = - j \\ m \ne 0}}^{+j}
     \frac{\prod_{i = 1}^{2}(b_i^m - b_i^{-m})}{(q^{m/2} - q^{-m/2})^2}
     \ ,
  \quad
  \langle W_{j \in \mathbb{Z} + \frac{1}{2}}\rangle = 0 \ .
\end{align}
The result is symmetric in $b_1, b_2$ as expected. Note that the first term is clearly the vacuum character of the associated chiral algebra of $\mathcal{T}[\Sigma_{1,2}]$. The factor $\eta(\tau^2)/\prod_{i = 1}^2 \vartheta_1(2 \mathfrak{b}_i)$ arises as the unique\footnote{One can try different nested residues, but they are either zero or proportional to $\eta(\tau^2)/\prod_{i = 1}^2 \vartheta_1(2 \mathfrak{b}_i)$. } nested residue of $\mathcal{Z}_{1,2}(a)$,
\begin{align}
  \operatorname{Res}_{\mathfrak{a}_2 = - \frac{\mathfrak{b}_1 - \mathfrak{b}_2}{2}}\operatorname{Res}_{\mathfrak{a}_1 = \mathfrak{a}_2 + \mathfrak{b}_1 + \frac{\tau}{2}} \mathcal{Z}_{1,2}(\mathfrak{a}_{1,2}) = \frac{\eta(\tau)^2}{8 \vartheta_1(2 \mathfrak{b}_1)\vartheta_1(2 \mathfrak{b}_2)} \ ,
\end{align}
and is also expected to be a linear combination of non-vacuum module character, since it has been shown to satisfy a set of flavored modular differential equations that should annihilate all module characters \cite{zhu1996modular,Zheng:2022zkm}. For example, at weight-two there are two equations
\begin{align}
  0 = \Bigg[
  D_q^{(1)}
  - \frac{1}{4} \sum_{i = 1,2} D_{b_i}^2
  -\frac{1}{4}& \ \sum_{\alpha_i = \pm} E_1 \begin{bmatrix}
    1 \\ b_1^{\alpha_1}b_2^{\alpha_2}
  \end{bmatrix}
  \sum_{i = 1,2}\alpha_i D_{b_i}
  - \sum_{i = 1,2} E_1 \begin{bmatrix}
    1 \\ b_i^2
  \end{bmatrix}D_{b_i} \\
  & \ + 2 \bigg(
  E_2 + \frac{1}{2} \sum_{\alpha_i = \pm}E_2 \begin{bmatrix}
    1 \\ b_1^{\alpha_1}b_2^{\alpha_2}
  \end{bmatrix}
  + \sum_{i = 1,2} E_2 \begin{bmatrix}
    1 \\ b_i^2
  \end{bmatrix}
  \bigg) \Bigg] \mathcal{I}_{1,2} \ ,
\end{align}
and
\begin{align}
  0 = \left(D_{b_1}^2 + 4 E_1 \begin{bmatrix}
    1 \\ b_1^2 
  \end{bmatrix}
  - 8 E_2 \begin{bmatrix}
    1 \\ b_1^2
  \end{bmatrix}\right) \mathcal{I}_{1,2}
  = \left(D_{b_2}^2 + 4 E_1 \begin{bmatrix}
    1 \\ b_2^2 
  \end{bmatrix}
  - 8 E_2 \begin{bmatrix}
    1 \\ b_2^2
  \end{bmatrix}\right)\mathcal{I}_{1,2} \ .
\end{align}

\subsection{\texorpdfstring{Type-1 half Wilson line index in $\mathcal{T}[\Sigma_{g,n}]$}{}}

\begin{figure}
  \centering

  \tikzset{every picture/.style={line width=0.75pt}} %set default line width to 0.75pt        

  \begin{tikzpicture}[x=0.75pt,y=0.75pt,yscale=-1,xscale=1]
  %uncomment if require: \path (0,300); %set diagram left start at 0, and has height of 300

  %Shape: Circle [id:dp7306756594567125] 
  \draw   (84,125) .. controls (84,97.39) and (106.39,75) .. (134,75) .. controls (161.61,75) and (184,97.39) .. (184,125) .. controls (184,152.61) and (161.61,175) .. (134,175) .. controls (106.39,175) and (84,152.61) .. (84,125) -- cycle ;
  %Shape: Circle [id:dp7246516054531755] 
  \draw   (214,125) .. controls (214,97.39) and (236.39,75) .. (264,75) .. controls (291.61,75) and (314,97.39) .. (314,125) .. controls (314,152.61) and (291.61,175) .. (264,175) .. controls (236.39,175) and (214,152.61) .. (214,125) -- cycle ;
  %Shape: Circle [id:dp5769167622777016] 
  \draw  [fill={rgb, 255:red, 0; green, 0; blue, 0 }  ,fill opacity=1 ] (108.5,125) .. controls (108.5,123.62) and (109.62,122.5) .. (111,122.5) .. controls (112.38,122.5) and (113.5,123.62) .. (113.5,125) .. controls (113.5,126.38) and (112.38,127.5) .. (111,127.5) .. controls (109.62,127.5) and (108.5,126.38) .. (108.5,125) -- cycle ;
  %Curve Lines [id:da46424721337134023] 
  \draw    (151.47,97) .. controls (167.27,107.41) and (237.67,108.21) .. (251.47,97) ;
  %Curve Lines [id:da2579581974443781] 
  \draw    (152.27,117.8) .. controls (172.07,108.21) and (237.27,107.41) .. (252.27,117.8) ;
  %Curve Lines [id:da5073486218187262] 
  \draw    (152.67,135.8) .. controls (168.47,146.21) and (238.87,147.01) .. (252.67,135.8) ;
  %Curve Lines [id:da8332827197053869] 
  \draw    (153.47,156.6) .. controls (173.27,147.01) and (238.47,146.21) .. (253.47,156.6) ;
  %Shape: Circle [id:dp9261534340770863] 
  \draw   (334,125) .. controls (334,97.39) and (356.39,75) .. (384,75) .. controls (411.61,75) and (434,97.39) .. (434,125) .. controls (434,152.61) and (411.61,175) .. (384,175) .. controls (356.39,175) and (334,152.61) .. (334,125) -- cycle ;
  %Curve Lines [id:da42718402547321155] 
  \draw    (274.47,115.5) .. controls (290.27,125.91) and (360.67,126.71) .. (374.47,115.5) ;
  %Curve Lines [id:da35029435888855875] 
  \draw    (275.27,136.3) .. controls (295.07,126.71) and (360.27,125.91) .. (375.27,136.3) ;
  %Shape: Circle [id:dp6164799703887061] 
  \draw  [fill={rgb, 255:red, 0; green, 0; blue, 0 }  ,fill opacity=1 ] (383.73,91) .. controls (383.73,89.62) and (384.85,88.5) .. (386.23,88.5) .. controls (387.61,88.5) and (388.73,89.62) .. (388.73,91) .. controls (388.73,92.38) and (387.61,93.5) .. (386.23,93.5) .. controls (384.85,93.5) and (383.73,92.38) .. (383.73,91) -- cycle ;
  %Shape: Circle [id:dp8450845002249934] 
  \draw  [fill={rgb, 255:red, 0; green, 0; blue, 0 }  ,fill opacity=1 ] (386.73,144) .. controls (386.73,142.62) and (387.85,141.5) .. (389.23,141.5) .. controls (390.61,141.5) and (391.73,142.62) .. (391.73,144) .. controls (391.73,145.38) and (390.61,146.5) .. (389.23,146.5) .. controls (387.85,146.5) and (386.73,145.38) .. (386.73,144) -- cycle ;
  %Shape: Circle [id:dp41624953377634855] 
  \draw   (83.67,243) .. controls (83.67,215.39) and (106.05,193) .. (133.67,193) .. controls (161.28,193) and (183.67,215.39) .. (183.67,243) .. controls (183.67,270.61) and (161.28,293) .. (133.67,293) .. controls (106.05,293) and (83.67,270.61) .. (83.67,243) -- cycle ;
  %Shape: Circle [id:dp8178156098645999] 
  \draw   (213.67,243) .. controls (213.67,215.39) and (236.05,193) .. (263.67,193) .. controls (291.28,193) and (313.67,215.39) .. (313.67,243) .. controls (313.67,270.61) and (291.28,293) .. (263.67,293) .. controls (236.05,293) and (213.67,270.61) .. (213.67,243) -- cycle ;
  %Shape: Circle [id:dp516661554372567] 
  \draw  [fill={rgb, 255:red, 0; green, 0; blue, 0 }  ,fill opacity=1 ] (108.17,243) .. controls (108.17,241.62) and (109.29,240.5) .. (110.67,240.5) .. controls (112.05,240.5) and (113.17,241.62) .. (113.17,243) .. controls (113.17,244.38) and (112.05,245.5) .. (110.67,245.5) .. controls (109.29,245.5) and (108.17,244.38) .. (108.17,243) -- cycle ;
  %Curve Lines [id:da002419392347458027] 
  \draw    (151.14,215) .. controls (166.94,225.41) and (237.34,226.21) .. (251.14,215) ;
  %Curve Lines [id:da6514806906240194] 
  \draw    (151.94,235.8) .. controls (171.74,226.21) and (236.94,225.41) .. (251.94,235.8) ;
  %Curve Lines [id:da8063085564418857] 
  \draw    (152.34,253.8) .. controls (168.14,264.21) and (238.54,265.01) .. (252.34,253.8) ;
  %Curve Lines [id:da07013326573581935] 
  \draw    (153.14,274.6) .. controls (172.94,265.01) and (238.14,264.21) .. (253.14,274.6) ;
  %Shape: Circle [id:dp549496212606494] 
  \draw   (333.67,243) .. controls (333.67,215.39) and (356.05,193) .. (383.67,193) .. controls (411.28,193) and (433.67,215.39) .. (433.67,243) .. controls (433.67,270.61) and (411.28,293) .. (383.67,293) .. controls (356.05,293) and (333.67,270.61) .. (333.67,243) -- cycle ;
  %Curve Lines [id:da8720232557446674] 
  \draw    (274.14,233.5) .. controls (289.94,243.91) and (360.34,244.71) .. (374.14,233.5) ;
  %Curve Lines [id:da2568601455252375] 
  \draw    (274.94,254.3) .. controls (294.74,244.71) and (359.94,243.91) .. (374.94,254.3) ;
  %Shape: Circle [id:dp09723681495541392] 
  \draw  [fill={rgb, 255:red, 0; green, 0; blue, 0 }  ,fill opacity=1 ] (383.4,209) .. controls (383.4,207.62) and (384.52,206.5) .. (385.9,206.5) .. controls (387.28,206.5) and (388.4,207.62) .. (388.4,209) .. controls (388.4,210.38) and (387.28,211.5) .. (385.9,211.5) .. controls (384.52,211.5) and (383.4,210.38) .. (383.4,209) -- cycle ;
  %Shape: Circle [id:dp6755778742218745] 
  \draw  [fill={rgb, 255:red, 0; green, 0; blue, 0 }  ,fill opacity=1 ] (415.06,248.67) .. controls (415.06,247.29) and (416.18,246.17) .. (417.56,246.17) .. controls (418.95,246.17) and (420.06,247.29) .. (420.06,248.67) .. controls (420.06,250.05) and (418.95,251.17) .. (417.56,251.17) .. controls (416.18,251.17) and (415.06,250.05) .. (415.06,248.67) -- cycle ;
  %Shape: Arc [id:dp9262236212985635] 
  \draw  [draw opacity=0][line width=1.5]  (203.11,116.67) .. controls (202.38,120.58) and (201.19,123.13) .. (199.85,123.13) .. controls (197.64,123.13) and (195.85,116.23) .. (195.85,107.72) .. controls (195.85,99.2) and (197.64,92.3) .. (199.85,92.3) .. controls (201.19,92.3) and (202.38,94.85) .. (203.11,98.76) -- (199.85,107.72) -- cycle ; \draw  [line width=1.5]  (203.11,116.67) .. controls (202.38,120.58) and (201.19,123.13) .. (199.85,123.13) .. controls (197.64,123.13) and (195.85,116.23) .. (195.85,107.72) .. controls (195.85,99.2) and (197.64,92.3) .. (199.85,92.3) .. controls (201.19,92.3) and (202.38,94.85) .. (203.11,98.76) ;  
  %Shape: Arc [id:dp11863653886155245] 
  \draw  [draw opacity=0][line width=1.5]  (203.11,234) .. controls (202.38,237.91) and (201.19,240.46) .. (199.85,240.46) .. controls (197.64,240.46) and (195.85,233.56) .. (195.85,225.05) .. controls (195.85,216.54) and (197.64,209.64) .. (199.85,209.64) .. controls (201.19,209.64) and (202.38,212.19) .. (203.11,216.1) -- (199.85,225.05) -- cycle ; \draw  [line width=1.5]  (203.11,234) .. controls (202.38,237.91) and (201.19,240.46) .. (199.85,240.46) .. controls (197.64,240.46) and (195.85,233.56) .. (195.85,225.05) .. controls (195.85,216.54) and (197.64,209.64) .. (199.85,209.64) .. controls (201.19,209.64) and (202.38,212.19) .. (203.11,216.1) ;  
  %Shape: Circle [id:dp4712486612083133] 
  \draw  [fill={rgb, 255:red, 0; green, 0; blue, 0 }  ,fill opacity=1 ] (364.4,274) .. controls (364.4,272.62) and (365.52,271.5) .. (366.9,271.5) .. controls (368.28,271.5) and (369.4,272.62) .. (369.4,274) .. controls (369.4,275.38) and (368.28,276.5) .. (366.9,276.5) .. controls (365.52,276.5) and (364.4,275.38) .. (364.4,274) -- cycle ;
  %Shape: Circle [id:dp3662515832010249] 
  \draw  [fill={rgb, 255:red, 0; green, 0; blue, 0 }  ,fill opacity=1 ] (123.06,276.67) .. controls (123.06,275.29) and (124.18,274.17) .. (125.56,274.17) .. controls (126.95,274.17) and (128.06,275.29) .. (128.06,276.67) .. controls (128.06,278.05) and (126.95,279.17) .. (125.56,279.17) .. controls (124.18,279.17) and (123.06,278.05) .. (123.06,276.67) -- cycle ;
  %Shape: Arc [id:dp26582533678833964] 
  \draw  [draw opacity=0] (415.74,230.42) .. controls (416.54,233.73) and (409.1,239.55) .. (398.78,243.63) .. controls (388.48,247.7) and (379.09,248.53) .. (377.38,245.58) -- (396.46,237.73) -- cycle ; \draw   (415.74,230.42) .. controls (416.54,233.73) and (409.1,239.55) .. (398.78,243.63) .. controls (388.48,247.7) and (379.09,248.53) .. (377.38,245.58) ;  
  %Shape: Arc [id:dp5829571194985446] 
  \draw  [draw opacity=0] (386.66,247.29) .. controls (385.3,243.96) and (389.83,238.57) .. (396.93,235.12) .. controls (404.02,231.68) and (411.05,231.46) .. (412.84,234.58) -- (399.84,241.12) -- cycle ; \draw   (386.66,247.29) .. controls (385.3,243.96) and (389.83,238.57) .. (396.93,235.12) .. controls (404.02,231.68) and (411.05,231.46) .. (412.84,234.58) ;  
  %Shape: Circle [id:dp3784388258013136] 
  \draw  [fill={rgb, 255:red, 0; green, 0; blue, 0 }  ,fill opacity=1 ] (122.06,220.5) .. controls (122.06,219.12) and (123.18,218) .. (124.56,218) .. controls (125.95,218) and (127.06,219.12) .. (127.06,220.5) .. controls (127.06,221.88) and (125.95,223) .. (124.56,223) .. controls (123.18,223) and (122.06,221.88) .. (122.06,220.5) -- cycle ;

  % Text Node
  \draw (113,128.4) node [anchor=north west][inner sep=0.75pt]    {$b_{1}$};
  % Text Node
  \draw (388.23,94.4) node [anchor=north west][inner sep=0.75pt]    {$b_{2}$};
  % Text Node
  \draw (391.23,147.4) node [anchor=north west][inner sep=0.75pt]    {$b_{3}$};

  \end{tikzpicture}
  \caption{Some (half) Wilson line operators of type-1. When the tube where the loop resides is cut, the Riemann surface remains connected. \label{Wilson-loop-type-1}}

\end{figure}
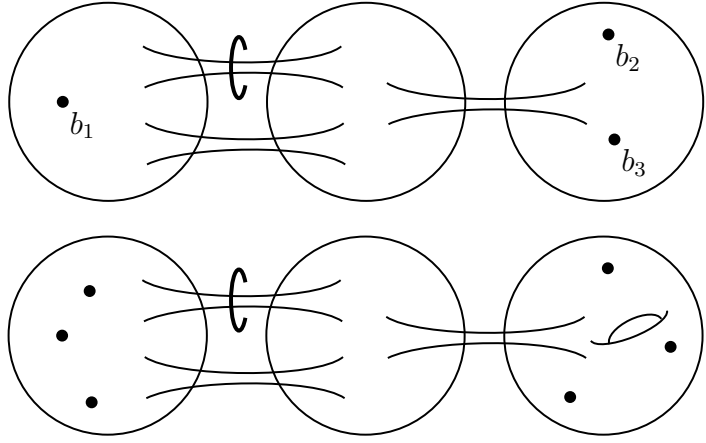

Now we are ready to consider more general type $A_1$ class-$\mathcal{S}$ theories $\mathcal{T}[\Sigma_{g,n}]$. Any such theory usually admits several weak-coupling limits as different supersymmetric gauge theories. With respect to each gauge theory description, we can introduce a half Wilson operator associated to one of the $SU(2)$ gauge groups. In general one can introduce Wilson line charged under multiple $SU(2)$ gauge groups in the weak-coupling description, however, we leave the study of their index and correlation functions to future work.

Let us build on top of the previous $\langle W_j\rangle_{1,2}$ by extending the corresponding Riemann surface to the left and right, while maintaining the location of the Wilson line operator. We simply refer to such a construction of Wilson line operator as type-$1$. The resulting configuration is shown in Figure \ref{Wilson-loop-type-1}, and it is clear from the figure that type-1 Wilson line operator encircles a tube that when cut the Riemann surface $\Sigma_{g, n}$ remain connected. Put differently, a type-1 Wilson operator can be constructed from a single connected Riemann surface $\Sigma_{g, n + 2}$ by gluing two existing punctures and simultaneously inserting a Wilson operator at the tube. In this subsection we will prove that the index of type-1 Wilson line operator in the spin-$j$ representation is given by
\begin{align}\label{Wilson-index-1-general}
  \langle W_{j \in \mathbb{Z}}\rangle^{(1)}_{g \ge 1, n}
  = & \ \mathcal{I}_{g,n}
  - \frac{1}{2}\left[
    \prod_{i = 1}^{n} \frac{i \eta(\tau)}{\vartheta_1(2 \mathfrak{b}_i)}
  \right]
    \sum_{\substack{m = - j\\m \ne 0}}^{+ j}
    \left[\frac{\eta(\tau)}{q^{m/2} - q^{-m /2}}\right]^{2g - 2}
    \prod_{i = 1}^{n} \frac{b_i^m - b_i^{-m}}{q^{m/2} - q^{- m /2}} \ ,\\
  \langle W_{j \in \mathbb{Z} + \frac{1}{2}}\rangle^{(1)}_{g \ge 1, n} = & \ 0 \ .
\end{align}
In particular, when $n = 0$, the products $\prod_{i = 1}^n$ simply return $1$.

Before discussing the proof, here are a few remarks. Although in any given gauge theory description of $\mathcal{T}[\Sigma_{g,n}]$ there may be different choices of $SU(2)$ gauge groups to support a half Wilson line, the final index is actually independent of the choice, as long as they are all type-1. Also we emphasize that type-1 Wilson line exists only for genus $g \ge 1$.

The factor $\eta(\tau)^{2g - 2}\prod_{i = 1}^n \frac{\eta(\tau)}{\vartheta_1(2 \mathfrak{b}_i)}$ can be shown to be the unique\footnote{Up to some numerical factors.} nested residue of the integrand $\mathcal{Z}_{g,n}$ that computes the original Schur index. The uniqueness is only true for $g \ge 1$, as we have already encountered four different residues $R_i$ in the $\mathcal{T}[\Sigma_{0,4}]$ computation; in this sense, class-$\mathcal{S}$ theories at $g \ge 1$ seem to enjoy some nicer properties than the $g = 0$ counterparts\footnote{See also \cite{Satoshi:2023}, where Landau-Ginzburg description can be found for $g \ge 1$ $\mathcal{N} = (0,2)$ and $(0,4)$ class-$\mathcal{S}$ theories in two dimensions. It might suggest some subtle difference in the representation theory of associated chiral algebras of the $g = 0$ and $g \ge 1$ cases. It will be interesting to clarify this issue in the future.}. Extrapolating from the discussions in \cite{Zheng:2022zkm,Pan:2021ulr}, it is natural to expect that this factor is a solution to the set of flavored modular differential equations that annihilate the Schur index, namely, the vacuum character of the associated chiral algebra $\chi(\mathcal{T}[\Sigma_{g,n}])$ of $\mathcal{T}[\Sigma_{g,n}]$, and therefore a linear combination (with constant coefficients) of the vacuum and non-vacuum module characters. For example, when $n = 0$, the relevant factor is simply $\eta(\tau)^{2g - 2}$, and it has been explicitly checked for $g = 2, 3, 4$ that $\eta(\tau)^{2g - 2}$ and the original Schur index $\mathcal{I}_{g, n = 0}$ solve the same modular differential equations. For general $g \ge 2$, $\eta(\tau)^{2g - 2}$ is a particular linear combination of the vortex defect indices $\mathcal{I}_{g, n = 0}^\text{def}(k)$. These observations suggest that the Wilson line index of type-1 is also a linear combination of $\chi(\mathcal{T}[\Sigma_{g,n}])$ characters, with rational coefficient
\begin{align}
  \sum_{\substack{m = - j\\m \ne 0}}^{+ j}
  \left[\frac{1}{q^{m/2} - q^{-m /2}}\right]^{2g - 2}
  \prod_{j = 1}^{n} \frac{b_j^m - b_j^{-m}}{q^{m/2} - q^{- m /2}}\  .
\end{align}

Moreover, the closed-form expression $\langle W_{j}\rangle_{g, n}^{(1)}$ is essentially a finite sum (over $m$) of products of contributions $\frac{b_i^m - b_i^{-m}}{q^{m/2} - q^{-m/2}}$ from the $n$ punctures and a ``three point function'' contribution $\frac{\eta(\tau)}{q^{m/2} - q^{-m/2}}$, which closely resembles that of the $q$-deformed Yang-Mills partition function on $\Sigma_{g, n}$. It would be interesting to match our result in detail with those from the punctured network \cite{Watanabe:2016bwr,Watanabe:2017bmi}, and understand our formula from the perspective of 2d $q$-deformed Yang-Mills.

The proof of the index formula (\ref{Wilson-index-1-general}) can be done recursively by assuming at some $g \ge 1, n\ge 0$ the index $\langle W_j\rangle^{(1)}_{g,n}$ is given by the ansatz (\ref{Wilson-index-1-general}). We already know that the ansatz works for the $g = 1, n = 2$ case which provides a good starting point. We can compute $\langle W_j\rangle_{g, n + 1}^{(1)}$ by gauging,
\begin{align}
  \langle W_j \rangle_{g, n + 1}^{(1)} = \oint \frac{da}{2\pi i a} \langle W_j\rangle^{(1)}_{g, n}(\mathfrak{b}_1, \ldots, \mathfrak{b}_{n - 2}, \mathfrak{a}) \mathcal{I}_\text{VM}(\mathfrak{a}) \mathcal{I}_{0,3}( - \mathfrak{a}, \mathfrak{b}_{n - 1}, \mathfrak{b}_n) \ .
\end{align}
Let us compute
\begin{align}
  & \ \langle W_{j \in \mathbb{Z}}\rangle_{g, n + 1} \nonumber\\
  = & \ \oint \frac{da}{2\pi i a}\bigg[\mathcal{I}_{g, n}(\mathfrak{b}_1, \ldots, \mathfrak{b}_{n - 1}, \mathfrak{a}) \mathcal{I}_\text{VM}(\mathfrak{a}) \mathcal{I}_{0,3}(-\mathfrak{a}, \mathfrak{b}_n, \mathfrak{b}_{n + 1}) \nonumber \\
  & \ - \frac{i^n \eta(\tau)^n}{2
    \vartheta_1(2 \mathfrak{a})\prod_{j = 1}^{n - 1} \vartheta_1(2 \mathfrak{b}_j)
  }
    \\
  & \ \qquad \times \sum_{\substack{m = - j \\ m \ne 0}}^{+j}
  \left[\frac{\eta(\tau)}{q^{m/2} - q^{-m/2}}\right]^{2g - 2}
  \frac{\prod_{j = 1}^{n - 1}(b_j^m - b_j^{-m})}{(q^{m/2} - q^{-m/2})^n}
  (a^m - a^{-m})\mathcal{I}_\text{VM}(\mathfrak{a}) \mathcal{I}_{0,3}(-\mathfrak{a}, \mathfrak{b}_n, \mathfrak{b}_{n + 1})  \bigg]\ . \nonumber
\end{align}
The first term clearly gives $\mathcal{I}_{g, n + 1}$. The second integral is of the form (up to irrelevant factors pulled out of the integral)
\begin{align}
  \oint \frac{da}{2\pi i a} \frac{a^m - a^{-m}}{\vartheta_1(2\mathfrak{a})}
  \mathcal{I}_\text{VM}(a)\mathcal{I}_{0,3}(-\mathfrak{a}, \mathfrak{b}_n, \mathfrak{b}_{n + 1})\ .
\end{align}
It is easy to check that 
\begin{align}
  \frac{\mathcal{I}_\text{VM}(a)}{\vartheta_1(2\mathfrak{a})}\mathcal{I}_{0,3}(- \mathfrak{a}, \mathfrak{b}_n, \mathfrak{b}_{n + 1})
\end{align}
is elliptic in $\mathfrak{a}$. Therefore \eqref{integration-formula-monomial} implies that
\begin{align}
  & \ \oint \frac{da}{2\pi i a}(a^m - a^{-m})\frac{\mathcal{I}_\text{VM}(a)}{\vartheta_1(2\mathfrak{a})}\mathcal{I}_{0,3}(- \mathfrak{a}, \mathfrak{b}_1, \mathfrak{b}_2)
  =
  \frac{i \eta(\tau)}{\prod_{j = 1}^2\vartheta_1(2 \mathfrak{b}_j)} \frac{\prod_{j = 1}^{2}(b_j^m - b_j^{-m})}{(q^{m/2} - q^{-m/2})} \ .
\end{align}
In other words we have verified $\langle W_j\rangle_{g, n + 1}^{(1)}$ also satisfies (\ref{Wilson-index-1-general}),
\begin{align}
  \langle W_{j \in \mathbb{Z}}\rangle _{g, n + 1}
  = \mathcal{I}_{g, n + 1}
    - \frac{i^{n + 1}\eta(\tau)^{n + 1}}{2\prod_{j = 1}^{n + 1}\vartheta_1(2 \mathfrak{b}_j)} \sum_{\substack{m = -j\\m \ne 0}}^{+j}
    \left[\frac{\eta(\tau)}{q^{m/2} - q^{-m/2}}\right]^{2g - 2}
    \frac{\prod_{j = 1}^{n+1}(b_j^m - b_j^{-m})}{(q^{m/2} - q^{-m/2})^{n + 1}} \ .
\end{align}

In the direction of increasing genus $g$, one can glue pairs of punctures to obtain Wilson line operator index $\langle W_j\rangle_{g + 1, n}^{(1)}$ for theories of higher genus $g + 1$,
\begin{align}
  \langle W_j \rangle^{(1)}_{g + 1, n} = \oint \frac{da}{2\pi i a} \mathcal{I}_\text{VM}(a) \langle W_j \rangle^{(1)}_{g, n + 2}(\mathfrak{b}_1, \ldots, \mathfrak{b}_n, \mathfrak{a}, - \mathfrak{a}) \ .
\end{align}
Assuming the ansatz holds at genus $g$, we have
\begin{align}
  & \ \langle W_j\rangle^{(1)}_{g + 1, n} \nonumber \\
   = & \ \mathcal{I}_{g + 1, n}
   - \oint \frac{da}{2\pi i a} \frac{1}{2} \frac{i^n \eta(\tau)^n}{\prod_{j = 1}^{n}\vartheta_1(2\mathfrak{b}_j)}
   \frac{i^2 \eta(\tau)^2}{\vartheta_1(\pm 2 \mathfrak{a})}\left(-\frac{1}{2} \vartheta_1(\pm 2 \mathfrak{a})\right)\\
   & \ \qquad \qquad\qquad\times \sum_{\substack{m = - j\\ m \ne 0}}^{+ j}
   \left[\frac{\eta(\tau)}{q^{m/2} - q^{-m /2}}\right]^{2g - 2}
   \frac{(a^{m} - a^{-m})(a^{- m} - a^{+m})}{(q^{m/2} - q^{-m/2})^2}
   \prod_{j = 1}^{n}\frac{b_j^{m} - b_j^{-m}}{q^{m/2} - q^{-m/2}} \ .\nonumber
\end{align}
The two $\vartheta_1(\pm 2 \mathfrak{a})$ factors are cancelled, while
\begin{align}
  {(a^{m} - a^{-m})(a^{- m} - a^{+m})} = - a^{2m} - a^{-2m} + 2 \ .
\end{align}
Only the $+2$ survives the $a$-integration since $m \ne 0$. Hence,
\begin{align}
  \langle W_j\rangle^{(1)}_{g + 1, n}
  = \mathcal{I}_{g + 1, n}
  - \frac{1}{2}\prod_{j = 1}^{n}\frac{i \eta(\tau)}{\vartheta_1(2 \mathfrak{b}_j)}
    \sum_{\substack{m = - j\\ m \ne 0}}^{+ j}
     \left[\frac{\eta(\tau)}{q^{m/2} - q^{- m /2}}\right]^{2(g+1) - 2}
     \prod_{j = 1}^{n}\frac{b_j^{m} - b_j^{-m}}{q^{m/2} - q^{-m/2}} \ ,
\end{align}
proving the index formula (\ref{Wilson-index-1-general}).

\vspace{2em}

The Type-1 Wilson index $\langle W_j\rangle_{g \ge 1,n}^{(1)}$ can be computed in a different approach, by gluing two existing punctures and simultaneously insert a half Wilson operator,
\begin{align}
  \langle W_j\rangle_{g \ge 1, n}
  = \oint \frac{da}{2\pi i a} \chi_j(a) \mathcal{I}_{g - 1, n + 2}(\mathfrak{b}_1, \ldots, \mathfrak{b}_n, \mathfrak{a}, - \mathfrak{a}) \mathcal{I}_\text{VM}(\mathfrak{a}) \ .
\end{align}
Recall that for $g \ge 0, n > 0$, the $A_1$ Schur index is given by \cite{Pan:2021mrw}
\begin{align}\label{Ign}
  \mathcal{I}_{g, n} = & \ \frac{i^n}{2} \frac{\eta(\tau)^{n + 2g - 2}}{\prod_{j = 1}^{n}\vartheta_1(2 \mathfrak{b}_j)}
  \sum_{\vec\alpha = \pm}\Big(  \prod_{j = 1}^{n}\alpha_j  \Big)\sum_{k = 1}^{n + 2g - 2}\lambda_k^{(n + 2g - 2)} E_k\left[\begin{matrix}
    (-1)^n \\ \prod_{j = 1}^{n}b_j^{\alpha_j}
  \end{matrix}\right] \ .
\end{align}
After identifying $\mathfrak{b}_{n + 1} = \mathfrak{a}$, $\mathfrak{b}_{n + 2} = - \mathfrak{a}$ and multiplying the vector multiplet contribution $\mathcal{I}_\text{VM}(\mathfrak{a})$, all the $\vartheta_1(2\mathfrak{a})$ factors cancel out, and the integration variable $a$ is only present inside the Eisenstein series. When $j \in \mathbb{Z}$, the constant term in $\chi_j(a)$ leads to an additive term $\mathcal{I}_{g, n}$. For the terms in $\chi_j(a)$ with non-zero $m$, we proceed with the integration,
\begin{align}
  \oint \frac{da}{2\pi i a} a^{2m} \frac{i^{n + 2}}{2} \frac{\eta(\tau)^{2g - 2 + n}}{\prod_{j = 1}^{n}\vartheta_1(2\mathfrak{b}_j)}
  \sum_{\vec \alpha = \pm 1} \left(\prod_{j = 1}^{n + 2}\alpha_j\right)
  \sum_{k = 1}^{2g - 2 + n}
  \lambda_k^{(2g - 2 + n)}
  E_k \begin{bmatrix}
    (-1)^n \\
    \prod_{j = 1}^{n + 2}b_j^{\alpha_j}
  \end{bmatrix}_{\substack{b_{n + 1} = a\\b_{n + 2} = 1/a}} \ .
\end{align}
Only the terms with $\alpha_{n + 1} = - \alpha_{n + 2} \coloneqq \beta$, such that $b_{n + 1}^{\alpha_{n + 1}}b_{n + 2}^{\alpha_{n + 2}} = a^{2\beta}$, survives the integration since $2m \ne 0$. 

Let us look at cases with even $n$, where the integral becomes
\begin{align}
  % & \ \frac{i^{n}}{2} \frac{\eta(\tau)^{2g - 2 + n}}{\prod_{j = 1}^{n}\vartheta_1(2\mathfrak{b}_j)}
  % \oint \frac{da}{2\pi i a} a^{2m}
  % \sum_{\beta = \pm}\sum_{\vec \alpha = \pm 1} \left(
  % \prod_{j = 1}^{n}\alpha_j\right)
  % \sum_{k = 1}^{2g - 2 + n}
  % \lambda_{k}^{(2g - 2 + n)}E_k \begin{bmatrix}
  %   (-1)^n \\
  %   a^{2\beta}\prod_{j = 1}^{n}b_j^{\alpha_j}
  % \end{bmatrix}\\
  = & \ - \frac{i^{n}}{2} \frac{\eta(\tau)^{2g - 2 + n}}{\prod_{j = 1}^{n}\vartheta_1(2\mathfrak{b}_j)}
  \sum_{k = 1}^{2g - 2 + n}
    \lambda_k^{(2g - 2 + n)}
    \frac{q^m}{(k-1)!}
    \frac{\text{Eu}_{k - 1}(q^m)}{(1 - q^m)^k}
  \prod_{j = 1}^{n}(b_j^{m} - b_j^{-m})\ . \nonumber
\end{align}
where we applied integration formula (\ref{integration-formula-zE}). Note that since $n$ is even, $k$ must also be even in order for the rational numbers $\lambda$ to be non-zero, and
\begin{align}
  \sum_{\vec \alpha = \pm}\left(\prod_{j = 1}^{n}\alpha_j\right)
  \left(
  \frac{1}{\prod_{j = 1}^n b_j^{m\alpha_j}}
  + \prod_{j = 1}^{n}b_j^{m \alpha_j}
  \right)
  = 2 \prod_{j = 1}^{n}(b_j^{m} - b_j^{-m}) \ .
\end{align}
Therefore,
\begin{align}
  \langle W_j\rangle_{g, n}^{(1)}
  = & \ \mathcal{I}_{g, n}\delta_{j \in \mathbb{Z}}
  - \sum_{\substack{m = - j\\m \ne 0}}^{+j} \frac{i^{n}}{2} \frac{\eta(\tau)^{2g - 2 + n}}{\prod_{j = 1}^{n}\vartheta_1(2\mathfrak{b}_j)}\prod_{j = 1}^{n}(b_j^{m} - b_j^{-m})
    \sum_{k = 1}^{2g - 2 + n}
      \lambda_k^{(2g - 2 + n)}
      \frac{q^m}{(k-1)!}
      \frac{\text{Eu}_{k - 1}(q^m)}{(1 - q^m)^k} \nonumber \\
  = & \ \mathcal{I}_{g, n}\delta_{j \in \mathbb{Z}}
  - \frac{1}{2}
    \prod_{i = 1}^{n} \frac{i \eta(\tau)^n}{\vartheta_2(\mathfrak{b}_j)}
    \sum_{\substack{m = - j\\m \ne 0}}^{+j} 
    \frac{\eta(\tau)^{2g - 2}}{(q^{m/2} - q^{-m/2})^{2g - 2}}\prod_{j = 1}^{n}\frac{b_j^{m} - b_j^{-m}}{q^{m/2} - q^{-m/2}} \ ,
\end{align}
where in the second equality we apply the identity (for even $n$)
\begin{align}
  \sum_{k = 1}^{2g -2 + n}\lambda_k^{(2g - 2 + n)} \frac{q^m}{(k-1)!} \frac{\text{Eu}_{k - 1}(q^m)}{(1 - q^m)^k}
  = \frac{1}{(q^{m/2} - q^{-m/2})^{2g - 2 + n}} \ .
\end{align}

A similar computation can be carried out with odd $n$. Again, an $m \ne 0$ term integrates to
\begin{align}
  = + \frac{i^n}{2} \frac{\eta(\tau)^{2g - 2 + n}}{\prod_{j = 1}^{n} \vartheta_1(2 \mathfrak{b}_j)}
  \sum_{k = 1}^{2g - 2 + n} 
  \lambda_k^{(2g - 2 + n)} \frac{q^{m/2}}{(k - 1)!}\Phi(q^m, 1 - k, \frac{1}{2})\prod_{j - 1}^{n}(b_j^m - b_j^{-m}) \ ,
\end{align}
where we used for odd $n$,
\begin{align}
  \sum_{\vec \alpha = \pm } \left(\prod_{j = 1}^{n}\alpha_j\right)
  \left(\prod_{j = 1}^{n}b_j^{-m\alpha_j} - \prod_{j = 1}^{n} b_j^{m \alpha_j}\right)
  = -2 \prod_{j = 1}^{n}(b_j^m - b_j^{-m})\ .
\end{align}
For odd $n$ we continue to have the same formula as the even $n$ case,
\begin{align}
  \langle W_j\rangle^{(1)}_{g,n}
  = & \ \mathcal{I}_{g, n}\delta_{j \in \mathbb{Z}}
  - \frac{1}{2}
    \prod_{i = 1}^{n} \frac{i \eta(\tau)^n}{\vartheta_1(\mathfrak{b}_j)}
    \sum_{\substack{m = - j\\m \ne 0}}^{+j} 
    \frac{\eta(\tau)^{2g - 2}}{(q^{m/2} - q^{-m/2})^{2g - 2}}\prod_{j = 1}^{n}\frac{b_j^{m} - b_j^{-m}}{q^{m/2} - q^{-m/2}} \ ,
\end{align}
thanks to the curious identity for odd $n$,
\begin{align}
  \sum_{k = 1}^{2g - 2 + n}\lambda_k^{(2g - 2 + n)} \frac{q^{m/2}}{(k - 1)!}
  \Phi(q^m, 1 - k, \frac{1}{2}) = - \frac{1}{(q^{m/2} - q^{-m/2})^{2 g - 2 + n}} \ .
\end{align}

\subsection{\texorpdfstring{Type-2 half Wilson line index in $\mathcal{T}[\Sigma_{g,n}]$}{}}

Next we consider another type of half Wilson operator index, which can be built on top of that of the $SU(2)$ SQCD by extending the relevant Riemann surface on either sides (but not further connecting the two sides). Put differently, we consider a half Wilson operator sitting at a tube that separates the Riemann surface into two disconnected pieces $\Sigma_{g_1, n_1}$ and $\Sigma_{g_2, n_2}$. See Figure \ref{fig:type-2-Wilson-line}. Let us denote such a Wilson index by $\langle W_j\rangle^{(2)}_{g_1, n_1; g_2, n_2}$. In this notation, the previous Wilson index $\langle W_j\rangle_{0,4}$ of the $SU(2)$ SQCD can be denoted as $\langle W\rangle_{0,3;0;3}^{(2)}$.
\begin{figure}
  \centering

  \tikzset{every picture/.style={line width=0.75pt}} %set default line width to 0.75pt        

  \begin{tikzpicture}[x=0.75pt,y=0.75pt,yscale=-1,xscale=1]
  %uncomment if require: \path (0,300); %set diagram left start at 0, and has height of 300

  %Shape: Circle [id:dp7144195152447967] 
  \draw   (69.67,126) .. controls (69.67,98.39) and (92.05,76) .. (119.67,76) .. controls (147.28,76) and (169.67,98.39) .. (169.67,126) .. controls (169.67,153.61) and (147.28,176) .. (119.67,176) .. controls (92.05,176) and (69.67,153.61) .. (69.67,126) -- cycle ;
  %Shape: Circle [id:dp8567893143217518] 
  \draw   (199.67,126) .. controls (199.67,98.39) and (222.05,76) .. (249.67,76) .. controls (277.28,76) and (299.67,98.39) .. (299.67,126) .. controls (299.67,153.61) and (277.28,176) .. (249.67,176) .. controls (222.05,176) and (199.67,153.61) .. (199.67,126) -- cycle ;
  %Shape: Circle [id:dp8561209098410414] 
  \draw  [fill={rgb, 255:red, 0; green, 0; blue, 0 }  ,fill opacity=1 ] (94.17,126) .. controls (94.17,124.62) and (95.29,123.5) .. (96.67,123.5) .. controls (98.05,123.5) and (99.17,124.62) .. (99.17,126) .. controls (99.17,127.38) and (98.05,128.5) .. (96.67,128.5) .. controls (95.29,128.5) and (94.17,127.38) .. (94.17,126) -- cycle ;
  %Curve Lines [id:da28699377599032916] 
  \draw    (137.14,98) .. controls (152.94,108.41) and (223.34,109.21) .. (237.14,98) ;
  %Curve Lines [id:da7670143051247209] 
  \draw    (137.94,118.8) .. controls (157.74,109.21) and (222.94,108.41) .. (237.94,118.8) ;
  %Curve Lines [id:da6515835728151329] 
  \draw    (138.34,136.8) .. controls (154.14,147.21) and (224.54,148.01) .. (238.34,136.8) ;
  %Curve Lines [id:da020510483455377315] 
  \draw    (139.14,157.6) .. controls (158.94,148.01) and (224.14,147.21) .. (239.14,157.6) ;
  %Shape: Circle [id:dp8057855878899254] 
  \draw   (319.67,126) .. controls (319.67,98.39) and (342.05,76) .. (369.67,76) .. controls (397.28,76) and (419.67,98.39) .. (419.67,126) .. controls (419.67,153.61) and (397.28,176) .. (369.67,176) .. controls (342.05,176) and (319.67,153.61) .. (319.67,126) -- cycle ;
  %Curve Lines [id:da5743362121284936] 
  \draw    (260.14,116.5) .. controls (275.94,126.91) and (346.34,127.71) .. (360.14,116.5) ;
  %Curve Lines [id:da6483774151354644] 
  \draw    (260.94,137.3) .. controls (280.74,127.71) and (345.94,126.91) .. (360.94,137.3) ;
  %Shape: Circle [id:dp4869581807050487] 
  \draw  [fill={rgb, 255:red, 0; green, 0; blue, 0 }  ,fill opacity=1 ] (369.4,92) .. controls (369.4,90.62) and (370.52,89.5) .. (371.9,89.5) .. controls (373.28,89.5) and (374.4,90.62) .. (374.4,92) .. controls (374.4,93.38) and (373.28,94.5) .. (371.9,94.5) .. controls (370.52,94.5) and (369.4,93.38) .. (369.4,92) -- cycle ;
  %Shape: Circle [id:dp35908109076550554] 
  \draw  [fill={rgb, 255:red, 0; green, 0; blue, 0 }  ,fill opacity=1 ] (401.06,131.67) .. controls (401.06,130.29) and (402.18,129.17) .. (403.56,129.17) .. controls (404.95,129.17) and (406.06,130.29) .. (406.06,131.67) .. controls (406.06,133.05) and (404.95,134.17) .. (403.56,134.17) .. controls (402.18,134.17) and (401.06,133.05) .. (401.06,131.67) -- cycle ;
  %Shape: Arc [id:dp3554653474842462] 
  \draw  [draw opacity=0][line width=1.5]  (313.11,136.2) .. controls (312.38,140.11) and (311.19,142.66) .. (309.85,142.66) .. controls (307.64,142.66) and (305.85,135.76) .. (305.85,127.25) .. controls (305.85,118.74) and (307.64,111.84) .. (309.85,111.84) .. controls (311.19,111.84) and (312.38,114.39) .. (313.11,118.3) -- (309.85,127.25) -- cycle ; \draw  [line width=1.5]  (313.11,136.2) .. controls (312.38,140.11) and (311.19,142.66) .. (309.85,142.66) .. controls (307.64,142.66) and (305.85,135.76) .. (305.85,127.25) .. controls (305.85,118.74) and (307.64,111.84) .. (309.85,111.84) .. controls (311.19,111.84) and (312.38,114.39) .. (313.11,118.3) ;  
  %Shape: Circle [id:dp22315860908474194] 
  \draw  [fill={rgb, 255:red, 0; green, 0; blue, 0 }  ,fill opacity=1 ] (350.4,157) .. controls (350.4,155.62) and (351.52,154.5) .. (352.9,154.5) .. controls (354.28,154.5) and (355.4,155.62) .. (355.4,157) .. controls (355.4,158.38) and (354.28,159.5) .. (352.9,159.5) .. controls (351.52,159.5) and (350.4,158.38) .. (350.4,157) -- cycle ;
  %Shape: Circle [id:dp42567241897590136] 
  \draw  [fill={rgb, 255:red, 0; green, 0; blue, 0 }  ,fill opacity=1 ] (109.06,159.67) .. controls (109.06,158.29) and (110.18,157.17) .. (111.56,157.17) .. controls (112.95,157.17) and (114.06,158.29) .. (114.06,159.67) .. controls (114.06,161.05) and (112.95,162.17) .. (111.56,162.17) .. controls (110.18,162.17) and (109.06,161.05) .. (109.06,159.67) -- cycle ;
  %Shape: Arc [id:dp08988621963874444] 
  \draw  [draw opacity=0] (398.86,110.58) .. controls (400.64,113.59) and (394.14,120.74) .. (384.33,126.56) .. controls (374.52,132.37) and (365.12,134.65) .. (363.34,131.64) -- (381.1,121.11) -- cycle ; \draw   (398.86,110.58) .. controls (400.64,113.59) and (394.14,120.74) .. (384.33,126.56) .. controls (374.52,132.37) and (365.12,134.65) .. (363.34,131.64) ;  
  %Shape: Arc [id:dp6067809912780477] 
  \draw  [draw opacity=0] (371.13,132.83) .. controls (371.13,132.83) and (371.13,132.83) .. (371.13,132.83) .. controls (369.25,129.67) and (373.33,123.78) .. (380.25,119.68) .. controls (387.16,115.58) and (394.29,114.82) .. (396.17,117.98) -- (383.65,125.41) -- cycle ; \draw   (371.13,132.83) .. controls (371.13,132.83) and (371.13,132.83) .. (371.13,132.83) .. controls (369.25,129.67) and (373.33,123.78) .. (380.25,119.68) .. controls (387.16,115.58) and (394.29,114.82) .. (396.17,117.98) ;  
  %Shape: Circle [id:dp5812392349387883] 
  \draw  [fill={rgb, 255:red, 0; green, 0; blue, 0 }  ,fill opacity=1 ] (108.06,103.5) .. controls (108.06,102.12) and (109.18,101) .. (110.56,101) .. controls (111.95,101) and (113.06,102.12) .. (113.06,103.5) .. controls (113.06,104.88) and (111.95,106) .. (110.56,106) .. controls (109.18,106) and (108.06,104.88) .. (108.06,103.5) -- cycle ;
  %Shape: Arc [id:dp21287770889324586] 
  \draw  [draw opacity=0] (103,158.24) .. controls (103,158.24) and (103,158.24) .. (103,158.24) .. controls (100.24,160.39) and (92.32,154.85) .. (85.31,145.86) .. controls (78.29,136.87) and (74.84,127.83) .. (77.6,125.68) -- (90.3,141.96) -- cycle ; \draw   (103,158.24) .. controls (103,158.24) and (103,158.24) .. (103,158.24) .. controls (100.24,160.39) and (92.32,154.85) .. (85.31,145.86) .. controls (78.29,136.87) and (74.84,127.83) .. (77.6,125.68) ;  
  %Shape: Arc [id:dp666175663461585] 
  \draw  [draw opacity=0] (77.41,133.56) .. controls (77.41,133.56) and (77.41,133.56) .. (77.41,133.56) .. controls (80.31,131.29) and (86.67,134.6) .. (91.61,140.94) .. controls (96.56,147.28) and (98.22,154.25) .. (95.31,156.51) -- (86.36,145.04) -- cycle ; \draw   (77.41,133.56) .. controls (77.41,133.56) and (77.41,133.56) .. (77.41,133.56) .. controls (80.31,131.29) and (86.67,134.6) .. (91.61,140.94) .. controls (96.56,147.28) and (98.22,154.25) .. (95.31,156.51) ;  
  %Shape: Arc [id:dp46087749348293006] 
  \draw  [draw opacity=0] (111.15,136.98) .. controls (111.15,136.98) and (111.15,136.98) .. (111.15,136.98) .. controls (108.3,134.95) and (111.36,125.78) .. (117.98,116.5) .. controls (124.61,107.22) and (132.28,101.34) .. (135.13,103.37) -- (123.14,120.18) -- cycle ; \draw   (111.15,136.98) .. controls (111.15,136.98) and (111.15,136.98) .. (111.15,136.98) .. controls (108.3,134.95) and (111.36,125.78) .. (117.98,116.5) .. controls (124.61,107.22) and (132.28,101.34) .. (135.13,103.37) ;  
  %Shape: Arc [id:dp17472318212088545] 
  \draw  [draw opacity=0] (127.52,105.43) .. controls (127.52,105.43) and (127.52,105.43) .. (127.52,105.43) .. controls (130.52,107.56) and (129.16,114.6) .. (124.49,121.15) .. controls (119.82,127.69) and (113.61,131.26) .. (110.61,129.12) -- (119.07,117.27) -- cycle ; \draw   (127.52,105.43) .. controls (127.52,105.43) and (127.52,105.43) .. (127.52,105.43) .. controls (130.52,107.56) and (129.16,114.6) .. (124.49,121.15) .. controls (119.82,127.69) and (113.61,131.26) .. (110.61,129.12) ;  

  \end{tikzpicture}
  \caption{A half Wilson operator of type 2, where it is inserted at a tube which separates the Riemann surface into the left and right pieces, respectively with $(g_1 = 2, n_1 = 4)$ and $g_2 = 1, n_2 = 4$.\label{fig:type-2-Wilson-line}}
\end{figure}
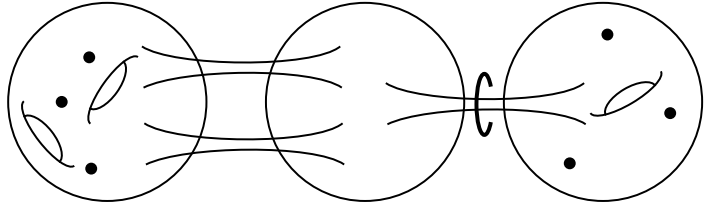

\begin{figure}
  \centering

  \tikzset{every picture/.style={line width=0.75pt}} %set default line width to 0.75pt        

  \begin{tikzpicture}[x=0.75pt,y=0.75pt,yscale=-1,xscale=1]
  %uncomment if require: \path (0,300); %set diagram left start at 0, and has height of 300

  %Shape: Circle [id:dp405377649208158] 
  \draw   (100,150) .. controls (100,122.39) and (122.39,100) .. (150,100) .. controls (177.61,100) and (200,122.39) .. (200,150) .. controls (200,177.61) and (177.61,200) .. (150,200) .. controls (122.39,200) and (100,177.61) .. (100,150) -- cycle ;
  %Shape: Circle [id:dp7049737804672216] 
  \draw   (230,150) .. controls (230,122.39) and (252.39,100) .. (280,100) .. controls (307.61,100) and (330,122.39) .. (330,150) .. controls (330,177.61) and (307.61,200) .. (280,200) .. controls (252.39,200) and (230,177.61) .. (230,150) -- cycle ;
  %Shape: Circle [id:dp6884266201570455] 
  \draw  [fill={rgb, 255:red, 0; green, 0; blue, 0 }  ,fill opacity=1 ] (284,168) .. controls (284,166.62) and (285.12,165.5) .. (286.5,165.5) .. controls (287.88,165.5) and (289,166.62) .. (289,168) .. controls (289,169.38) and (287.88,170.5) .. (286.5,170.5) .. controls (285.12,170.5) and (284,169.38) .. (284,168) -- cycle ;
  %Curve Lines [id:da7874625252090193] 
  \draw    (167.47,141) .. controls (183.27,151.41) and (253.67,152.21) .. (267.47,141) ;
  %Curve Lines [id:da4503868869054204] 
  \draw    (168.27,161.8) .. controls (188.07,152.21) and (253.27,151.41) .. (268.27,161.8) ;
  %Shape: Arc [id:dp8350371639808809] 
  \draw  [draw opacity=0][line width=1.5]  (217.44,160.33) .. controls (216.72,164.25) and (215.53,166.8) .. (214.18,166.8) .. controls (211.97,166.8) and (210.18,159.9) .. (210.18,151.38) .. controls (210.18,142.87) and (211.97,135.97) .. (214.18,135.97) .. controls (215.53,135.97) and (216.72,138.52) .. (217.44,142.43) -- (214.18,151.38) -- cycle ; \draw  [line width=1.5]  (217.44,160.33) .. controls (216.72,164.25) and (215.53,166.8) .. (214.18,166.8) .. controls (211.97,166.8) and (210.18,159.9) .. (210.18,151.38) .. controls (210.18,142.87) and (211.97,135.97) .. (214.18,135.97) .. controls (215.53,135.97) and (216.72,138.52) .. (217.44,142.43) ;  
  %Shape: Circle [id:dp6883808412679802] 
  \draw  [fill={rgb, 255:red, 0; green, 0; blue, 0 }  ,fill opacity=1 ] (284,127.5) .. controls (284,126.12) and (285.12,125) .. (286.5,125) .. controls (287.88,125) and (289,126.12) .. (289,127.5) .. controls (289,128.88) and (287.88,130) .. (286.5,130) .. controls (285.12,130) and (284,128.88) .. (284,127.5) -- cycle ;
  %Shape: Arc [id:dp9682143977791058] 
  \draw  [draw opacity=0] (130.47,170.88) .. controls (126.97,170.83) and (124.25,161.55) .. (124.39,150.15) .. controls (124.54,138.75) and (127.49,129.54) .. (130.99,129.59) -- (130.73,150.23) -- cycle ; \draw   (130.47,170.88) .. controls (126.97,170.83) and (124.25,161.55) .. (124.39,150.15) .. controls (124.54,138.75) and (127.49,129.54) .. (130.99,129.59) ;  
  %Shape: Arc [id:dp8123033878648791] 
  \draw  [draw opacity=0] (125.91,135.61) .. controls (129.59,135.66) and (132.49,142.21) .. (132.39,150.25) .. controls (132.29,158.29) and (129.22,164.77) .. (125.54,164.72) .. controls (125.54,164.72) and (125.54,164.72) .. (125.54,164.72) -- (125.73,150.17) -- cycle ; \draw   (125.91,135.61) .. controls (129.59,135.66) and (132.49,142.21) .. (132.39,150.25) .. controls (132.29,158.29) and (129.22,164.77) .. (125.54,164.72) .. controls (125.54,164.72) and (125.54,164.72) .. (125.54,164.72) ;  

  % Text Node
  \draw (288.5,171.4) node [anchor=north west][inner sep=0.75pt]    {$b_{4}$};
  % Text Node
  \draw (288.5,130.9) node [anchor=north west][inner sep=0.75pt]    {$b_{3}$};

  \end{tikzpicture}
  \caption{A type-2 Wilson line operator in the genus-one theory with two punctures.} \label{fig:genus-one-type-2}
\end{figure}
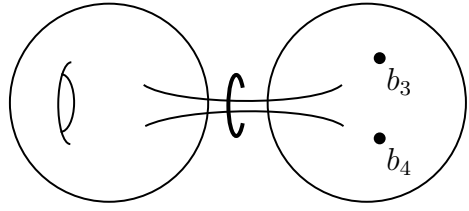

\begin{figure}
  \centering

  \tikzset{every picture/.style={line width=0.75pt}} %set default line width to 0.75pt        

  \begin{tikzpicture}[x=0.75pt,y=0.75pt,yscale=-1,xscale=1]
  %uncomment if require: \path (0,300); %set diagram left start at 0, and has height of 300

  %Shape: Circle [id:dp02150655988764094] 
  \draw   (69.67,126) .. controls (69.67,98.39) and (92.05,76) .. (119.67,76) .. controls (147.28,76) and (169.67,98.39) .. (169.67,126) .. controls (169.67,153.61) and (147.28,176) .. (119.67,176) .. controls (92.05,176) and (69.67,153.61) .. (69.67,126) -- cycle ;
  %Shape: Circle [id:dp6912788531503886] 
  \draw   (199.67,126) .. controls (199.67,98.39) and (222.05,76) .. (249.67,76) .. controls (277.28,76) and (299.67,98.39) .. (299.67,126) .. controls (299.67,153.61) and (277.28,176) .. (249.67,176) .. controls (222.05,176) and (199.67,153.61) .. (199.67,126) -- cycle ;
  %Curve Lines [id:da266039649041943] 
  \draw    (137.14,98) .. controls (152.94,108.41) and (223.34,109.21) .. (237.14,98) ;
  %Curve Lines [id:da9268798873351114] 
  \draw    (137.94,118.8) .. controls (157.74,109.21) and (222.94,108.41) .. (237.94,118.8) ;
  %Curve Lines [id:da9446438963192818] 
  \draw    (138.34,136.8) .. controls (154.14,147.21) and (224.54,148.01) .. (238.34,136.8) ;
  %Curve Lines [id:da6722118484507553] 
  \draw    (139.14,157.6) .. controls (158.94,148.01) and (224.14,147.21) .. (239.14,157.6) ;
  %Shape: Circle [id:dp19052808592893022] 
  \draw   (319.67,126) .. controls (319.67,98.39) and (342.05,76) .. (369.67,76) .. controls (397.28,76) and (419.67,98.39) .. (419.67,126) .. controls (419.67,153.61) and (397.28,176) .. (369.67,176) .. controls (342.05,176) and (319.67,153.61) .. (319.67,126) -- cycle ;
  %Curve Lines [id:da504661812391844] 
  \draw    (260.14,116.5) .. controls (275.94,126.91) and (346.34,127.71) .. (360.14,116.5) ;
  %Curve Lines [id:da715043667228572] 
  \draw    (260.94,137.3) .. controls (280.74,127.71) and (345.94,126.91) .. (360.94,137.3) ;
  %Shape: Circle [id:dp2945781337730151] 
  \draw  [fill={rgb, 255:red, 0; green, 0; blue, 0 }  ,fill opacity=1 ] (370,100.4) .. controls (370,99.02) and (371.12,97.9) .. (372.5,97.9) .. controls (373.88,97.9) and (375,99.02) .. (375,100.4) .. controls (375,101.78) and (373.88,102.9) .. (372.5,102.9) .. controls (371.12,102.9) and (370,101.78) .. (370,100.4) -- cycle ;
  %Shape: Arc [id:dp02945759979795537] 
  \draw  [draw opacity=0][line width=1.5]  (313.11,136.2) .. controls (312.38,140.11) and (311.19,142.66) .. (309.85,142.66) .. controls (307.64,142.66) and (305.85,135.76) .. (305.85,127.25) .. controls (305.85,118.74) and (307.64,111.84) .. (309.85,111.84) .. controls (311.19,111.84) and (312.38,114.39) .. (313.11,118.3) -- (309.85,127.25) -- cycle ; \draw  [line width=1.5]  (313.11,136.2) .. controls (312.38,140.11) and (311.19,142.66) .. (309.85,142.66) .. controls (307.64,142.66) and (305.85,135.76) .. (305.85,127.25) .. controls (305.85,118.74) and (307.64,111.84) .. (309.85,111.84) .. controls (311.19,111.84) and (312.38,114.39) .. (313.11,118.3) ;  
  %Shape: Circle [id:dp3039645319894617] 
  \draw  [fill={rgb, 255:red, 0; green, 0; blue, 0 }  ,fill opacity=1 ] (89.17,126) .. controls (89.17,124.62) and (90.29,123.5) .. (91.67,123.5) .. controls (93.05,123.5) and (94.17,124.62) .. (94.17,126) .. controls (94.17,127.38) and (93.05,128.5) .. (91.67,128.5) .. controls (90.29,128.5) and (89.17,127.38) .. (89.17,126) -- cycle ;
  %Shape: Circle [id:dp8939837034416016] 
  \draw  [fill={rgb, 255:red, 0; green, 0; blue, 0 }  ,fill opacity=1 ] (370,156.4) .. controls (370,155.02) and (371.12,153.9) .. (372.5,153.9) .. controls (373.88,153.9) and (375,155.02) .. (375,156.4) .. controls (375,157.78) and (373.88,158.9) .. (372.5,158.9) .. controls (371.12,158.9) and (370,157.78) .. (370,156.4) -- cycle ;

  % Text Node
  \draw (93.67,129.4) node [anchor=north west][inner sep=0.75pt]    {$b_{1}$};
  % Text Node
  \draw (374.5,103.8) node [anchor=north west][inner sep=0.75pt]    {$b_{2}$};
  % Text Node
  \draw (374.5,153) node [anchor=south west] [inner sep=0.75pt]    {$b_{3}$};

  \end{tikzpicture}
  \caption{A simple example of type-2 Wilson operator for a genus-one theory with three punctures.\label{Wilson-type-2-example-1}}
\end{figure}
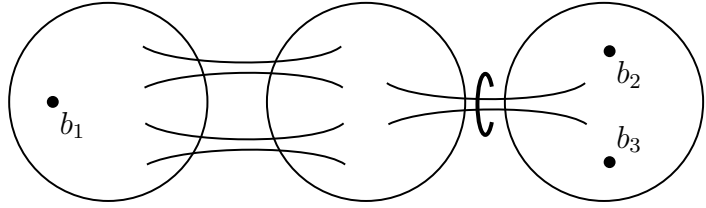

\subsubsection{Simple type-2 examples}

We begin our analysis by looking at a simple genus-one configuration in Figure \ref{fig:genus-one-type-2}. It can be constructed from the $SU(2)$ SQCD by gauging the diagonal of the $SU(2)_{b_1} \times SU(2)_{b_2}$. The Wilson index can be computed by
\begin{align}
  \langle W_j \rangle^{(2)}_{1, 2}
  = \oint \frac{da}{2\pi i a} \langle W\rangle_{0,3;0;3}^{(2)}\Big|_{\substack{b_1 = a\\b_2 = 1/a}} \left(-\frac{1}{2}\right)\vartheta_1(\pm 2 \mathfrak{a}) \ .
\end{align}
Recall that (\ref{Wilson-index-SQCD})
\begin{align}
  \langle W\rangle_{0,3;0;3}^{(2)}
  = & \ \mathcal{I}_{0,4}\delta_{j \in \mathbb{Z}}
  - \sum_{i = 1}^{4} \left(\sum_{\substack{m = - j \\ m \ne 0}}^{+j} \frac{M_i^{2m} - M_i^{-2m}}{q^{m} - q^{-m}}\right)R_i \ ,
\end{align}
where $M_1 = b_1b_2$, $M_2 = b_1/b_2$, $M_3 = b_3 b_4$ and $M_4 = b_3 / b_4$. Obviously as $b_1 = a, b_1 = 1/a$, the $i = 1$ term does not contribute. Therefore, the Wilson index reads (where we have renamed $b_3, b_4 \to b_1, b_2$),
\begin{align}
  \langle W_j\rangle^{(2)}_{1,1; 0,3}
  = & \ \delta_{j \in \mathbb{Z}}\mathcal{I}_{1,2}
  - \frac{\eta(\tau)^2}{\prod_{i = 1}^2\vartheta_1(2 \mathfrak{b}_i)}
    \sum_{\substack{m = j \\ m\ne 0}}^{+j} (q^m + q^{-m})\prod_{i =1,2}\frac{b_i^{2m} - b_i^{-2m}}{q^m - q^{-m}} \nonumber \\
  & \ - \frac{\eta(\tau)^2}{2 \prod_{i=1,2} \vartheta_1(2 \mathfrak{b}_i)}
  \sum_{\alpha = \pm} \bigg(\alpha
  E_1 \begin{bmatrix}
    1 \\ b_1 b_2^\alpha  
  \end{bmatrix}
  \sum_{\substack{m = -j\\m \ne 0}}^{+j}
  \frac{(b_1b_2^\alpha)^{2m} - (b_1b_2^\alpha)^{- 2m}}{q^m - q^{-m}}
  \bigg) \ .
\end{align}
There are four major terms in this half Wilson index, which are proportional respectively to four linearly independent expressions,
\begin{align}
  \mathcal{I}_{1,2}, \qquad
  \frac{\eta(\tau)^2}{\prod_{i = 1}^2 \vartheta_1(2 \mathfrak{b}_i)} , \qquad
  \frac{\eta(\tau)^2}{\prod_{i = 1}^2 \vartheta_1(2 \mathfrak{b}_i)}
    E_1 \begin{bmatrix}
      1 \\ b_1 b_2^\pm
    \end{bmatrix} \ ,
\end{align}
with rational coefficients in $b_i, q$. The first two expressions have appeared previously in section \ref{section:genus-one-two-punctures}, both being solutions to the flavored modular differential equations \cite{Zheng:2022zkm}. It turns out that the two new expressions containing $E_1$ are also additional solutions to the same set of differential equations, and therefore the type-2 index $\langle W_j\rangle^{(2)}_{1,1; 0,3}$ should also be a linear combinations of $\chi(\mathcal{T}[\Sigma_{1,2}])$ characters with rational coefficients.

Next we consider a Wilson operator as demonstrated in Figure \ref{Wilson-type-2-example-1}. There are different ways to compute the index, and the most straightforward way is through the contour integral
\begin{align}
  \langle W_j\rangle_{1,2; 0,3}^{(2)} = & \ \oint \prod_{i = }^{3}\frac{da_i}{2\pi i a_i}\left[\sum_{m = -j}^{+j}a_3^{2m}\right]
  \prod_{\pm\pm}\frac{\eta(\tau)}{
    \vartheta_4(\mathfrak{b}_1 \pm \mathfrak{a}_1 \pm \mathfrak{a}_2)
  }
  \prod_{\pm\pm}\frac{\eta(\tau)}{
    \vartheta_4(\mathfrak{a}_3 \mp \mathfrak{a}_1 \mp \mathfrak{a}_2)
  }\nonumber\\
  & \ \qquad \times \prod_{\pm \pm}\frac{\eta(\tau)}{\vartheta_4(- \mathfrak{a}_3 \pm \mathfrak{b}_2 \pm \mathfrak{b}_3)}
  \prod_{i = 1}^{3}\left(- \frac{1}{2}\vartheta_1(\pm 2 \mathfrak{a}_i)\right) \ .
\end{align}
We choose to evaluate first the $a_3$-integral, and then $a_1, a_2$-integral. The computation is fairly tedious, and we only show the end result,
\begin{align}
  & \ \langle W_j\rangle_{1,2; 0,3}^{(2)} = \mathcal{I} \delta_{j \in \mathbb{Z}} \nonumber \\
  & \ + \sum_{\alpha, \beta = \pm}\sum_{\substack{m = -j \\ m \ne 0}}^j \frac{i \eta(\tau)^3}{8 \prod_{i = }^{3}\vartheta_1(2\mathfrak{b}_i)} \bigg[
  - \frac{4\alpha \beta b_2^{2m \alpha}b_3^{2m \beta}}{q^m - q^{-m}}
  \sum_{\gamma, \delta = \pm} \delta E_2 \begin{bmatrix}
    1 \\ q^{\frac{\gamma}{2}}  b_1^\delta b_2^\alpha b_3^\beta
  \end{bmatrix}(2\tau) \nonumber \\
  & \ \qquad\qquad\qquad\qquad\qquad + \frac{\alpha \beta b_2^{2m \alpha}b_3^{2m \beta}}{q^m - q^{-m}}
  \sum_{\gamma, \delta = \pm} \delta \gamma E_1 \begin{bmatrix}
    1 \\ q^{\frac{\gamma}{2}} b_1^\delta b_2^\alpha b_3^\beta
  \end{bmatrix}(2\tau)\\
  & \ \qquad\qquad\qquad\qquad\qquad - \frac{2 \alpha \beta b_2^{2m \alpha}b_3^{2m \beta}}{q^m - q^{-m}}
  \sum_{\delta = \pm} \delta E_2 \begin{bmatrix}
    -1 \\ b_1^\delta b_2^\alpha b_3^\beta  
  \end{bmatrix} \nonumber\\
  & \ \qquad\qquad\qquad\qquad\qquad + \frac{1}{q^m - q^{-m}}\frac{1}{1 - q^{-2m\alpha}} \sum_{\kappa, \gamma, \delta = \pm}b_2^{2m\gamma \alpha}b_3^{2m \delta \alpha} \alpha \gamma \delta \kappa E_1 \begin{bmatrix}
    -1 \\ b_1^\kappa b_2^{\gamma \alpha \beta}  b_3^{\delta \alpha \beta}
  \end{bmatrix}
  \bigg] \nonumber \\
  & \ + \sum_{\alpha, \beta = \pm} \sum_m'
    \frac{b_1^{2m \alpha} + b_1^{-2m\alpha}}{(q^m - q^{-m})(q^{m \alpha} - q^{-m \alpha})}
    \frac{\alpha \eta(\tau)^6}{8 \prod_{\pm \pm}\vartheta_4(\mathfrak{b}_1 \pm \mathfrak{b}_2 \pm \mathfrak{b}_3)} \ . \nonumber
\end{align}
Note that the Eisenstein series in the first two lines depend on $2\tau$ instead of just $\tau$, a price to pay for simplifying the result using the following identities,
\begin{align}
  \sum_{\pm}E_k\left[\begin{matrix}
    \phi \\ \pm z
  \end{matrix}\right](\tau) = & \ 2 E_k\left[\begin{matrix}
    \phi \\ z^2
  \end{matrix}\right](2\tau) \ , \nonumber \\
  \sum_{\pm} \pm E_k\left[\begin{matrix}
    \phi \\ \pm z
  \end{matrix}\right](\tau)
  = & \ -2 E_k\left[\begin{matrix}
    \phi \\ z^2
  \end{matrix}\right](2\tau)
   + 2 E_k\left[\begin{matrix}
    \phi \\ z
   \end{matrix}\right](\tau)\ , \nonumber
  \\
  E_k\left[\begin{matrix}
    + 1\\z
  \end{matrix}\right](2\tau)
  + E_k\left[\begin{matrix}
    - 1\\z
  \end{matrix}\right](2\tau) = & \ 
  \frac{2}{2^k}E_k\left[\begin{matrix}
    + 1 \\ z
  \end{matrix}\right] \ ,\\
  \sum_{\pm \pm} E_k\left[\begin{matrix}
    \pm 1 \\ \pm z
  \end{matrix}\right](\tau) = & \ \frac{4}{2^k}E_k\left[
  \begin{matrix}
    + 1 \\ z^2
  \end{matrix}\right](\tau)\ . \nonumber
\end{align}

\subsubsection{General type-2 Wilson index}

From the above two examples, it is somewhat clear that the Wilson index of type-2 is significantly more complex than the type-1 index. Moreover, unlike that in type-1, the Wilson index with spin $j \in \mathbb{Z}+\frac{1}{2}$ is nontrivial. Let us compute the type-2 index from another perspective. We consider gluing two Schur indices $\mathcal{I}_{g_i, n_i}$ and insert a Wilson operator at the connecting tube,
\begin{align}
  \langle W\rangle_{g_1, n_1; g_2, n_2}^{(2)}
  = \oint \frac{da}{2\pi i a} \chi_j(a) \mathcal{I}_{g_1, n_1}(\mathfrak{b}_1, \ldots, \mathfrak{b}_{n_1 - 1}, \mathfrak{a}) \mathcal{I}_\text{VM}(\mathfrak{a})
  \mathcal{I}_{g_2, n_2}( - \mathfrak{a}, \tilde{\mathfrak{b}}_1, \ldots, \tilde{\mathfrak{b}}_{n_2 - 1}) \ .
\end{align}
For this we can apply the closed-form expressions (\ref{Ign}) for $\mathcal{I}_{g, n}$ \cite{Pan:2021mrw}, and the above becomes
\begin{align}
  - \frac{1}{2}\oint\frac{da}{2\pi i a}
  & \ \chi_j(a)
  \frac{i^{n_1 + n_2}}{4}
  \frac{\eta(\tau)^{n_1 + 2g_1 - 2}}{\prod_{j = 1}^{n_1 - 1}\vartheta_1(2 \mathfrak{b}_j)}
  \frac{\eta(\tau)^{n_2 + 2g_2 - 2}}{\prod_{j = 1}^{n_2 - 1}\vartheta_1(2 \tilde{\mathfrak{b}}_j)} \nonumber \\
  & \ \times \sum_{\vec\alpha,\vec \beta} \left(\prod_{j = 1}^{n_1}\alpha_j\right)
  \left(\prod_{j = 1}^{n_2}\beta_j\right)
  \sum_{k = 1}^{n_1 + 2g_1 - 2}\sum_{\ell = 1}^{n_2 + 2g_2 - 2}
  \lambda_k^{(n_1 + 2g_1 - 2)}
  \lambda_\ell^{(n_2 + 2g_2 - 2)}\\
  & \ \qquad\qquad E_k \begin{bmatrix}
    (-1)^{n_1}  \\ a^{\alpha_{n_1}} \prod_{j = 1}^{n_1 - 1}b_j^{\alpha_j}
  \end{bmatrix}
  E_\ell \begin{bmatrix}
      (-1)^{n_2}  \\ a^{ - \beta_{n_2}} \prod_{j = 1}^{n_2 - 1}\tilde b_j^{\beta_j}
    \end{bmatrix} \ . \nonumber
\end{align}
Note that the vector multiplet factor has cancelled the $\vartheta_1(2 \mathfrak{a})\vartheta_1( - 2 \mathfrak{a})$ in the denominator. Therefore, the integration boils down to computing
\begin{align}
  \oint \frac{da}{2\pi i z}\chi_j(z) E_k \begin{bmatrix}
    \pm 1 \\
    z a
  \end{bmatrix}
  E_\ell \begin{bmatrix}
    \pm 1 \\
    z b
  \end{bmatrix} \ .
\end{align}

\begin{figure}
  \centering
  
  \tikzset{every picture/.style={line width=0.75pt}} %set default line width to 0.75pt        

  \begin{tikzpicture}[x=0.75pt,y=0.75pt,yscale=-1,xscale=1]
  %uncomment if require: \path (0,300); %set diagram left start at 0, and has height of 300

  %Shape: Circle [id:dp10491492343022712] 
  \draw   (100,150) .. controls (100,122.39) and (122.39,100) .. (150,100) .. controls (177.61,100) and (200,122.39) .. (200,150) .. controls (200,177.61) and (177.61,200) .. (150,200) .. controls (122.39,200) and (100,177.61) .. (100,150) -- cycle ;
  %Shape: Circle [id:dp37565702714649296] 
  \draw   (230,150) .. controls (230,122.39) and (252.39,100) .. (280,100) .. controls (307.61,100) and (330,122.39) .. (330,150) .. controls (330,177.61) and (307.61,200) .. (280,200) .. controls (252.39,200) and (230,177.61) .. (230,150) -- cycle ;
  %Curve Lines [id:da22117651263756288] 
  \draw    (167.47,141) .. controls (183.27,151.41) and (253.67,152.21) .. (267.47,141) ;
  %Curve Lines [id:da12992385838227616] 
  \draw    (168.27,161.8) .. controls (188.07,152.21) and (253.27,151.41) .. (268.27,161.8) ;
  %Shape: Arc [id:dp3255180467542469] 
  \draw  [draw opacity=0][line width=1.5]  (217.44,160.33) .. controls (216.72,164.25) and (215.53,166.8) .. (214.18,166.8) .. controls (211.97,166.8) and (210.18,159.9) .. (210.18,151.38) .. controls (210.18,142.87) and (211.97,135.97) .. (214.18,135.97) .. controls (215.53,135.97) and (216.72,138.52) .. (217.44,142.43) -- (214.18,151.38) -- cycle ; \draw  [line width=1.5]  (217.44,160.33) .. controls (216.72,164.25) and (215.53,166.8) .. (214.18,166.8) .. controls (211.97,166.8) and (210.18,159.9) .. (210.18,151.38) .. controls (210.18,142.87) and (211.97,135.97) .. (214.18,135.97) .. controls (215.53,135.97) and (216.72,138.52) .. (217.44,142.43) ;  
  %Shape: Arc [id:dp8159484951032729] 
  \draw  [draw opacity=0] (130.47,170.88) .. controls (126.97,170.83) and (124.25,161.55) .. (124.39,150.15) .. controls (124.54,138.75) and (127.49,129.54) .. (130.99,129.59) -- (130.73,150.23) -- cycle ; \draw   (130.47,170.88) .. controls (126.97,170.83) and (124.25,161.55) .. (124.39,150.15) .. controls (124.54,138.75) and (127.49,129.54) .. (130.99,129.59) ;  
  %Shape: Arc [id:dp6318509431106123] 
  \draw  [draw opacity=0] (125.91,135.61) .. controls (129.59,135.66) and (132.49,142.21) .. (132.39,150.25) .. controls (132.29,158.29) and (129.22,164.77) .. (125.54,164.72) .. controls (125.54,164.72) and (125.54,164.72) .. (125.54,164.72) -- (125.73,150.17) -- cycle ; \draw   (125.91,135.61) .. controls (129.59,135.66) and (132.49,142.21) .. (132.39,150.25) .. controls (132.29,158.29) and (129.22,164.77) .. (125.54,164.72) .. controls (125.54,164.72) and (125.54,164.72) .. (125.54,164.72) ;  
  %Shape: Arc [id:dp2886937973849866] 
  \draw  [draw opacity=0] (304.14,130.55) .. controls (307.63,130.66) and (310.17,139.99) .. (309.81,151.38) .. controls (309.44,162.78) and (306.31,171.93) .. (302.82,171.82) -- (303.48,151.18) -- cycle ; \draw   (304.14,130.55) .. controls (307.63,130.66) and (310.17,139.99) .. (309.81,151.38) .. controls (309.44,162.78) and (306.31,171.93) .. (302.82,171.82) ;  
  %Shape: Arc [id:dp894818109300004] 
  \draw  [draw opacity=0] (308.01,165.89) .. controls (308.01,165.89) and (308.01,165.89) .. (308.01,165.89) .. controls (304.33,165.77) and (301.56,159.16) .. (301.81,151.13) .. controls (302.07,143.09) and (305.26,136.67) .. (308.94,136.79) -- (308.48,151.34) -- cycle ; \draw   (308.01,165.89) .. controls (308.01,165.89) and (308.01,165.89) .. (308.01,165.89) .. controls (304.33,165.77) and (301.56,159.16) .. (301.81,151.13) .. controls (302.07,143.09) and (305.26,136.67) .. (308.94,136.79) ;  

  \end{tikzpicture}
  \caption{A type-2 Wilson operator in the genus-two theory $\mathcal{T}[\Sigma_{2,0}]$. \label{fig:type-2-genus-two}}
\end{figure}
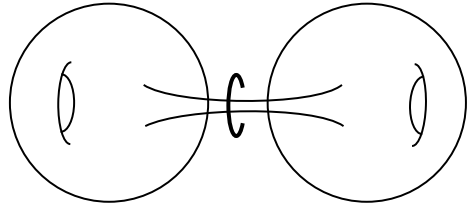

For the special case of $n_1 = n_2 = 1$, $g_1 = g_2 = 1$ corresponding to a Wilson line in the genus-two theory as illustrated in Figure \ref{fig:type-2-genus-two}, we can easily compute the type-2 Wilson index by applying the two identities
\begin{align}
  \oint \frac{dz}{2\pi i z} E_1 \begin{bmatrix}
      + 1 \\ z
  \end{bmatrix}^2 = \frac{q^n( (n - 2) - n q^n )}{(1 - q^n)^2}, \quad
  \oint \frac{dz}{2\pi i z} E_1 \begin{bmatrix}
      - 1 \\ z
  \end{bmatrix}^2 = \frac{q^{n/2}( (n - 1) - (n + 1) q^n )}{(1 - q^n)^2} \ . \nonumber
\end{align}
The index then reads,
\begin{align}
  \langle W_j\rangle^{(2)}_{1,1;1,1}
  = & \ \oint \frac{da}{2\pi i a}
  \chi_j(a)
  \frac{i \eta(\tau)}{\vartheta_1(2 \mathfrak{a})}
  \frac{i \eta(\tau)}{\vartheta_1(-2 \mathfrak{a})}
  \left(- \frac{1}{2}\vartheta_1(\pm 2 \mathfrak{a})\right)
  E_1 \begin{bmatrix}
    -1 \\ a  
  \end{bmatrix}
  E_1 \begin{bmatrix}
    -1 \\ a^{-1}
  \end{bmatrix} \nonumber\\
  = & \ \frac{1}{2} \bigg(
     \delta_{j \in \mathbb{Z}}\eta(\tau)^2\left(E_2(\tau) + \frac{1}{12}\right)
    - \eta(\tau)^2\sum_{\substack{m = - j \\ m\ne 0 }}^{+j}
    \frac{ (2m - 1)q^{-m} - (2m + 1)q^{m}}{(q^m - q^{-m})^2}
    \bigg) \ . \nonumber
\end{align}
We note that the two factors $\eta(\tau)^2$ and $\eta(\tau)^2 (E_2 + \frac{1}{12})$  appearing in the above index are solutions to the modular differential equation that annihilates the genus two Schur index $\mathcal{I}_{2,0}$ \cite{Beem:2017ooy,Zheng:2022zkm},
\begin{align}
  0 = \Big[D_q^{(6)} - 305 E_4 D_q^{(4)} - 4060E_6 D_q^{(3)}
      + 20275E_4^2 & \ D_q^{(2)} + 2100E_4 E_6 D_q^{(1)} \nonumber \\
      & \ - 68600(E_6^2 - 49125E_4^3) \Big]\mathcal{I}_{2,0} \ ,
\end{align}
and therefore the above Wilson index $\langle W_j\rangle^{(2)}_{1,1;1,1}$ is also expected to be a linear combination of characters of the chiral algebra $\chi(\mathcal{T}[\Sigma_{2,0}])$.

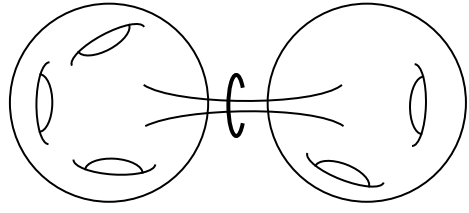
\begin{figure}
  \centering

  \tikzset{every picture/.style={line width=0.75pt}} %set default line width to 0.75pt        

  \begin{tikzpicture}[x=0.75pt,y=0.75pt,yscale=-1,xscale=1]
  %uncomment if require: \path (0,300); %set diagram left start at 0, and has height of 300

  %Shape: Circle [id:dp24248155922862158] 
  \draw   (100,150) .. controls (100,122.39) and (122.39,100) .. (150,100) .. controls (177.61,100) and (200,122.39) .. (200,150) .. controls (200,177.61) and (177.61,200) .. (150,200) .. controls (122.39,200) and (100,177.61) .. (100,150) -- cycle ;
  %Shape: Circle [id:dp6211587114409196] 
  \draw   (230,150) .. controls (230,122.39) and (252.39,100) .. (280,100) .. controls (307.61,100) and (330,122.39) .. (330,150) .. controls (330,177.61) and (307.61,200) .. (280,200) .. controls (252.39,200) and (230,177.61) .. (230,150) -- cycle ;
  %Curve Lines [id:da5567704660931434] 
  \draw    (167.47,141) .. controls (183.27,151.41) and (253.67,152.21) .. (267.47,141) ;
  %Curve Lines [id:da40049403488601976] 
  \draw    (168.27,161.8) .. controls (188.07,152.21) and (253.27,151.41) .. (268.27,161.8) ;
  %Shape: Arc [id:dp8118071741712987] 
  \draw  [draw opacity=0][line width=1.5]  (217.44,160.33) .. controls (216.72,164.25) and (215.53,166.8) .. (214.18,166.8) .. controls (211.97,166.8) and (210.18,159.9) .. (210.18,151.38) .. controls (210.18,142.87) and (211.97,135.97) .. (214.18,135.97) .. controls (215.53,135.97) and (216.72,138.52) .. (217.44,142.43) -- (214.18,151.38) -- cycle ; \draw  [line width=1.5]  (217.44,160.33) .. controls (216.72,164.25) and (215.53,166.8) .. (214.18,166.8) .. controls (211.97,166.8) and (210.18,159.9) .. (210.18,151.38) .. controls (210.18,142.87) and (211.97,135.97) .. (214.18,135.97) .. controls (215.53,135.97) and (216.72,138.52) .. (217.44,142.43) ;  
  %Shape: Arc [id:dp5619765699552235] 
  \draw  [draw opacity=0] (119.47,170.88) .. controls (115.97,170.83) and (113.25,161.55) .. (113.39,150.15) .. controls (113.54,138.75) and (116.49,129.54) .. (119.99,129.59) -- (119.73,150.23) -- cycle ; \draw   (119.47,170.88) .. controls (115.97,170.83) and (113.25,161.55) .. (113.39,150.15) .. controls (113.54,138.75) and (116.49,129.54) .. (119.99,129.59) ;  
  %Shape: Arc [id:dp8986151991497267] 
  \draw  [draw opacity=0] (114.91,135.61) .. controls (118.59,135.66) and (121.49,142.21) .. (121.39,150.25) .. controls (121.29,158.29) and (118.22,164.77) .. (114.54,164.72) .. controls (114.54,164.72) and (114.54,164.72) .. (114.54,164.72) -- (114.73,150.17) -- cycle ; \draw   (114.91,135.61) .. controls (118.59,135.66) and (121.49,142.21) .. (121.39,150.25) .. controls (121.29,158.29) and (118.22,164.77) .. (114.54,164.72) .. controls (114.54,164.72) and (114.54,164.72) .. (114.54,164.72) ;  
  %Shape: Arc [id:dp16492775265686044] 
  \draw  [draw opacity=0] (304.14,130.55) .. controls (307.63,130.66) and (310.17,139.99) .. (309.81,151.38) .. controls (309.44,162.78) and (306.31,171.93) .. (302.82,171.82) -- (303.48,151.18) -- cycle ; \draw   (304.14,130.55) .. controls (307.63,130.66) and (310.17,139.99) .. (309.81,151.38) .. controls (309.44,162.78) and (306.31,171.93) .. (302.82,171.82) ;  
  %Shape: Arc [id:dp41911216428269826] 
  \draw  [draw opacity=0] (308.01,165.89) .. controls (308.01,165.89) and (308.01,165.89) .. (308.01,165.89) .. controls (304.33,165.77) and (301.56,159.16) .. (301.81,151.13) .. controls (302.07,143.09) and (305.26,136.67) .. (308.94,136.79) -- (308.48,151.34) -- cycle ; \draw   (308.01,165.89) .. controls (308.01,165.89) and (308.01,165.89) .. (308.01,165.89) .. controls (304.33,165.77) and (301.56,159.16) .. (301.81,151.13) .. controls (302.07,143.09) and (305.26,136.67) .. (308.94,136.79) ;  
  %Shape: Arc [id:dp6450212139609126] 
  \draw  [draw opacity=0] (173.31,181.14) .. controls (173.31,181.14) and (173.31,181.14) .. (173.31,181.14) .. controls (173.31,181.14) and (173.31,181.14) .. (173.31,181.14) .. controls (173.1,184.63) and (163.71,186.92) .. (152.33,186.24) .. controls (140.95,185.57) and (131.89,182.19) .. (132.09,178.7) -- (152.7,179.92) -- cycle ; \draw   (173.31,181.14) .. controls (173.31,181.14) and (173.31,181.14) .. (173.31,181.14) .. controls (173.31,181.14) and (173.31,181.14) .. (173.31,181.14) .. controls (173.1,184.63) and (163.71,186.92) .. (152.33,186.24) .. controls (140.95,185.57) and (131.89,182.19) .. (132.09,178.7) ;  
  %Shape: Arc [id:dp4007637991670845] 
  \draw  [draw opacity=0] (137.87,184.05) .. controls (137.87,184.05) and (137.87,184.05) .. (137.87,184.05) .. controls (137.87,184.05) and (137.87,184.05) .. (137.87,184.05) .. controls (138.09,180.38) and (144.77,177.78) .. (152.8,178.26) .. controls (160.83,178.73) and (167.15,182.1) .. (166.94,185.77) -- (152.41,184.91) -- cycle ; \draw   (137.87,184.05) .. controls (137.87,184.05) and (137.87,184.05) .. (137.87,184.05) .. controls (137.87,184.05) and (137.87,184.05) .. (137.87,184.05) .. controls (138.09,180.38) and (144.77,177.78) .. (152.8,178.26) .. controls (160.83,178.73) and (167.15,182.1) .. (166.94,185.77) ;  
  %Shape: Arc [id:dp5156573933861222] 
  \draw  [draw opacity=0] (131.31,131.28) .. controls (131.31,131.28) and (131.31,131.28) .. (131.31,131.28) .. controls (131.31,131.28) and (131.31,131.28) .. (131.31,131.28) .. controls (129.66,128.19) and (136.46,121.32) .. (146.51,115.93) .. controls (156.55,110.54) and (166.04,108.66) .. (167.69,111.75) -- (149.5,121.51) -- cycle ; \draw   (131.31,131.28) .. controls (131.31,131.28) and (131.31,131.28) .. (131.31,131.28) .. controls (131.31,131.28) and (131.31,131.28) .. (131.31,131.28) .. controls (129.66,128.19) and (136.46,121.32) .. (146.51,115.93) .. controls (156.55,110.54) and (166.04,108.66) .. (167.69,111.75) ;  
  %Shape: Arc [id:dp7247976636792555] 
  \draw  [draw opacity=0] (159.96,110.22) .. controls (159.96,110.22) and (159.96,110.22) .. (159.96,110.22) .. controls (159.96,110.22) and (159.96,110.22) .. (159.96,110.22) .. controls (161.7,113.46) and (157.37,119.17) .. (150.29,122.98) .. controls (143.21,126.78) and (136.05,127.23) .. (134.31,123.99) -- (147.14,117.11) -- cycle ; \draw   (159.96,110.22) .. controls (159.96,110.22) and (159.96,110.22) .. (159.96,110.22) .. controls (159.96,110.22) and (159.96,110.22) .. (159.96,110.22) .. controls (161.7,113.46) and (157.37,119.17) .. (150.29,122.98) .. controls (143.21,126.78) and (136.05,127.23) .. (134.31,123.99) ;  
  %Shape: Arc [id:dp5677614072471426] 
  \draw  [draw opacity=0] (288.51,190.29) .. controls (287.24,193.56) and (277.6,192.85) .. (266.97,188.73) .. controls (256.34,184.6) and (248.75,178.61) .. (250.02,175.35) -- (269.26,182.82) -- cycle ; \draw   (288.51,190.29) .. controls (287.24,193.56) and (277.6,192.85) .. (266.97,188.73) .. controls (256.34,184.6) and (248.75,178.61) .. (250.02,175.35) ;  
  %Shape: Arc [id:dp11521933547929897] 
  \draw  [draw opacity=0] (253.88,182.21) .. controls (253.88,182.21) and (253.88,182.21) .. (253.88,182.21) .. controls (253.88,182.21) and (253.88,182.21) .. (253.88,182.21) .. controls (255.22,178.78) and (262.37,178.36) .. (269.87,181.27) .. controls (277.36,184.18) and (282.36,189.32) .. (281.02,192.75) -- (267.45,187.48) -- cycle ; \draw   (253.88,182.21) .. controls (253.88,182.21) and (253.88,182.21) .. (253.88,182.21) .. controls (253.88,182.21) and (253.88,182.21) .. (253.88,182.21) .. controls (255.22,178.78) and (262.37,178.36) .. (269.87,181.27) .. controls (277.36,184.18) and (282.36,189.32) .. (281.02,192.75) ;  
  \end{tikzpicture}
  \caption{A generic type-2 Wilson line in a genus $g = g_1 + g_2$ theory without any puncture. \label{fig:genus-g-type-2}}
\end{figure}

The same structure of linear combination actually holds true for all type-2 index $\langle W_j\rangle^{(2)}_{g_1, 1; g_2, 1}$ illustrated in Figure \ref{fig:genus-g-type-2}. Indeed, the relevant integrals are of the form ($k_i \le 2g_i - 1$)
\begin{align}
  \oint \frac{da}{2\pi i a} E_{k_1} \begin{bmatrix}
      -1 \\ a
  \end{bmatrix}E_{k_2} \begin{bmatrix}
      -1 \\ a
  \end{bmatrix}
  \sim \text{linear combination of } E_{2}, E_4, \cdots, E_{2g - 2} \ ,
\end{align}
where we have used (\ref{integration-formula-zEE-1}), (\ref{integration-formula-zEE-2}). In the end, the Wilson index $\langle W_j\rangle^{(2)}_{g_1, 1; g_2, 1}$ is a linear combination of $\eta(\tau)^{2g - 2}, \eta(\tau)^{2g - 2} E_2(\tau), \cdots, \eta(\tau)^{2g - 2}E_{2g - 2}(\tau)$ with the coefficients being rational functions of $q$. This series of functions are conjectured to be solutions to the modular differential equations annihilating the Schur index $\mathcal{I}_{g, 0}$, as they are simply the Schur index of the vortex surface defects in the 4d theory $\mathcal{T}[\Sigma_{g,0}]$ \cite{Gaiotto:2012xa,Zheng:2022zkm}.

For more general $n_i, g_i$, we need to apply the integration formula \eqref{integration-formula-zEE-1}, \eqref{integration-formula-zEE-2} and their variants. For example, with both $n_1, n_2$ even, we have
\begin{align}
  & \ \langle W_j\rangle_{g_1, n_1; g_2, n_2}^{(2)} \nonumber \\
  = & \ \mathcal{I}_{g_1 + g_2, n_1 + n_2 - 2}\delta_{j \in \mathbb{Z}} \nonumber\\
  & \ + \frac{\eta(\tau)^{2(g_1 + g_2) + (n_1 + n_2 - 2) - 2}}{
    2\prod_{i = 1}^{n_1 + n_2 - 2}\vartheta_1(2 \mathfrak{b}_i)
  }\\
  & \ \qquad \times \sum_{\substack{m = - j \\ m\ne 0}}^j \sum_{\vec \alpha} \left(\prod_{i=1}^{n_1 + n_2 - 2}\alpha_i\right)
  \sum_{\ell = 0}^{\operatorname{max}(n_i + 2g_i - 2)}
    \Lambda_\ell^{(g_1, n_1; g_2, n_2)}(\mathbf{b}^{2m}, q^{2m})
    E_\ell \begin{bmatrix}
    1 \\ \prod_{i}^{n_1 + n_2 -2} b_i^{\alpha_i}
  \end{bmatrix} \ . \nonumber
\end{align}
Here we have merged the two sets of flavor fugacities $(b_1, \ldots, b_{n_1 - 1})$ and $(\tilde b_1, \ldots, \tilde b_{n_2 - 1})$ into a larger set $\mathbf{b} = (b_{i}, \ldots, b_{n_1 + n_2 - 2})$, and the corresponding signs $(\alpha_1, \ldots, \alpha_{n_1 - 1}, \beta_1, \ldots, \beta_{n_2 - 1})$ into $(\alpha_1, \ldots, \alpha_{n_1 + n_2 - 2})$. Finally, the $\Lambda$ are a set of rational functions of $b_i$ and $q$ coming from applying the integration formula (\ref{integration-formula-zEE-3}),
\begin{align}
  \Lambda_\ell^{(g_1, n_1; g_2, n_2)}(\mathbf{b}^{2m}, q^{2m})
  = \sum_{k_i = 0}^{n_i + 2g_i - 2}\frac{1}{\ell!} \frac{(-1)^{k_2 + 1} q^{2m}}{\prod_{i = 1}^{n_2 - 1}\tilde b_i^{2m \beta_i}} \lambda_{k_1}^{(n_1 + 2g_1 - 2)}\lambda_{k_2}^{(n_2 + 2g_2 - 2)} \mathcal{E}_{k_1, k_2; \ell}(\mathbf{b}^{2m \alpha}, q^{2m}) \ . \nonumber
\end{align}
Although it is a finite sum, unlike the beautiful result for the type-1 index formula, we are unable to reorganize the above type-2 result into a more elegant form. It would be interesting to further explore the relation between the type-2 Wilson line index and the characters of the associated chiral algebra $\chi(\mathcal{T}[\Sigma_{g,n}])$, and it is likely that the Wilson line index has access to new characters besides those from the surface defects index \cite{Zheng:2022zkm}.

\section{Line operator index in other gauge theories}

\subsection{\texorpdfstring{$\mathcal{N} = 4$ $SU(3)$ theory}{}}

The flavored $\mathcal{N} = 4$ $SU(N)$ Schur index in the presence of Wilson line operators is studied in \cite{Hatsuda:2023iwi} using the Fermi-gas formalism. In the following we also compute some simple examples using our integration formula. The relevant integral is of the form
\begin{align}
  \langle W_\mathcal{R}\rangle
  = - \frac{1}{N!} \frac{\eta(\tau)^{3N - 3}}{\vartheta_4(\mathfrak{b})^{N - 1}}\oint \prod_{A = 1}^{N - 1}  \frac{da_A}{2\pi i a_A}
  \chi_\mathcal{R}(a)
  \prod_{\substack{A, B = 1 \\ A\ne B}}^N \frac{\vartheta_1(\mathfrak{a}_A - \mathfrak{a}_B)}{\vartheta_4(\mathfrak{b} + \mathfrak{a}_A - \mathfrak{a}_B)} \ .
\end{align}

We will focus on $N = 3$. The $SU(3)$ character $\chi_\mathcal{R}(a)$ is a sum of monomials $a_1^{n_1} a_2^{n_2}$. Note that the ratio of the Jacobi theta functions is symmetric in $a_1 \leftrightarrow a_2$ and $\mathfrak{a}_A \to -\mathfrak{a}_A$, and therefore we can focus on monomials of the form $a_1^{n_1 > 0} a_2^{n_2}$; trivial monomial $a_1^0 a_2^0$ insertion simply integrates to the original $\mathcal{N} = 4$ Schur index. Now we compute
\begin{align}
  - \frac{1}{N!} \frac{\eta(\tau)^{3N - 3}}{\vartheta_4(\mathfrak{b})^{N - 1}}\oint \prod_{A = 1}^{N - 1}  \frac{da_A}{2\pi i a_A}
  a_1^{n_1} a_2^{n_2}
  \prod_{\substack{A, B = 1 \\ A\ne B}}^N \frac{\vartheta_1(\mathfrak{a}_A - \mathfrak{a}_B)}{\vartheta_4(\mathfrak{b} + \mathfrak{a}_A - \mathfrak{a}_B)} \ ,
\end{align}
by first integrating $a_1$ and then $a_2$. The $a_1$ integration is easy, leaving an $a_2$ integration of
\begin{align}
  - a_2^{n_2} \sum_{\pm}R^{(1)}_{1,\pm} \frac{a_2^{n_1}b^{\pm n_1}}{q^{n_1/2} - q^{-n_1/2}}
  & \ - a_2^{n_2}\sum_{\pm} R^{(1)}_{2, \pm}\frac{a_2^{-2n_1}b^{\pm n_1}}{q^{n_1/2} - q^{- n_1/2}} \nonumber \\
  & \ - a_2^{n_2}\sum_{\pm; k,\ell = 0,1}R^{(1)}_{3, \pm, k\ell} \frac{a_2^{- n_1/2 }b^{\pm n_1/2}q^{n_1/4} q^{\frac{k-1}{2} n_1} (-1)^{\ell n_1}}{q^{n_1/2} - q^{- n_1/2}}\ ,
\end{align}
where the poles are all imaginary with residues given in the following table.
\begin{center}
  \renewcommand{\arraystretch}{2}
  \begin{tabular}{c|c}
    $(\mathfrak{a}_1)^{(1)}_{1, \pm}$ & $\mathfrak{a}_2 \pm \mathfrak{b} + \tau/2$\\
    \hline
    $R_{1,\pm}^{(1)}$ & $\displaystyle \frac{i}{6}\eta(\tau)^3
    \frac{
      \vartheta_4(3 \mathfrak{a}_2 \pm \mathfrak{b})
      \vartheta_1(3 \mathfrak{a}_2 \pm 2 \mathfrak{b})
    }{
      \vartheta_1(\pm 2 \mathfrak{b})
      \vartheta_1(3 \mathfrak{a}_2)
      \vartheta_4(3 \mathfrak{a}_2 \pm 3 \mathfrak{b})
    }$\\
    \hline
    $(\mathfrak{a}_1)^{(1)}_{2, \pm}$ & $\mathfrak{a}_1 = - 2 \mathfrak{a}_2 \pm \mathfrak{b} + \tau/2$\\
    \hline
    $R_{2,\pm}^{(1)}$ & $\displaystyle
    \frac{i}{6} \eta(\tau)^3
    \frac{
      \vartheta_4(3 \mathfrak{a}_2 \mp \mathfrak{b})
      \vartheta_1(3 \mathfrak{a}_2 \mp 2 \mathfrak{b})
    }{
      \vartheta_1(\pm 2 \mathfrak{b})
      \vartheta_1(3 \mathfrak{a}_2)
      \vartheta_4(3 \mathfrak{a}_2 \mp 3 \mathfrak{b})}
    = - R_{1, \mp}^{(1)}
    $\\
    \hline
    $(\mathfrak{a}_1)^{(1)}_{3, \pm, k\ell}$ & $\displaystyle - \frac{\mathfrak{a}_2}{2} \pm \frac{\mathfrak{b}}{2} + \frac{\tau}{4} + \frac{k \tau}{2} + \frac{\ell}{2}$\\
    \hline
    $R^{(1)}_{3,\pm, k\ell}$ & $
    \displaystyle
    \frac{i}{12} \frac{\eta(\tau)^3}{\vartheta_1(\pm 2 \mathfrak{b})} \prod_{\gamma = \pm} \frac{\vartheta_1(\frac{3}{2} \gamma \mathfrak{a}_2 \pm \frac{1}{2}\mathfrak{b} + \frac{1}{4}\tau + \frac{k}{2}\tau + \frac{\ell}{2})^2}{
      \vartheta_4(\frac{3}{2} \gamma \mathfrak{a}_2 \pm \frac{3}{2}\mathfrak{b} + \frac{1}{4} \tau + \frac{k}{2}\tau + \frac{\ell}{2})
      \vartheta_4(\frac{3}{2} \gamma \mathfrak{a}_2 \mp \frac{1}{2}\mathfrak{b} + \frac{1}{4} \tau + \frac{k}{2}\tau + \frac{\ell}{2})
    }
    $
  \end{tabular}
\end{center}

% \begin{align}
%   R_{1,\pm}^{(1)} = & \ \frac{i}{6}\eta(\tau)^3
%   \frac{
%     \vartheta_4(3 \mathfrak{a}_2 \pm \mathfrak{b})
%     \vartheta_1(3 \mathfrak{a}_2 \pm 2 \mathfrak{b})
%   }{
%     \vartheta_1(\pm 2 \mathfrak{b})
%     \vartheta_1(3 \mathfrak{a}_2)
%     \vartheta_4(3 \mathfrak{a}_2 \pm 3 \mathfrak{b})
%   } \ , \\
%   R_{2,\pm}^{(1)} = & \ \frac{i}{6} \eta(\tau)^3
%     \frac{
%       \vartheta_4(3 \mathfrak{a}_2 \mp \mathfrak{b})
%       \vartheta_1(3 \mathfrak{a}_2 \mp 2 \mathfrak{b})
%     }{
%       \vartheta_1(\pm 2 \mathfrak{b})
%       \vartheta_1(3 \mathfrak{a}_2)
%       \vartheta_4(3 \mathfrak{a}_2 \mp 3 \mathfrak{b})}
%   = - R^{(1)}_{1, \mp} \ ,\\ 
%   R^{(1)}_{3,\pm, k\ell} = & \ \frac{i}{12} \frac{\eta(\tau)^3}{\vartheta_1(\pm 2 \mathfrak{b})} \prod_{\gamma = \pm} \frac{\vartheta_1(\frac{3}{2} \gamma \mathfrak{a}_2 \pm \frac{1}{2}\mathfrak{b} + \frac{1}{4}\tau + \frac{k}{2}\tau + \frac{\ell}{2})^2}{
%     \vartheta_4(\frac{3}{2} \gamma \mathfrak{a}_2 \pm \frac{3}{2}\mathfrak{b} + \frac{1}{4} \tau + \frac{k}{2}\tau + \frac{\ell}{2})
%     \vartheta_4(\frac{3}{2} \gamma \mathfrak{a}_2 \mp \frac{1}{2}\mathfrak{b} + \frac{1}{4} \tau + \frac{k}{2}\tau + \frac{\ell}{2})
%   } \ .  \nonumber
% \end{align}

It can be shown that,
\begin{align}
  - \oint \frac{da_2}{2\pi i a_2} a_2^n R^{(1)}_{1,\pm} = 0 \ , \qquad
  \text{if } n \not \in 3 \mathbb{Z} \ .
\end{align}
Therefore, we only focus on $n_1 + n_2 = 3p \ge 0$. Note also that $n_2 - 2n_1 = 3(p - n_1)$ in the second sum is also a multiple of $3$. With this assumption,
\begin{align}
  & \ - \sum_{\pm}\frac{b^{\pm n_1}}{q^{n_1/2} - q^{-n_1/2}} \oint \frac{da_2}{2\pi i a_2} a_2^{3p} R^{(1)}_{1,\pm}
  = - \sum_{\pm}\frac{b^{\pm n_1}}{q^{n_1/2} - q^{-n_1/2}} \oint \frac{da_2}{2\pi i a_2} a_2^{p} \left[R^{(1)}_{1,\pm}\right]_{3\mathfrak{a}_2 \to \mathfrak{a}_2} \nonumber \\
  = & \ - \delta_{n_1 + n_2 = 0} \sum_\pm
  \frac{b^{\pm n_1}}{q^{\frac{n_1}{2}} - q^{- \frac{n_1}{2}}}
  \frac{1}{6}\frac{\vartheta_4(\mathfrak{b})}{\vartheta_4(3 \mathfrak{b})}
  \left(
  E_1 \begin{bmatrix}
    -1 \\ b^{\pm 2} q^{\frac{1}{2}}
  \end{bmatrix}
  - E_1 \begin{bmatrix}
    -1 \\ b^{\mp}  
  \end{bmatrix}
  \right) \\
  & - \delta_{n_1 + n_2 \ne 0} \sum_{\pm}
  \frac{b^{\pm n_1}}{q^{\frac{n_1}{2}} - q^{- \frac{n_1}{2}}}
  \frac{1}{6}\frac{\vartheta_4(\mathfrak{b})}{\vartheta_4(3 \mathfrak{b})}
  \left(
  \frac{q^{\frac{1}{2}p} - b^{\mp 3 p}}{q^{p/2} - q^{-p/2}}
  \right) \ . \nonumber
\end{align}
Similarly
\begin{align}
  & \ - \oint \frac{da_2}{2\pi i a_2} a_2^{n_2}\sum_{\pm} R^{(1)}_{2, \pm}\frac{a_2^{-2n_1}b^{\pm n_1}}{q^{n_1/2} - q^{- n_1/2}} \nonumber \\
  = & \ - \delta_{n_2 - 2n_1 = 0}
  \frac{1}{6}\frac{\vartheta_4(\mathfrak{b})}{\vartheta_4(3 \mathfrak{b})}
  \sum_{\pm} \frac{b^{\pm n_1}}{q^{n_1/2} - q^{- n_1/2}}
  \left(
    E_1 \begin{bmatrix}
      -1 \\ b^{\mp 2}q^{\frac{1}{2}}
    \end{bmatrix}
    - E_1 \begin{bmatrix}
      -1 \\ b^{\pm}
    \end{bmatrix}
  \right)\\
  & \ + \delta_{n_2 - 2n_1 \ne 0}
  \frac{1}{6}
  \frac{\vartheta_4(\mathfrak{b})}{\vartheta_4(3 \mathfrak{b})}
  \sum_{\pm} \frac{b^{\pm n_1}}{q^{n_1/2} - q^{- n_1/2}}
  \left(
  \frac{q^{\frac{n_2 - 2n_1}{6}} - b^{\pm (n_2 - 2n_1)}}{q^{\frac{n_2 - 2n_1}{6}} - q^{- \frac{n_2 - 2n_1}{6}}}
  \right) \ . \nonumber
\end{align}
Lastly, one can also check that
\begin{align}
  \oint \frac{da_2}{2\pi i a_2}a_2^n R^{(1)}_{3, \pm, k\ell} = 0, \qquad
  \text{if } n \not \in \frac{3}{2} \mathbb{Z} \ .
\end{align}
Therefore, since $n_1 + n_2 $ is an integer, we may assume $n_2 - n_1/2 = n_1 + n_2 - \frac{3}{2}n_1 = 3p - \frac{3n_1}{2}$ with $p \in \mathbb{Z}$ in order for the integral to be non-zero,
\begin{align}
  & \ - \sum_{\pm, k,\ell} \frac{b^{\pm n_1/2}q^{n_1/4} q^{\frac{k-1}{2} n_1} (-1)^{\ell n_1}}{q^{n_1/2} - q^{- n_1/2}}
  \oint \frac{da_2}{2\pi i a_2} R^{(1)}_{3, \pm, k\ell} a_2^{n_2 - \frac{n_1}{2}} \nonumber \\
  = & \ \delta_{n_2 \ne \frac{1}{2}n_1} \frac{\vartheta_4(\mathfrak{b})}{12\vartheta_4(3 \mathfrak{b})}
  \sum_{\alpha, \gamma = \pm}\sum_{k,\ell = 0,1}
  \gamma
  \frac{
    b^{\frac{\alpha}{2}( (1 + \gamma) n_1 - 2\gamma n_2)}
    q^{- \frac{1}{12}(2k - 1)( (\gamma - 1)n_1 - 2\gamma n_2 )}
  }{
    (q^{\frac{n_1}{2}} - q^{- \frac{n_1}{2}})
    (q^{\frac{1}{6}(2n_2 - 1)}
        - q^{ - \frac{1}{6}(2n_2 - 1)})
  } \\
  & \ - \delta_{n_2 = \frac{1}{2}n_1}
  \frac{\vartheta_4(\mathfrak{b})}{12\vartheta_4(3 \mathfrak{b})}
  \sum_{\alpha, \gamma = \pm}\sum_{k,\ell = 0}^{1}
  \gamma
  \frac{
    b^{\alpha \frac{n_1}{2}}
    q^{\frac{1}{4}n_1(2k - 1)}
    (-1)^{\ell n_1}
  }{
    q^{n_1/2} - q^{- n_1/2}
  }
  E_1 \begin{bmatrix}
    -1\\
    b^{ - \frac{1}{2}\alpha(3\gamma + 1)}  
    q^{- \frac{1}{4}(2k(\gamma - 1) - (\gamma + 1))}
  \end{bmatrix} \ . \nonumber
\end{align}

In the above we have used the poles and residues of the $R$-factors listed in the following table.
{
\renewcommand{\arraystretch}{1.5}
\begin{table}[h!]
\centering
  \begin{tabular}{c|c|c}
    factor & poles & residues\\
    \hline
    $R^{(1)}_{1,\pm}$ & $\mathfrak{a}_2 = 0$ & $ - \frac{i}{6\eta(\tau)} \frac{\vartheta_4( \mathfrak{b})}{\vartheta_4( 3 \mathfrak{b})}$\\
                      & $\mathfrak{a}_2 = \mp 3 \mathfrak{b} + \frac{\tau}{2}$ & $ + \frac{i}{6\eta(\tau)} \frac{\vartheta_4( \mathfrak{b})}{\vartheta_4( 3 \mathfrak{b})}$\\
    \hline
    $R^{(1)}_{2,\pm}$ & $\mathfrak{a}_2 = 0$ & $ + \frac{i}{6\eta(\tau)} \frac{\vartheta_4( \mathfrak{b})}{\vartheta_4( 3 \mathfrak{b})}$\\
                      & $\mathfrak{a}_2 = \pm 3 \mathfrak{b} + \frac{\tau}{2}$ & $ - \frac{i}{6\eta(\tau)} \frac{\vartheta_4( \mathfrak{b})}{\vartheta_4( 3 \mathfrak{b})}$\\
    \hline
    $R^{(1)}_{3,\pm,k\ell}$ & $\mathfrak{a}_2 = \mp \frac{3}{2} \gamma \mathfrak{b} + \frac{\tau}{2} + \frac{1}{4}(2k - 1)\gamma \tau + \frac{\ell}{2}$, $\gamma = \pm 1$ & $ \gamma \frac{\vartheta_4(\mathfrak{b})}{12 \vartheta_4(3 \mathfrak{b})}$
  \end{tabular}
\end{table}
}

Putting all the above together, we have
\begin{align}
  & \ - \frac{1}{N!} \frac{\eta(\tau)^{3N - 3}}{\vartheta_4(\mathfrak{b})^{N - 1}}\oint \prod_{A = 1}^{N - 1}  \frac{da_A}{2\pi i a_A}
  a_1^{n_1} a_2^{n_2}
  \prod_{\substack{A, B = 1 \\ A\ne B}}^N \frac{\vartheta_1(\mathfrak{a}_A - \mathfrak{a}_B)}{\vartheta_4(\mathfrak{b} + \mathfrak{a}_A - \mathfrak{a}_B)}\nonumber\\
  = & \ 0 \qquad \text{ if } n_1 + n_2 \ne 0 \mod 3 ,  \\
  \text{else} = & \ 
  + \delta_{n_1 + n_2 = 0} 
    \frac{1}{6}\frac{\vartheta_4(\mathfrak{b})}{\vartheta_4(3 \mathfrak{b})}
    \sum_\pm
    \frac{b^{\pm n_1}}{q^{\frac{n_1}{2}} - q^{- \frac{n_1}{2}}}
    \left(
    E_1 \begin{bmatrix}
      -1 \\ b^{\pm 2} q^{\frac{1}{2}}
    \end{bmatrix}
    - E_1 \begin{bmatrix}
      -1 \\ b^{\mp}  
    \end{bmatrix}
    \right) \nonumber\\
    & - \delta_{n_1 + n_2 \ne 0} 
    \frac{1}{6}\frac{\vartheta_4(\mathfrak{b})}{\vartheta_4(3 \mathfrak{b})}
    \sum_{\pm}
    \frac{b^{\pm n_1}}{q^{\frac{n_1}{2}} - q^{- \frac{n_1}{2}}}
    \left(
    \frac{q^{\frac{1}{2}p} - b^{\mp 3 p}}{q^{p/2} - q^{-p/2}}
    \right) \nonumber\\
    & \ - \delta_{n_2 = 2n_1}
      \frac{1}{6}\frac{\vartheta_4(\mathfrak{b})}{\vartheta_4(3 \mathfrak{b})}
      \sum_{\pm} \frac{b^{\pm n_1}}{q^{n_1/2} - q^{- n_1/2}}
      \left(
        E_1 \begin{bmatrix}
          -1 \\ b^{\mp 2}q^{\frac{1}{2}}
        \end{bmatrix}
        - E_1 \begin{bmatrix}
          -1 \\ b^{\pm}
        \end{bmatrix}
      \right) \nonumber\\
      & \ + \delta_{n_2 \ne 2n_1}
      \frac{1}{6}
      \frac{\vartheta_4(\mathfrak{b})}{\vartheta_4(3 \mathfrak{b})}
      \sum_{\pm} \frac{b^{\pm n_1}}{q^{n_1/2} - q^{- n_1/2}}
      \left(
      \frac{q^{\frac{n_2 - 2n_1}{6}} - b^{\pm (n_2 - 2n_1)}}{q^{\frac{n_2 - 2n_1}{6}} - q^{- \frac{n_2 - 2n_1}{6}}}
      \right)\nonumber \\
    & \ - \delta_{n_2 = \frac{1}{2}n_1}
    \frac{\vartheta_4(\mathfrak{b})}{12\vartheta_4(3 \mathfrak{b})}
    \sum_{\alpha, \gamma = \pm}\sum_{k,\ell = 0}^{1}
    \gamma
    \frac{
      b^{\alpha \frac{n_1}{2}}
      q^{\frac{1}{4}n_1(2k - 1)}
      (-1)^{\ell n_1}
    }{
      q^{n_1/2} - q^{- n_1/2}
    }
    E_1 \begin{bmatrix}
      -1\\
      b^{ - \frac{1}{2}\alpha(3\gamma + 1)}  
      q^{- \frac{1}{4}(2k(\gamma - 1) - (\gamma + 1))}
    \end{bmatrix} \nonumber\\
    & \ + \delta_{n_2 \ne \frac{1}{2}n_1} \frac{\vartheta_4(\mathfrak{b})}{12\vartheta_4(3 \mathfrak{b})}
      \sum_{\alpha, \gamma = \pm}\sum_{k,\ell = 0,1}
      \gamma
      \frac{
        b^{\frac{\alpha}{2}( (1 + \gamma) n_1 - 2\gamma n_2)}
        q^{- \frac{1}{12}(2k - 1)( (\gamma - 3)n_1 - 2\gamma n_2 )}
      }{
        (q^{\frac{n_1}{2}} - q^{- \frac{n_1}{2}})
        (q^{\frac{1}{6}(2n_2 - n_1)}
            - q^{ - \frac{1}{6}(2n_2 - n_1)})
      } \nonumber \ .
\end{align}
The formula above implies the following symmetry which can be used to simplify computations,
\begin{align}
  \mathcal{I}(n_1, n_2) = & \ \mathcal{I}(n_2, n_1) = \mathcal{I}(- n_1, - n_2), \\
  \mathcal{I}(n_1, n_2) = & \ \mathcal{I}(n_1, n_1 - n_2) = \mathcal{I}(n_2 - n_1, n_2) \ .
\end{align}
With this formula, one can compute any Wilson line index in any $SU(3)$ representation $\mathcal{R}$ in closed-form. For example,
\begin{align}
  \langle W_{[1,1]} \rangle
  = & \ 2\mathcal{I}_{\mathcal{N} = 4 \ SU(3)} + 6 \mathcal{I}_{1,2} \nonumber \\
  = & \ 2 \mathcal{I}_{\mathcal{N}=4 \ SU(3)}\\
  & \ + \frac{\vartheta_4(\mathfrak{b})}{\vartheta_4(3 \mathfrak{b})}
  \Bigg[
    \frac{b\sqrt{q} - (1+b^4)q + b^3 q^{\frac{3}{2}}}{b^2 (1 - q)^2}
    + \frac{(b^2 -1)\sqrt{q}}{b(q-1)}
    \left(
    E_1 \begin{bmatrix}
      -1 \\ b  
    \end{bmatrix}
    + E_1 \begin{bmatrix}
      -1 \\ b^2 q^{\frac{1}{2}}
    \end{bmatrix}
    \right)
  \Bigg] \ . \nonumber
\end{align}

\begin{align}
  \langle W_{[2,2]} \rangle
  = & \ 3\mathcal{I}_{\mathcal{N} = 4 \ SU(3)} + 12 \mathcal{I}_{1,2}
  + 6 \mathcal{I}_{2,4} + 6 \mathcal{I}_{3,0} \nonumber \\
  = & \ 3 \mathcal{I}_{\mathcal{N}=4 \ SU(3)} \nonumber\\
  & \ + \frac{\vartheta_4(\mathfrak{b})}{\vartheta_4(3 \mathfrak{b})}
  \Bigg[
    \frac{\sqrt{q} (b^3 q^{\frac{1}{2}}-1) (-b^5 q-2 b^4 q^{\frac{1}{2}} (q+1)-b^3 (q (q+4)+2))}{b^4 \left(q^2-1\right)^2}
  \Bigg] \\
  & \ + \frac{\vartheta_4(\mathfrak{b})}{\vartheta_4(3 \mathfrak{b})}
  \Bigg[
    \frac{\sqrt{q} (+b^2 (2 q (q+2)+1) q^{\frac{1}{2}}+2 b (q+1) q+q^{3/2})}{b^4 \left(q^2-1\right)^2}
  \Bigg] \nonumber \\
  & \ + \frac{\vartheta_4(\mathfrak{b})}{\vartheta_4(3 \mathfrak{b})}
  \frac{\sqrt{q}(b^2 -1) \Big[(b^2+1) \sqrt{q}+2 b q+2 b \Big]}{b^2(q^2 - 1)}
  \left(
  E_1 \begin{bmatrix}
    -1 \\ b  
  \end{bmatrix}
  + E_1 \begin{bmatrix}
    -1 \\ b^2\sqrt{q}  
  \end{bmatrix}
  \right) \ . \nonumber
\end{align}

\begin{align}
  \langle W_{[3,3]} \rangle
  = & \ 4\mathcal{I}_{\mathcal{N} = 4 \ SU(3)} + 18 \mathcal{I}_{1,2}
  + 12 \mathcal{I}_{2,4} + 12 \mathcal{I}_{3,0} + 12 \mathcal{I}_{4,5}
  + 6 \mathcal{I}_{3,6} \ .
\end{align}

\subsection{\texorpdfstring{$\mathcal{N} = 2$ $SU(3)$ SQCD}{}}

Let us also consider Wilson operator in the $SU(3)$ SQCD. The relevant integral reads
\begin{align}
  \mathcal{I}_{SU(3) \ \text{SQCD}} = & \ - \frac{1}{3!} \eta(\tau)^{16} \oint \prod_{A = 1}^2 \frac{da_A}{2\pi i a_A}
  \chi_\mathcal{R}(a)
  \frac{\prod_{A \ne B} \vartheta_1(\mathfrak{a}_A - \mathfrak{a}_B)}{\prod_{A = 1}^3 \prod_{i = 1}^{6} \vartheta_4(\mathfrak{a}_A - \mathfrak{m}_i)} \nonumber\\
  \coloneqq & \ \oint \prod_{A = 1}^2 \frac{da_A}{2\pi i a_A}
  \chi_\mathcal{R}(a) \mathcal{Z}(\mathfrak{a}, \mathfrak{m}) \ .
\end{align}

\subsubsection{Fundamental representation}

As the simplest example, we consider the fundamental representation
\begin{align}
  \chi (a) = a_1 + a_2 + \frac{1}{a_1 a_2} \ .
\end{align}
First we note that
\begin{align}
  \oint \prod_{A = 1}^2 \frac{da_A}{2\pi i a_A} a_1 \mathcal{Z}(\mathfrak{a}, \mathfrak{m})
  = \oint \prod_{A = 1}^2 \frac{da_A}{2\pi i a_A} a_2 \mathcal{Z}(\mathfrak{a}, \mathfrak{m}) \ .
\end{align}
Therefore we simply compute the one with $a_1$ insertion. The relevant poles when performing the $a_1$ integration are all imaginary given by
\begin{align}
  \mathfrak{a}_1 = \mathfrak{m}_{j_1} + \frac{\tau}{2}, \qquad
  \mathfrak{a}_1 = - \mathfrak{a}_2 - \mathfrak{m}_{j_1} + \frac{\tau}{2} \ ,
\end{align}
with the respective residues
\begin{align}
  R_{j_1} \coloneqq - \frac{1}{6}
  \frac{
    \eta(\tau)^{13}q^{\frac{1}{8}}
    \prod_{A\ne B} \vartheta_1(\mathfrak{a}_A - \mathfrak{a}_B)|_{\mathfrak{a}_1 = \mathfrak{m}_{j_1} + \frac{\tau}{2}}
  }{
    \prod_i\vartheta_4(\mathfrak{a}_2 - \mathfrak{m}_i)
    \prod_i\vartheta_4(\mathfrak{a}_2 + \mathfrak{m}_{j_1} + \mathfrak{m}_i + \frac{\tau}{2})
    \prod_{i \ne j_1}\vartheta_4(\mathfrak{m}_i - \mathfrak{m}_{j_1} - \frac{\tau}{2})
  } \ ,  \quad -R_{j_1} \ . \nonumber
\end{align}
Therefore, after the $a_1$ integral we are left with
\begin{align}
  \oint \frac{da_2}{2\pi i a_2} \left[- \sum_{j_1 = 1}^{6}R_{j_1} \frac{1}{q^1 - 1} (m_{j_1}q^{\frac{1}{2}})
      + \sum_{j_1 = 1}^{6}R_{j_1} \frac{1}{q^1 - 1} (a_2^{-1} m^{-1}_{j_1}q^{\frac{1}{2}})\right] \ .
\end{align}

Next we perform the $a_2$ integral. Each residue $R_{j_1}$ is an elliptic function with respect to $\mathfrak{a}_2$, with imaginary and real poles
\begin{align}
  \mathfrak{a}_2 = & \ + \mathfrak{m}_{j_2} + \frac{\tau}{2}, & j_2 \ne & \ j_1 \\
  \text{or}, \qquad = & \ - \mathfrak{m}_{j_1} - \mathfrak{m}_{j_2} \ ,  & j_2 \ne & \ j_1 \ .
\end{align}
The corresponding residues are, respectively,
\begin{align}
  R_{j_1 j_2} = \frac{
      \eta(\tau)^{10}
      \vartheta_4(2 \mathfrak{m}_{j_1} + \mathfrak{m}_{j_2})
      \vartheta_4(\mathfrak{m}_{j_1} + 2\mathfrak{m}_{j_2})}{
    6
    \prod_{i\ne j_1, j_2}\vartheta_1(\mathfrak{m}_{j_1} - \mathfrak{m}_i)\vartheta_1(\mathfrak{m}_{j_2} - \mathfrak{m}_i)
    \prod_{i \ne j_1, j_2} \vartheta_4(\mathfrak{m}_{j_1}+ \mathfrak{m}_{j_2} + \mathfrak{m}_i)
  }, \quad
  - R_{j_1 j_2} \ .
\end{align}
We also set $R_{j_1 j_2} = 0$ when $j_1 = j_2$. With this, we have by applying (\ref{integration-formula-f})
\begin{align}
  \oint \frac{da_2}{2\pi i a_2} R_{j_1} = R_{j_1}(\mathfrak{a}_2 = 0)
  + \sum_{j_2 = 1}^{6} R_{j_1 j_2}E_1 \begin{bmatrix}
    -1 \\ m_{j_2}  
  \end{bmatrix}
  + R_{j_1 j_2}E_1 \begin{bmatrix}
    -1 \\ m_{j_1}m_{j_2}q^{-\frac{1}{2}}
  \end{bmatrix}\ ,
\end{align}
where we have picked $\mathfrak{a}_2 = 0$ as the reference point, and
\begin{align}
  \oint \frac{da_2}{2\pi i a_2}a_2^{-1}R_{j_1}
  = & \ + \sum_{j_2 = 1}^6 R_{j_1 j_2} \frac{1}{1 - q}m_{j_1}m_{j_2}
  - \sum_{j_2 = 1}^{6}R_{j_1 j_2} \frac{1}{q^{-1} - 1} (m_{j_2}q^{\frac{1}{2}})^{-1} \\
  = & \ + \sum_{j_2 = 1}^6 R_{j_1 j_2} \frac{m_{j_1}m_{j_2} - m_{j_2}^{-1}q^{\frac{1}{2}}}{1 - q} \ .
\end{align}
Collecting the results, the integral with $a_1$-insertion reads
\begin{align}
  & \ \oint \prod_{A = }^{2} \frac{da_A}{2\pi i a_A} a_1 \mathcal{Z} \\
  = & \ \frac{q^{\frac{1}{2}}}{1 - q} \sum_{j_1 = 1}^{6}m_{j_1}\left(
  R_{j_1}(\mathfrak{a}_2 = 0)
  + \sum_{j_2 = 1}^{6} R_{j_1 j_2}E_1 \begin{bmatrix}
    -1 \\ m_{j_2}  
  \end{bmatrix}
  + R_{j_1 j_2}E_1 \begin{bmatrix}
    -1 \\ m_{j_1}m_{j_2}q^{-\frac{1}{2}}
  \end{bmatrix}
  \right) \\
  & \ - \frac{1}{(1-q)^2} \sum_{j_1 = 1}^{6}
  \sum_{j_2 = 1}^{6}R_{j_1 j_2}(m_{j_2}q^{\frac{1}{2}} - m_{j_1}^{-1}m_{j_2}^{-1}q) \ .
\end{align}

Next we compute the integral
\begin{align}
  \oint \prod_{A = 1}^{2} \frac{da_A}{2\pi i a_A} \frac{1}{a_1 a_2} \mathcal{Z} \ .
\end{align}
Similar to the previous computation, we first integrate $a_1$ with poles $\mathfrak{a}_1 = \mathfrak{m}_{j_1} + \frac{\tau}{2}$ and $\mathfrak{a}_1 = - \mathfrak{a}_2 - \mathfrak{m}_{j_2} + \frac{\tau}{2}$,
\begin{align}
  & \ \oint \frac{da_2}{2\pi i a_2} \frac{1}{a_2} \left[- \sum_{j_1 = 1}^{6}R_{j_1} \frac{1}{q^{ - 1} - 1} (m_{j_1}q^{\frac{1}{2}})^{-1}
      + \sum_{j_1 = 1}^{6}R_{j_1} \frac{1}{q^{-1} - 1} (a_2^{-1} m^{-1}_{j_1}q^{\frac{1}{2}})^{-1}\right]\\
  = & \ \oint \frac{da_2}{2\pi i a_2} \left[- \sum_{j_1 = 1}^{6}R_{j_1} \frac{1}{q^{ - 1} - 1} (a_2^{-1} m_{j_1}^{-1}q^{-\frac{1}{2}})
      + \sum_{j_1 = 1}^{6}R_{j_1} \frac{1}{q^{-1} - 1} m_{j_1}q^{ - \frac{1}{2}}\right] \ .
\end{align}
Carrying out the $a_2$ integration, we have
\begin{align}
  & \ \oint \frac{da_1}{2\pi i a_1}\frac{da_2}{2\pi i a_2} \frac{1}{a_1 a_2} \mathcal{Z}\\
  = & \ + \frac{q^{\frac{1}{2}}}{1 - q} \sum_{j_1 = 1}^{6} m_{j_1} \left(
  R_{j_1}(\mathfrak{a}_2 = 0)
  + \sum_{j_2 = 1}^{6}R_{j_1 j_2} E_1 \begin{bmatrix}
    -1 \\ m_{j_2}  
  \end{bmatrix}
  + R_{j_1 j_2} E_1 \begin{bmatrix}
    -1 \\ m_{j_1} m_{j_2} q^{-1/2}  
  \end{bmatrix}
  \right) \\
  & \ - \frac{1}{(1 - q)^2} \sum_{j_1 = 1}^{6}
  \sum_{j_2 = 1}^{6} R_{j_1 j_2} (m_{j_2}q^{+ \frac{1}{2}} - m_{j_1}^{-1}m_{j_2}^{-1}q) \ .
\end{align}
Actually, this is identical to the previous result,
\begin{align}
  \oint \frac{da_1}{2\pi i a_1}\frac{da_2}{2\pi i a_2} \frac{1}{a_1 a_2} \mathcal{Z}
  = \oint \frac{da_1}{2\pi i a_1}\frac{da_2}{2\pi i a_2} a_1 \mathcal{Z}
  = \oint \frac{da_1}{2\pi i a_1}\frac{da_2}{2\pi i a_2} a_2 \mathcal{Z} \ .
\end{align}

Combining the integration of all three terms in the fundamental characters, we therefore have a fairly simple result,
\begin{align}
  \langle W_{\mathbf{3}} \rangle_{SU(3) \ \text{SQCD}}
  = & \ \frac{3q^{\frac{1}{2}}}{1 - q}\sum_{j_1 = 1}^{6}\left(
  R_{j_10} + \sum_{j_2 = 1}^{6}R_{j_1 j_2} \left(E_1 \begin{bmatrix}
    -1 \\ m_{j_2}  
  \end{bmatrix}
  + E_1 \begin{bmatrix}
    -1 \\ m_{j_1} m_{j_2}q^{-\frac{1}{2}}  
  \end{bmatrix}
  \right)
  \right) \\
  & \ - \frac{3}{(1-q)^2} \sum_{j_1, j_2 = 1}^{6}R_{j_1 j_2} (m_{j_2} q^{\frac{1}{2}} - m_{j_1}^{-1}m_{j_2}^{-1}q) \ .
\end{align}
where we abbreviate
\begin{align}
  R_{j_1 0} \coloneqq R_{j_1}(\mathfrak{a}_2 = 0) \ .
\end{align}

\subsubsection{General representation}

The above computation can be generalized to insertion of all half Wilson operators  in any representation of the gauge group $SU(3)$. The basic building block is of course a monomial $a_1^{n_1} a_2^{n_2}$. Let us therefore compute the basic integral
\begin{align}
  \oint \frac{da_1}{2\pi i a_1}\frac{da_2}{2\pi i a_2} a_1^{n_1} a_2^{n_2} \mathcal{Z} \ .
\end{align}
Note that
\begin{align}
  \oint \frac{da_1}{2\pi i a_1}\frac{da_2}{2\pi i a_2}a_2^{n_2}\mathcal{Z}
  = \oint \frac{da_1}{2\pi i a_1}\frac{da_2}{2\pi i a_2}a_1^{n_2}\mathcal{Z} \ .
\end{align}
Therefore, without loss of generality we assume $n_1 \in \mathbb{Z}_{\ne0}$, and we first perform $a_1$ and then $a_2$ integration. The first step picks up the imaginary poles $\mathfrak{a}_1 = \mathfrak{m}_{j_1} + \frac{\tau}{2}$ and $- \mathfrak{a}_2 - \mathfrak{m}_{j_1} + \frac{\tau}{2}$, which produces
\begin{align}
  & \ \oint \frac{da_2}{2\pi i a_2} a_2^{n_2} \left[- \sum_{j_1}^{6}R_{j_1} \frac{1}{q^{n_1} - 1}(m_{j_1} q^{\frac{1}{2}})^{n_1}
  - \sum_{j_1 = 1}^{6}(-R_{j_1}) \frac{1}{q^{n_1} - 1}(a_2^{-1} m_{j_1}^{-1}q^{\frac{1}{2}})^{n_1}
  \right] \\
  = & \ \oint \frac{da_2}{2\pi i a_2} \left[- \sum_{j_1}^{6} a_2^{n_2} R_{j_1} \frac{1}{q^{n_1} - 1}(m_{j_1} q^{\frac{1}{2}})^{n_1}
  - \sum_{j_1 = 1}^{6}(-R_{j_1})a_2^{n_2 - n_1} \frac{1}{q^{n_1} - 1}(m_{j_1}^{-1}q^{\frac{1}{2}})^{n_1}
  \right] \ . \nonumber
\end{align}
Depending on whether $n_2 = 0$ or $n_2 - n_1 = 0$ or a generic $n_2$, the $a_2$-integration of the two terms take different form.

For the first term, if $n_2 = 0$, then the integral picks up contributions from the imaginary poles $\mathfrak{m}_{j_2} + \frac{\tau}{2}$ and the real poles $- \mathfrak{m}_{j_1} - \mathfrak{m}_{j_2}$, which reads
\begin{align}
  & \ \oint \frac{da_2}{2\pi i a_2} \left[- \sum_{j_1}^{6} a_2^{n_2 = 0} R_{j_1} \frac{1}{q^{n_1} - 1}(m_{j_1} q^{\frac{1}{2}})^{n_1}\right] \\
  = & \ - \sum_{j_1 = 1}^{6}\frac{(m_{j_1}q^{1/2})^{n_1}}{q^{n_1} - 1}\left(
  R_{j_10} + R_{j_1 j_2} E_1 \begin{bmatrix}
    -1 \\ m_{j_2}  
  \end{bmatrix}
  + R_{j_1 j_2} E_1 \begin{bmatrix}
    -1 \\ m_{j_1} m_{j_2}q^{-1/2}  
  \end{bmatrix}
  \right) \ .
\end{align}
However, if $n_2 \ne 0$, then
\begin{align}
  & \ \oint \frac{da_2}{2\pi i a_2} \left[- \sum_{j_1}^{6} a_2^{n_2} R_{j_1} \frac{1}{q^{n_1} - 1}(m_{j_1} q^{\frac{1}{2}})^{n_1}\right] \\
  = & \ - \sum_{j_1 = 1}^{6} \frac{(m_{j_1}q^{\frac{1}{2}})^{n_1}}{q^{n_1} - 1}
  \left(
  - \sum_{j_2 = 1}^{6}R_{j_1 j_2} \frac{1}{q^{n_2} - 1}(m_{j_2}q^{\frac{1}{2}})^{n_2}
  - \sum_{j_2 = 1}^{6} (- R_{j_1 j_2}) \frac{1}{1 - q^{-m_2}} (m_{j_1}^{-1} m_{j_2}^{-1})^{n_2}
  \right) \ . \nonumber
\end{align}

For the second term, if $n_2 - n_1 = 0$, then
\begin{align}
  & \ \oint \frac{da_2}{2\pi i a_2}\left[
    \sum_{j_1 = 1}^{6}R_{j_1} a_2^{n_2 - n_1} \frac{1}{q^{n_1} - 1} (m_{j_1}^{-1} q^{\frac{1}{2}})^{n_1}
  \right]\\
  = & \ \sum_{j_1 = 1}^{6} \frac{m_{j_1}^{-n_1}q^{\frac{n_1}{2}}}{q^{n_1} - 1} \left(
  R_{j_10} + R_{j_1 j_2} E_1 \begin{bmatrix}
    -1 \\ m_{j_2}  
  \end{bmatrix}
  + R_{j_1 j_2} E_1 \begin{bmatrix}
    -1 \\ m_{j_1} m_{j_2}q^{-1/2}  
  \end{bmatrix}
  \right) \ .
\end{align}
On the other hand, if $n_2 - n_1 \ne 0$ then 
\begin{align}
  & \ \oint \frac{da_2}{2\pi i a_2}\left[
    \sum_{j_1 = 1}^{6}R_{j_1} a_2^{n_2 - n_1} \frac{1}{q^{n_1} - 1} (m_{j_1}^{-1} q^{\frac{1}{2}})^{n_1}
  \right]\\
  = & \ \sum_{j_1 = 1}^{6} \frac{m_{j_1}^{-n_1} q^{\frac{n_1}{2}}}{q^{n_1} - 1}
  \left(
  - \sum_{j_2 = 1}^{6} R_{j_1j_2} \frac{m_{j_2}^{n_2 - n_1}q^{\frac{1}{2}(n_2 - n_1)}}{q^{n_1 - n_2} - 1}
  + \sum_{j_2 = 1}^{6} R_{j_1 j_2} \frac{(m_{j_1}^{-1} m_{j_2}^{-1})^{n_2 - n_1}}{1 - q^{-(n_2 - n_1)}}
  \right) \ .
\end{align}

Putting all terms together, we have
\begin{align}
  & \ \oint \frac{da_1}{2\pi i a_1}\frac{da_2}{2\pi i a_2}a_1^{n_1 \ne 0} a_2^{n_2} \mathcal{Z} \nonumber\\
  = & \ - \delta_{n_2 = 0} \sum_{j_1 = 1}^{6}\frac{m_{j_1}^{n_1}q^{\frac{1}{2} n_1}}{q^{n_1} - 1}\left(
  R_{j_10} + R_{j_1 j_2} E_1 \begin{bmatrix}
    -1 \\ m_{j_2}  
  \end{bmatrix}
  + R_{j_1 j_2} E_1 \begin{bmatrix}
    -1 \\ m_{j_1} m_{j_2}q^{-1/2}  
  \end{bmatrix}
  \right) \nonumber\\
  & \ + \delta_{n_2 \ne 0}\sum_{j_1, j_2 = 1}^{6}
    R_{j_1j_2}\frac{m_{j_1}^{n_1}q^{\frac{1}{2} n_1}}{q^{n_1} - 1}
    \frac{(m_{j_2}q^{\frac{1}{2}})^{n_2}-(m_{j_1}^{-1}m_{j_2}^{-1})^{n_2}q^{n_2}}{q^{n_2}-1} \nonumber \\
  & \ + \delta_{n_2 = n_1}\sum_{j_1 = 1}^{6} \frac{m_{j_1}^{-n_1}q^{\frac{n_1}{2}}}{q^{n_1} - 1} \left(
  R_{j_10} + R_{j_1 j_2} E_1 \begin{bmatrix}
    -1 \\ m_{j_2}  
  \end{bmatrix}
  + R_{j_1 j_2} E_1 \begin{bmatrix}
    -1 \\ m_{j_1} m_{j_2}q^{-1/2}  
  \end{bmatrix}
  \right) \\
  & \ - \delta_{n_2 \ne n_1}\sum_{j_1, j_2 = 1}^{6}
    R_{j_1j_2}
    \frac{m_{j_1}^{-n_1} q^{\frac{n_1}{2}}}{q^{n_1} - 1}
    \frac{m_{j_2}^{n_2-n_1}q^{\frac{1}{2}(n_2-n_1)}-(m_{j_1}^{-1}m_{j_2}^{-1})^{n_2-n_1}q^{n_2-n_1}}{q^{n_2-n_1}-1} \ . \nonumber
\end{align}

For example, for the Wilson operator in the anti-fundamental representation,
\begin{align}
  \langle W_{\overline {\mathbf{3}}}\rangle_{SU(3) \ \text{SQCD}}
  = & \ 3 \oint \frac{da_1}{2\pi i a_1}\frac{da_2}{2\pi i a_2} a_1^{-1} \mathcal{Z} \nonumber \\
  = & \  - 3 \frac{q^{\frac{1}{2}}}{1 - q} \sum_{j_1 = 1}^{6}m_{j_1}^{-1}\left(
  R_{j_10} + R_{j_1 j_2} E_1 \begin{bmatrix}
    -1 \\ m_{j_2}  
  \end{bmatrix}
  + R_{j_1 j_2} E_1 \begin{bmatrix}
    -1 \\ m_{j_1} m_{j_2}q^{-1/2}  
  \end{bmatrix}
  \right) \nonumber \\
  & \  + 3 \frac{1}{1-q} \sum_{j_1, j_2 = 1}^{6}
    R_{j_1j_2}
    \frac{m_{j_2}^{-1}q^{\frac{1}{2}} - m_{j_1} m_{j_2}q}{q-1}
\end{align}

\subsection{\texorpdfstring{ $\mathcal{N} = 4$ $SO(4)$ SYM}{}}

The Lie algebra $\mathfrak{so}(4)$ is isomorphic to $\mathfrak{su}(2)^2$. The Schur index of a Lagrangian theory is only sensitive to the gauge Lie algebra, and therefore the $\mathcal{N} = 4$ $SO(4)$ and $SU(2)^2$ gauge theory share an identical Schur index,
\begin{align}
  \mathcal{I}_{SU(2)^2} = \mathcal{I}_{SO(4)}
  = \frac{1}{4}\eta(\tau)^{4} \frac{\eta(\tau)^2}{\vartheta_4(\mathfrak{b})^2} & \ \oint \prod_{A = 1}^{2} \frac{da_A}{2\pi i a_A} \prod_{\alpha, \beta = \pm}\prod_{A < B} \frac{\vartheta_1(\alpha \mathfrak{a}_A + \beta \mathfrak{a}_B)}{\vartheta_4 (\alpha \mathfrak{a}_A + \beta\mathfrak{a}_B + \mathfrak{b})} \nonumber \\
  \coloneqq & \ \oint \prod_{A = 1}^{2} \frac{da_A}{2\pi i a_A} \mathcal{Z}(\mathfrak{a}_1, \mathfrak{a}_2) \ .
\end{align}
In the following we will compute a few full Wilson operator index and compare it with the $S$-dual 't Hooft operator index using the formula in \cite{Gang:2012yr}.

We first analyze the index of a full Wilson operator associated to the vector representation $\mathbf{4}$ and its $S$-dual. The full Wilson index reads
\begin{align}
  \langle W^\text{full}_{\mathbf{4}}\rangle_{SO(4) \ \mathcal{N} = 4} = \oint \prod_{A = 1}^{2} \frac{da_A}{2\pi i a_A} (a_1 + \frac{1}{a_1} + a_2 + \frac{1}{a_2})^2 \mathcal{Z} \ .
\end{align}
By a change of variables $\mathfrak{a}_1' \coloneqq \mathfrak{a}_1 + \mathfrak{a}_2$ and $\mathfrak{a}'_2 \coloneqq \mathfrak{a}_1 - \mathfrak{a}_2$,  the Wilson index factorized into a product of two identical integrals,
\begin{align}
  \langle W^\text{full}_{\mathbf{4}}\rangle_{SO(4) \ \mathcal{N} = 4}
  = \left[- \frac{1}{2} \oint \frac{da'}{2\pi i a'} \frac{(a' + 1)^2}{a'} \frac{\vartheta_1(\pm \mathfrak{a}')}{\vartheta_4(\pm \mathfrak{a}' + \mathfrak{b})} \frac{\eta(\tau)^3}{\vartheta_4(\mathfrak{b})}\right]^2 \ ,
\end{align}
which is identical to
\begin{align}
  (\langle W_{j = 1/2}^\text{full} \rangle_{SU(2) \ \mathcal{N} = 4})^2 \ .
\end{align}
The vector representation of $SO(4)$ is minuscule, and the S-dual 't Hooft index is safe from monopole bulling, given by
\begin{align}
  \langle H\rangle_{SO(4) \ \mathcal{N} = 4}
  = \oint \prod_{A = 1}^{2} \frac{da_A}{2\pi i a_A}
  \frac{4q^{\frac{1}{4}} (ba_1 - a_2)(-a_1 + ba_2)(b - a_1 a_2)(-1 + b a_1 a_2)}{b^2(\sqrt{q}a_1 - a_2)(\sqrt{q}a_2 - a_1)(\sqrt{q} - a_1 a_2)(-1 + \sqrt{q}a_1 a_2 ) } \mathcal{Z}' \ , \nonumber
\end{align}
where
\begin{align}
  \mathcal{Z}' = \frac{1}{4} \eta(\tau)^4 \frac{\eta(\tau)^2}{\vartheta_4(\mathfrak{b})^2}\prod_{\alpha, \beta = \pm} \frac{\vartheta_4(\alpha \mathfrak{a}_1 + \beta \mathfrak{a}_2)}{\vartheta_1(\alpha \mathfrak{a}_1 + \beta \mathfrak{a}_2 + \mathfrak{b})} \ .
\end{align}
In terms of the $a'$ variables, the above factorizes into
\begin{align}
  \langle H\rangle_{SO(4) \ \mathcal{N} = 4} = \left[\oint \frac{da'}{2\pi i a'_1}\frac{q^{\frac{1}{8}}(b - a'_1)(-1 + b a'_1)}{b(\sqrt{q} - a'_1)(-1 + \sqrt{q}a'_1)}\frac{\eta(\tau)^3}{\vartheta_4(\mathfrak{b})} \frac{\vartheta_4(\pm \mathfrak{a}')}{\vartheta_1(\pm\mathfrak{a}' + \mathfrak{b})}\right]^2 \ .
\end{align}
Up to the square and some simple factors, the result is identical to that of the $U(2)$ minimal 't Hooft operator index (\ref{U2-t-hooft}) in section \ref{section:N4SU(2)}, and naturally
\begin{align}
  \langle H\rangle_{SO(4) \ \mathcal{N} = 4} = \langle W^\text{full}_{\mathbf{4}}\rangle_{SO(4) \ \mathcal{N} = 4} \ .
\end{align}

Next we consider the index of a full Wilson operator in chiral spinor representation $\mathbf{2}$. The corresponding character is
\begin{align}
  \chi_\mathbf{2}(a) = \frac{1}{\sqrt{a_1 a_2}} + \sqrt{a_1 a_2} \ ,
\end{align}
and the relevant index is given by
\begin{align}
  \langle W_\mathbf{2}^\text{f}\rangle_{SO(4) \ \mathcal{N} = 4} = & \ \oint \prod_{A = 1}^{2} \frac{da_A}{2\pi i a_A}
  \chi_{\mathbf{2}}(a)^2 
  \mathcal{Z}(\mathfrak{a}_1, \mathfrak{a}_2) \\
  = & \ \oint \prod_{A = 1}^{2} \frac{da_A}{2\pi i a_A}
  (1 + 1 + a_1 a_2 + \frac{1}{a_1 a_2})
  \mathcal{Z}(\mathfrak{a}_1, \mathfrak{a}_2) \ .
\end{align}
In terms of the $a'$ variable, the above factorizes
\begin{align}
  \langle W_\mathbf{2}^\text{f}\rangle_{SO(4) \ \mathcal{N} = 4}
  = & \ \left[\oint \frac{da'_1}{2\pi i a'_1}(\chi_{j = 0} + \chi_{j = 1})(a'_1) \left(- \frac{1}{2}\right)\frac{\eta(\tau)^3}{\vartheta_4(\mathfrak{b})} \frac{\vartheta_1(\pm\mathfrak{a}'_1)}{\vartheta_1(\pm\mathfrak{a}'_1 + \mathfrak{b})}\right] \mathcal{I}_{SU(2) \ \mathcal{N} = 4} \nonumber \\
  = & \ \left( \mathcal{I}_{\mathcal{N} = 4 \ SU(2)} + \langle W^\text{full}_{j = 1}\rangle_{\mathcal{N} = 4 \ SU(2)} \right) \mathcal{I}_{SU(2) \ \mathcal{N} = 4} \ .
\end{align}

The S-dual 't Hooft line index is given by
\begin{align}
  \langle H\rangle
  = \oint \prod_{A = 1}^{2} \frac{da_A}{2\pi i a_A}
  \frac{2(b - a_1 a_2)(-1 + b a_1a_2)}{bq^{\frac{1}{4}}(\sqrt{q} - a_1 a_2)(-1 + \sqrt{q} a_1 a_2)}
  \mathcal{Z}' \ , \nonumber
\end{align}
where
\begin{equation}
  \mathcal{Z}' = \frac{1}{4}\eta(\tau)^4 \frac{\eta(\tau)^2}{\vartheta_4(\mathfrak{b})^2}
  \frac{\vartheta_4(\pm (\mathfrak{a}_1 + \mathfrak{a}_2))}{\vartheta_1(\pm (\mathfrak{a}_1 + \mathfrak{a}_2) + \mathfrak{b})}
  \frac{\vartheta_1(\pm (\mathfrak{a}_1 - \mathfrak{a}_2))}{\vartheta_4(\pm (\mathfrak{a}_1 - \mathfrak{a}_2) + \mathfrak{b})}  \ .
\end{equation}
In terms of the $a'$ variables,
\begin{align}
  \langle H\rangle
  = \left[- \oint \frac{da'_1}{2\pi i a'_1} \frac{(b - a'_1)(-1 + b a_1' )}{bq^{1/4}(\sqrt{q} - a'_1)(-1 + \sqrt{q}a'_1)} \frac{\vartheta_4(\pm \mathfrak{a}_1)}{\vartheta_1(\pm \mathfrak{a}_1 + \mathfrak{b})} \frac{\eta(\tau)^3}{\vartheta_4(\mathfrak{b})}\right] \mathcal{I}_{\mathcal{N} = 4 \ SU(2)} \ .
\end{align}
The equality from S-duality also follows from the discussion in section \ref{section:N4SU(2)}.

\subsection{\texorpdfstring{$\mathcal{N} = 4$ $SO(5)$ SYM}{}}

Let us now consider $\mathcal{N} = 4$ $SO(5)$ SYM with insertion of a half Wilson operator in the fundamental representation
\begin{align}
  \oint \frac{da_1}{2\pi i a_1}\frac{da_2}{2\pi i a_2}\chi_{\mathbf{5}}(a)
  \mathcal{Z}(\mathfrak{a}_1, \mathfrak{a}_2) \ ,
\end{align}
where
\begin{equation}
\mathcal{Z}(\mathfrak{a}_1, \mathfrak{a}_2) = \frac{1}{8} \frac{\eta(\tau)^6}{\vartheta_4(\mathfrak{b})^2} \frac{
    - \vartheta_1(\mathfrak{a}_1)^2
    \vartheta_1(\mathfrak{a}_2)^2
    \vartheta_1(\mathfrak{a}_1 + \mathfrak{a}_2)^2
    \vartheta_1(\mathfrak{a}_1 - \mathfrak{a}_2)^2
  }{
    \vartheta_4(\mathfrak{a}_1 \pm \mathfrak{b})
    \vartheta_4(\mathfrak{a}_2 \pm \mathfrak{b})
    \vartheta_4(\mathfrak{a}_1 + \mathfrak{a}_2\pm \mathfrak{b})
    \vartheta_4(\mathfrak{a}_1 - \mathfrak{a}_2\pm \mathfrak{b})
  }\ ,
\end{equation}
and
\begin{align}
  \chi_{\mathbf{5}}(a) = a_1 + \frac{1}{a_1} + a_2 + \frac{1}{a_2} + 1 \ .
\end{align}

From the symmetry $\mathcal{Z}(\mathfrak{a}_1, \mathfrak{a}_2) = \mathcal{Z}(\mathfrak{a}_2, \mathfrak{a}_1)$, we only need to compute
\begin{align}
  \oint \frac{da_1}{2\pi i a_1}\frac{da_2}{2\pi i a_2}a_1^{\pm 1}
  \mathcal{Z}(\mathfrak{a}_1, \mathfrak{a}_2) \ .
\end{align}
Moreover, the symmetry $\mathcal{Z}(\mathfrak{a}_1, \mathfrak{a}_2) = \mathcal{Z}( - \mathfrak{a}_1, \mathfrak{a}_2)$ also implies
\begin{align}
  \oint \frac{da_1}{2\pi i a_1}\frac{da_2}{2\pi i a_2}a_1
  \mathcal{Z}(\mathfrak{a}_1, \mathfrak{a}_2)
  = \oint \frac{da_1}{2\pi i a_1}\frac{da_2}{2\pi i a_2}a_1^{-1}
  \mathcal{Z}(\mathfrak{a}_1, \mathfrak{a}_2) \ .
\end{align}

The $a_1$-integration picks up imaginary poles
\begin{align}
  \mathfrak{a}_1 = \alpha \mathfrak{b} + \frac{\tau}{2}, \qquad
  \mathfrak{a}_1 = \beta \mathfrak{a}_2 + \gamma \mathfrak{b} + \frac{\tau}{2} \ , \qquad \alpha, \beta, \gamma = \pm \ ,
\end{align}
with residues respectively
\begin{align}
  R_\alpha \coloneqq \frac{i}{8}\eta(\tau)^3 \frac{\vartheta_4(\mathfrak{a}_2 + \alpha \mathfrak{b})\vartheta_4(\mathfrak{a}_2 - \alpha \mathfrak{b})}{
        \vartheta_1(2 \alpha \mathfrak{b})
        \vartheta_1(\mathfrak{a}_2 + 2 \alpha \mathfrak{b})
        \vartheta_1(\mathfrak{a}_2 - 2 \alpha \mathfrak{b})} \ ,
\end{align}
and
\begin{align}
  R_{\beta \gamma} \coloneqq\frac{i}{8} \eta(\tau)^3 \frac{
    \vartheta_4(\mathfrak{a}_2 + \beta \gamma \mathfrak{b})
    \vartheta_1(\mathfrak{a}_2)
    \vartheta_4(2 \mathfrak{a}_2 + \beta \gamma \mathfrak{b})^2
  }{
    \vartheta_1(\mathfrak{a}_2 + 2 \beta \gamma \mathfrak{b})
    \vartheta_4(\mathfrak{a}_2 - \beta \gamma \mathfrak{b})
    \vartheta_1(2\mathfrak{a}_2 )
    \vartheta_1(2 \gamma \mathfrak{b}) \vartheta_1(2 \mathfrak{a}_2 + 2 \beta \gamma \mathfrak{b})
  } \ .
\end{align}
The $a_1$-integration leaves us with
\begin{align}
  \oint \frac{da_2}{2\pi i a_2} \left[
  - \sum_{\alpha = \pm} R_\alpha \frac{1}{q^\pm - 1} (b^\alpha q^{\frac{1}{2}})^\pm
  - \sum_{\beta \gamma = \pm} R_{\beta \gamma} \frac{1}{q^\pm - 1} (a_2^\beta b^\gamma q^{\frac{1}{2}})^\pm
  \right] \ .
\end{align}
The residues $R_\alpha$ and $R_{\beta \gamma}$ are all elliptic with respect to $\mathfrak{a}_2$, and therefore the $a_2$-integration of both terms can be carried out. In $R_\alpha$, there are poles and residues
\begin{align}
  \mathfrak{a}_2 = 2 \alpha \delta \mathfrak{b}, \qquad
  \mathop{\operatorname{Res}}_{\mathfrak{a}_2 = 2\alpha \delta \mathfrak{b}}R_\alpha = - \frac{\delta}{8}\frac{\vartheta_4(3 \mathfrak{b}) \vartheta_4(\mathfrak{b})}{\vartheta_1(2 \mathfrak{b}) \vartheta_1(4 \mathfrak{b})} \ .
\end{align}
Hence
\begin{align}
  & \ - \oint \frac{da_2}{2\pi i a_2} \sum_{\alpha = \pm}R_\alpha \frac{1}{q^{\pm} - 1} (b^\alpha q^{\frac{1}{2}})^\pm \\
  = & \ - \sum_{\alpha = \pm} \frac{b^{\pm\alpha} q^{\pm \frac{1}{2}}}{q^{\pm} - 1}
  \left(
  R_\alpha(\mathfrak{a} = 0) + \sum_{\delta = \pm} \frac{- \delta}{8} \frac{\vartheta_4(3 \mathfrak{b}) \vartheta_4(\mathfrak{b})}{\vartheta_1(2 \mathfrak{b}) \vartheta_1(4 \mathfrak{b})} E_1
  \begin{bmatrix}
    -1 \\ b^{2\alpha \delta}q^{\frac{1}{2}}  
  \end{bmatrix}
  \right) \ .
\end{align}
By direct computation, one sees that the above is actually independent of $\pm$ sign in the $a_1^\pm$ insertion, consistent with the symmetry $\mathcal{Z}(\mathfrak{a}_1, \mathfrak{a}_2) = \mathcal{Z}( - \mathfrak{a}_1, \mathfrak{a}_2)$.

The term with $R_{\beta \gamma}$ can be carried using (\ref{integration-formula-monomial}),
\begin{align}
  \oint \frac{da_2}{2\pi i a_2}R_{\beta \gamma} a_2^{\pm \beta}
  = - \sum_{\operatorname{real} \ j} R_{\beta \gamma j} \frac{(a^{(\beta \gamma j)}_{2}q)^{\pm \beta}}{q^{\pm \beta} - 1}
  - \sum_{\operatorname{img} \ j} R_{\beta \gamma j}\frac{(a^{(\beta \gamma j)}_{2})^{\pm \beta}}{q^{\pm \beta} - 1} \ .
\end{align}
Here $a^{(\beta \gamma j)}_{2}$ denotes the simple poles of $R_{\beta \gamma}$ with respect to $a_2$, with the corresponding residue $R_{\beta \gamma j}$. We list the poles and their residues in Table \ref{poles-residues-SO(5)}.
{
\renewcommand{\arraystretch}{1.8}
\begin{table}[h!]
\centering
  \begin{tabular}{c|c|c}
    & poles $a_2^{(\beta \gamma j)}$ & residues $R_{\beta \gamma j}$ \\
    \hline
    Real & $\mathfrak{a}_2 = - 2\beta \gamma \mathfrak{b}$ & $ + \frac{\beta}{8} \frac{\vartheta_4( \mathfrak{b})\vartheta_4(3 \mathfrak{b})}{\vartheta_1(2  \mathfrak{b})\vartheta_1(4  \mathfrak{b})}$\\
    & $\mathfrak{a}_2 = - \beta \gamma \mathfrak{b} + \frac{1}{2}$ & $ + \frac{\beta \vartheta_4(\mathfrak{b})^2 \vartheta_3(0)}{16 \vartheta_1(2 \mathfrak{b})^2 \vartheta_3(2 \mathfrak{b})}$\\
    & $\mathfrak{a}_2 = \frac{1}{2}$ & $ - \frac{\beta \vartheta_4(\mathfrak{b})
    ^2 \vartheta_2(0)}{16 \vartheta_1(2 \mathfrak{b})^2 \vartheta_2(2 \mathfrak{b})}$\\
    & $\mathfrak{a}_2 = - \beta \gamma \mathfrak{b}$ & $ - \frac{\beta \vartheta_4(\mathfrak{b})^2 \vartheta_4(0)}{16 \vartheta_1(2 \mathfrak{b})^2 \vartheta_4(2 \mathfrak{b})}$\\
    \hline
    Imaginary & $\mathfrak{a}_2 = \beta \gamma \mathfrak{b} + \frac{1}{2}$ & $- \frac{\beta}{8} \frac{\vartheta_4( \mathfrak{b})\vartheta_4(3  \mathfrak{b})}{\vartheta_1(2 \mathfrak{b})\vartheta_1(4 \mathfrak{b})}$\\
    & $\mathfrak{a}_2 = \frac{\tau}{2}$ & $\frac{\beta \vartheta_4(\mathfrak{b})^2 \vartheta_4(0)}{16 \vartheta_1(2 \mathfrak{b})^2 \vartheta_4(2 \mathfrak{b})}$\\
    & $\mathfrak{a}_2 = \frac{1}{2} + \frac{\tau}{2}$ & $ - \frac{\beta \vartheta_4(\mathfrak{b})^2 \vartheta_3(0)}{16 \vartheta_1(2 \mathfrak{b})^2 \vartheta_3(2 \mathfrak{b})}$\\
    & $\mathfrak{a}_2 = - \beta \gamma \mathfrak{b} + \frac{1}{2} + \frac{\tau}{2}$ & $ + \frac{\beta \vartheta_4(\mathfrak{b})^2 \vartheta_2(0)}{16 \vartheta_1(2 \mathfrak{b})^2 \vartheta_2(2 \mathfrak{b})}$
  \end{tabular}
  \caption{Poles and residues of the elliptic functions $R_{\beta \gamma}$.\label{poles-residues-SO(5)}}
\end{table}
}

Performing the sum over $\beta, \gamma$,
\begin{align}
  & \ - \oint \frac{da_2}{2\pi i a_2}\sum_{\beta \gamma = \pm 1} R_{\beta \gamma} \frac{1}{q^{\pm} - 1} (a_2^\beta b^\gamma q^{\frac{1}{2}})^\pm \nonumber \\
  = & \ \frac{\sqrt{q}\left( b^2(q+1)-4b\sqrt{q}+q+1 \right)}{2}\frac{\vartheta _4(\mathfrak{b} )^2}{8b(q-1)^2\vartheta _1(2\mathfrak{b} )^2}\frac{\vartheta _2(0)}{\vartheta _2(2\mathfrak{b} )} \nonumber
  \\
  & \ + ( b^2q-b\sqrt{q}(q+1)+q ) \frac{\vartheta _4(\mathfrak{b} )^2}{8b(q-1)^2\vartheta _1(2\mathfrak{b} )^2}\left[ \frac{\vartheta _3(0)}{\vartheta _3(2\mathfrak{b} )}+\frac{\vartheta _4(0)}{\vartheta _4(2\mathfrak{b} )} \right]  \nonumber\\
  & \ +\frac{\sqrt{q} \left(-2 b^4 \sqrt{q}+b^3 (q+1)+b (q+1)-2 \sqrt{q}\right) }{8 b^2 (q-1)^2} \frac{\vartheta_4(3 \mathfrak{b}) \vartheta_4(\mathfrak{b})}{ \vartheta_1(2 \mathfrak{b}) \vartheta_1(4 \mathfrak{b})} \ ,
\end{align}
which is independent of the $\pm$ in $a_2^\pm$, consistent with the symmetry $\mathcal{Z}(\mathfrak{a}_1, \mathfrak{a}_2) = \mathcal{Z}( - \mathfrak{a}_1, \mathfrak{a}_2)$.

To summarize,
\begin{align}
  & \ \oint \frac{da_1}{2\pi i a_1}\frac{da_2}{2\pi i a_2}a_1
  \mathcal{Z}(\mathfrak{a}_1, \mathfrak{a}_2) \\
  = & \ \frac{i(b^2 - 1)\sqrt{q}}{8b(q-1)} \frac{\eta(\tau)^3 \vartheta_4(\mathfrak{b})^2}{\vartheta_1(2 \mathfrak{b})^3}\\
  & \ + \frac{\sqrt{q}(1 - 4b \sqrt{q} + q + b^2(1 + q))}{16 b(q -1)^2}
  \frac{\vartheta_4(\mathfrak{b})^2}{\vartheta_1(2 \mathfrak{b})^2}
  \frac{\vartheta_2(0)}{\vartheta_2(2 \mathfrak{b})}\\
  & \ + \frac{\sqrt{q}(b - \sqrt{q})(b\sqrt{q} - 1)}{8b (q-1)^2} \frac{\vartheta_4(\mathfrak{b})^2}{\vartheta_1(2 \mathfrak{b})^2}
  \left(
  \frac{\vartheta_3(0)}{\vartheta_3(2 \mathfrak{b})}
  + \frac{\vartheta_4(0)}{\vartheta_4(2 \mathfrak{b})}
  \right)\\
  & \ + \frac{\vartheta_4(\mathfrak{b})\vartheta_4(3 \mathfrak{b})}{\vartheta_1(2 \mathfrak{b}) \vartheta_1(4 \mathfrak{b})}
  \left(
  \frac{\sqrt{q}(b - \sqrt{q})(1 - b^3 \sqrt{q})}{4b^2 (q - 1)^2}
  + \frac{(b^2 -1)\sqrt{q}}{4b(q - 1)} E_1 \begin{bmatrix}
    -1 \\ b^2 q^{\frac{1}{2}}  
  \end{bmatrix}
  \right) \ .
\end{align}
Therefore,
\begin{align}
  \langle W_\mathbf{5}\rangle_{\mathcal{N} = 4 \ SO(5)}
  = & \ \mathcal{I}_{\mathcal{N} = 4 \ SO(5)}
  + \frac{i(b^2 - 1)\sqrt{q}}{2b(q-1)} \frac{\eta(\tau)^3 \vartheta_4(\mathfrak{b})^2}{\vartheta_1(2 \mathfrak{b})^3} \nonumber\\
  & \ + \frac{\sqrt{q}(1 - 4b \sqrt{q} + q + b^2(1 + q))}{4 b(q -1)^2}
  \frac{\vartheta_4(\mathfrak{b})^2}{\vartheta_1(2 \mathfrak{b})^2}
  \frac{\vartheta_2(0)}{\vartheta_2(2 \mathfrak{b})} \nonumber\\
  & \ + \frac{\sqrt{q}(b - \sqrt{q})(b\sqrt{q} - 1)}{2b (q-1)^2} \frac{\vartheta_4(\mathfrak{b})^2}{\vartheta_1(2 \mathfrak{b})^2}
  \left(
  \frac{\vartheta_3(0)}{\vartheta_3(2 \mathfrak{b})}
  + \frac{\vartheta_4(0)}{\vartheta_4(2 \mathfrak{b})}
  \right)\\
  & \ + \frac{\vartheta_4(\mathfrak{b})\vartheta_4(3 \mathfrak{b})}{\vartheta_1(2 \mathfrak{b}) \vartheta_1(4 \mathfrak{b})}
  \left(
  \frac{\sqrt{q}(b - \sqrt{q})(1 - b^3 \sqrt{q})}{b^2 (q - 1)^2}
  + \frac{(b^2 -1)\sqrt{q}}{b(q - 1)} E_1 \begin{bmatrix}
    -1 \\ b^2 q^{\frac{1}{2}}  
  \end{bmatrix}
  \right) \ , \nonumber
\end{align}
where the $\mathcal{I}_{\mathcal{N} = 4 \ SO(5)}$ is the original Schur index of the $SO(5)$ $\mathcal{N} = 4$ SYM.

\subsubsection{General representation}

Let us consider the $\mathfrak{so}(5)$ representations whose characters can be written as polynomials of $a_1, a_2$ with integral powers,
\begin{align}
  \chi_\mathcal{R}(a) = \sum_{n_1, n_2} c_{n_1 n_2} a_1^{n_1} a_2^{n_2} \ .
\end{align}
In particular, using the symmetry $a_1 \leftrightarrow a_2$, $a_i \leftrightarrow a_i^{-1}$, we can focus on the integrals of the following form
\begin{align}
  \oint \frac{da_1}{2\pi i a_1} \frac{da_2}{2\pi i a_2}
  a_1^{n_1 > 0}a_2^{n_2 \ge 0} \mathcal{Z} \ .
\end{align}
The $a_1$-integration leaves (recall that the $a_1$-integral picks up $6$ imaginary poles)
\begin{align}\label{a1integral}
  \oint \frac{da_2}{2\pi i a_2} a_2^{n_2} \left[
  - \sum_{\alpha = \pm}R_\alpha \frac{b^{\alpha n_1} q^{\frac{1}{2} n_1}}{q^{n_1} - 1}
  - \sum_{\beta \gamma = \pm}R_{\beta \gamma} \frac{a_2^{n_1\beta} b^{n_1\gamma} q^{\frac{1}{2} n_1}}{q^{n_1} - 1}
  \right] \ .
\end{align}
Depending on whether $n_2 = 0$ or $n_2 \ne 0$ in the first term, and whether $n_2 \pm n_1 = 0$ in the second term, the integral leads to different closed-form result. When $n_2 = 0$, the first term integrates to
\begin{align}\label{Rintegrate}
  = \delta_{n_2 = 0} \frac{1}{4}\frac{\vartheta_4(\mathfrak{b})}{\vartheta_1(2 \mathfrak{b})}
  \left(
  \frac{i \eta(\tau)^3 \vartheta_4(\mathfrak{b})}{2 \vartheta_1(2 \mathfrak{b})^2} + \frac{\vartheta_4(3 \mathfrak{b})}{\vartheta_1(4 \mathfrak{b})}
  E_1 \begin{bmatrix}
    1 \\ b^2  
  \end{bmatrix}
  \right)
  \frac{b^{n_1} - b^{-n_1}}{q^{n_1/2} - q^{- n_1/2}} \ ,
\end{align}
while when $n_2 > 0$, it integrates to
\begin{align}
  & \ - \delta_{n_2 > 0} \sum_{\alpha = \pm} \frac{b^{n_1 \alpha} q^{\frac{1}{2}n_1}}{q^{n_1} - 1}
  \sum_{\delta = \pm 1}
  \left(- \frac{\delta}{8} \frac{\vartheta_4(3 \mathfrak{b}) \vartheta_4( \mathfrak{b})}{\vartheta_1(2 \mathfrak{b})\vartheta_1(4 \mathfrak{b})}\right)
  \frac{(b^{2\alpha \delta} q^{\frac{1}{2}})^{n_2}}{q^{n_2/2} - q^{- n_2/2}} \ \nonumber\\
  = & \ - \delta_{n_2 > 0}\frac{\vartheta_4(\mathfrak{b}) \vartheta_4(3 \mathfrak{b})}{8 \vartheta_1(2 \mathfrak{b}) \vartheta_1(4 \mathfrak{b})} \frac{(b^{n_1} - b^{-n_1})(b^{2n_2} - b^{-2n_2})}{(q^{n_1/2} - q^{- n_1/2})(1 - q^{-n_2})} \ .
\end{align}

In the second term, when $0 < n_1 \ne n_2$, we have $n_2 + n_1 \beta \ne 0$ for either $\beta = \pm 1$. In this case,
\begin{align}\label{RalphabetaIntegrate}
  & \ - \delta_{n_1 \ne n_2} \oint \frac{da_2}{2\pi i a_2}\sum_{\beta \gamma = \pm} \frac{b^{n_1 \gamma} q^{\frac{1}{2}n_1}}{q^{n_1} - 1} R_{\beta \gamma}
  a_2^{n_2 + n_1 \beta} \\
  = & \ + \delta_{n_1 \ne n_2}\sum_{\beta \gamma = \pm}\frac{b^{n_1 \gamma} q^{\frac{1}{2}n_1}}{q^{n_1} - 1}  \sum_{\text{real/img} \ j} R_{\beta \gamma j} \frac{(a_2^{(\beta \gamma j)} q^{\pm \frac{1}{2}})^{n_2 + n_1 \beta}}{q^{\frac{1}{2}(n_2 + n_1 \beta)} - q^{-\frac{1}{2}(n_2 + n_1 \beta)} } \ .
\end{align}
On the other hand, when $n_1 = n_2 > 0$, we have $n_2 + n_1 \beta = 0$ for $\beta = -1$, and $n_2 + n_1 \beta = 2n_1 \ne 0$ for $\beta = 1$. In this situation,
\begin{align}
  & \ - \oint \frac{da_2}{2\pi i a_2}\sum_{\beta \gamma = \pm}R_{\beta \gamma} \frac{a_2^{n_2 + n_1 \beta} b^{n_1 \gamma} q^{\frac{1}{2}n_1}}{q^{n_1} - 1}
  \nonumber\\
  = & \ \delta _{n_1=n_2}\sum_{\gamma =\pm}{\left[ \sum_{\text{real}/\text{img}\ j}{R_{+\gamma j}\frac{( a_{2}^{\left( +\gamma j \right)})^{2n_1} q^{\pm \frac{1}{2}2n_1}}{q^{n_1}-q^{-n_1}}} \right]}\frac{b^{n_1\gamma}q^{\frac{1}{2}n_1}}{q^{n_1}-1}\\
  & \ -\delta _{n_1=n_2}\sum_{\gamma =\pm}
  \frac{b^{n_1\gamma}q^{\frac{1}{2}n_1}}{q^{n_1}-1}
  \left( R_{-\gamma}\left( \mathfrak{a} _2=\mathfrak{a}_2^{(0)} \right) +\sum_{\text{real}/\text{img} \ j}{R_{-\gamma j}}E_1\begin{bmatrix}
    -1\\
    \frac{a_{2}^{\left( -\gamma j \right)}}{a_2^{(0)}}q^{\pm \frac{1}{2}}  
  \end{bmatrix} \right) \nonumber \ ,
\end{align}
where $a_2^{(0)}$ is a generic reference value, for example, $a_2^{(0)} = b^3$. In the above, we have used the poles and residues in Table \ref{poles-residues-SO(5)}. Putting all the contributions together, we deduce that for $n_1 > 0, n_2 \ge 0$,
\begin{align}
  & \ \oint\prod_{i = }^{2}\frac{da_i}{2\pi i a_i}a_1^{n_1}a_2^{n_2} \mathcal{Z}\\
  = & \ \delta_{n_2 = 0}\frac{1}{4}\frac{\vartheta_4(\mathfrak{b})}{\vartheta_1(2 \mathfrak{b})}
  \left(
  \frac{i \eta(\tau)^3 \vartheta_4(\mathfrak{b})}{2 \vartheta_1(2 \mathfrak{b})^2} + \frac{\vartheta_4(3 \mathfrak{b})}{\vartheta_1(4 \mathfrak{b})}
  E_1 \begin{bmatrix}
    1 \\ b^2  
  \end{bmatrix}
  \right)
  \frac{b^{n_1} - b^{-n_1}}{q^{n_1/2} - q^{- n_1/2}} \\
  & \ - \delta_{n_2 > 0}\frac{\vartheta_4(\mathfrak{b}) \vartheta_4(3 \mathfrak{b})}{8 \vartheta_1(2 \mathfrak{b}) \vartheta_1(4 \mathfrak{b})} \frac{(b^{n_1} - b^{-n_1})(b^{2n_2} - b^{-2n_2})}{(q^{n_1/2} - q^{- n_1/2})(1 - q^{-n_2})} \\
  & \ + \delta_{n_1 \ne n_2}\sum_{\beta \gamma = \pm}\frac{b^{n_1 \gamma} q^{\frac{1}{2}n_1}}{q^{n_1} - 1}  \sum_{\text{real/img} \ j} R_{\beta \gamma j} \frac{(a_2^{(\beta \gamma j)} q^{\pm \frac{1}{2}})^{n_2 + n_1 \beta}}{q^{\frac{1}{2}(n_2 + n_1 \beta)} - q^{-\frac{1}{2}(n_2 + n_1 \beta)} }\\
  & \ + \delta _{n_2=n_1}\sum_{\gamma =\pm}{\left[ \sum_{\text{real}/\text{img}\ j}{R_{+\gamma j}\frac{( a_{2}^{\left( +\gamma j \right)})^{2n_1} q^{\pm \frac{1}{2}2n_1}}{q^{n_1}-q^{-n_1}}} \right]}\frac{b^{n_1\gamma}q^{\frac{1}{2}n_1}}{q^{n_1}-1}\\
    & \ -\delta _{n_2=n_1}\sum_{\gamma =\pm}
    \frac{b^{n_1\gamma}q^{\frac{1}{2}n_1}}{q^{n_1}-1}
    \left( R_{-\gamma}\left( \mathfrak{a} _2=\mathfrak{a}_2^{(0)} \right) +\sum_{\text{real}/\text{img} \ j}{R_{-\gamma j}}E_1\begin{bmatrix}
      -1\\
      \frac{a_{2}^{\left( -\gamma j \right)}}{a_2^{(0)}}q^{\pm \frac{1}{2}}  
    \end{bmatrix} \right) \ .
\end{align}
The Wilson index corresponding to the $SO(5)$ representations with Dynkin labels $[n, 0]$ can be computed using the above integration formula by simple substitution, since the corresponding character can be written as a sum of simple monomials,
\begin{align}
  \chi_{[n,0]}\left(a_1,a_2\right)
  = & \ \sum_{m=0}^n \sum_{j=0}^{m} \sum_{i=0}^{m}
  a_1^{j-i} a_2^{i+j-m} \\
  = & \ \lceil\frac{n+1}{2}\rceil+\sum_{m=0}^n\sum_{\substack{i=0\\i\neq m/2}}^m a_2^{2i-m}+\sum_{m=1}^n \sum_{\substack{i,j = 0 \\ i\ne j}}^n a_2^{i+j-m}a_1^{j-i} \\
  \sim & \ \lceil\frac{n+1}{2}\rceil+\sum_{m=0}^n\sum_{\substack{i=0\\i\neq m/2}}^m a_1^{|2i-m|}+\sum_{m=1}^n \sum_{\substack{i,j = 0 \\ i\ne j}}^n a_2^{|i+j-m|}a_1^{|j-i|} \ .
\end{align}
Here in the last line we have rewritten the expression using the symmetries $a_1 \leftrightarrow a_2$, $a_i \leftrightarrow a_i^{-1}$ of the integral, so that each term can be easily computed with the above integration formula. Although the Wilson line index can be computed straightforwardly simply by substitution, we are unfortunately unable to reorganize the final result in an elegant form, so we will refrain from presenting the final expression of $\langle W_{[n,0]}\rangle_{\mathcal{N} = 4 \ SO(5)}$ here.

\section*{Acknowledgments}
Y.P. is supported by the National Natural Science Foundation of China (NSFC) under Grant No. 11905301, the Fundamental Research Funds for the Central Universities, Sun Yat-sen University under Grant No. 2021qntd27.

\appendix

\section{Special Functions\label{app:special-functions}}

Throughout this appendix and the paper, we adopt the convention that fraktur letters $\mathfrak{a}, \mathfrak{b},$ etc., are related to the standard letters by
\begin{align}
  a = e^{2\pi i \mathfrak{a}}, \qquad
  b = e^{2 \pi i \mathfrak{b}}, \qquad \cdots, \qquad
  z = e^{2\pi i \mathfrak{z}} \ .
\end{align}
except for the standard notation $q = e^{2\pi i \tau}$.

\subsection{\texorpdfstring{The Weierstrass $\zeta$-function}{}}

The Weierstrass $\zeta$-function is defined by
\begin{align}\label{weierstrasszetadef}
  \zeta(z) \coloneqq \frac{1}{z} + \sum_{\substack{(m, n) \in \mathbb{Z}^2\\(m, n) \ne (0, 0)}}
  \left[\frac{1}{z - m - n \tau} + \frac{1}{m + n \tau} + \frac{z }{(m + n \tau)^2} \right]\ .
\end{align}
In the following and in the main text we will often abbreviate
\begin{align}
  \sum_{\substack{(m, n) \in \mathbb{Z}^2\\(m, n) \ne (0, 0)}} \to \ \ \sum'_{m, n} \ , \qquad \sum_{\substack{m \in \mathbb{Z}\\m \ne 0}} \to \sum_m '\ .
\end{align}
The $\zeta$ function is not elliptic, and under full period shift of $z$,
\begin{align}\label{shift-formula-zeta}
  \zeta(z + 1 | \tau) - \zeta(z| \tau) = & \ 2\eta_1(\tau)\\
  \zeta(z + \tau |\tau) - \zeta(z|\tau) = & \ 2 \eta_2(\tau) \equiv 2\tau \eta_1(\tau) - 2\pi i\ ,
\end{align}
where $\eta_1$ and $\eta_2$ are independent of $z$ and are both related to the Eisenstein series $E_2$. Note that $\zeta$ has a simple pole at each lattice point $m + n \tau$ with unit residue. The fact that $\zeta$ is not fully elliptic is due to the fact that meromorphic function on $T^2$ with only one simple pole doesn't exist.

% \begin{itemize}

  % \item The Weierstrass $\wp$-function
  % \begin{align}
  %   \wp(z) \coloneqq & \ \frac{1}{z^2} + \sum_{(m,n) \ne (0,0)} \left[\frac{1}{(z - m - n \tau)^2} - \frac{1}{(m + n \tau)^2}\right] \ .
  % \end{align}
  % This function is elliptic,
  % \begin{align}
  %   \wp(z) = \wp(z + 1) = \wp(z + \tau) \ .
  % \end{align}
  % Following from the simple fact that $\partial_z z^{-1} = - z^{-2}$, one has
  % \begin{align}
  %   \wp(z) = - \partial_z \zeta(z)\ .
  % \end{align}
  % By definition, $\wp$ has only one double pole on $T^2_\tau$.

  % \item The descendants $\partial_z^n \wp(z)$ are all elliptic functions, all with a single $n + 2$-th order pole on $T^2_\tau$.
% \end{itemize}

\subsection{Jacobi theta functions}

The standard Jacobi theta functions can be defined as infinite products of the $q$-Pochhammer symbol $(z;q) \coloneqq \prod_{k = 0}^{+\infty}(1 - zq)$,
\begin{align}\label{def:Jacobi-theta}
  \vartheta_1(\mathfrak{z}) \coloneqq & \ - i z^{\frac{1}{2}}q^{\frac{1}{8}}(q;q)(zq;q)(z^{-1};q),
  & \vartheta_2(\mathfrak{z}) \coloneqq & z^{1/2}q^{\frac{1}{8}}(q;q)(-zq;q)(-z^{-1};q) \ ,\\
  \vartheta_3(\mathfrak{z}) \coloneqq & \ (q;q)(-zq^{1/2};q)(-z^{-1}q^{1/2}),
  & \vartheta_4(\mathfrak{z}) \coloneqq & (q;q)(zq^{1/2};q)(z^{-1}q^{1/2};q) \ .
\end{align}
From the definition it is easy to read off their simple zeros, for example,
\begin{align}
  \vartheta_1(m + n \tau) = 0, \qquad
  \vartheta_4(m + n \tau + \frac{\tau}{2}) = 0 \ , \qquad
  m, n \in \mathbb{Z} \ .
\end{align}
The Jacobi theta functions can also be rewritten in infinite series in $q$, or Fourier series in $\mathfrak{z}$,
\begin{align}
  \vartheta_1(\mathfrak{z}) = & \ -i \sum_{r \in \mathbb{Z} + \frac{1}{2}} (-1)^{r-\frac{1}{2}} e^{2\pi i r \mathfrak{z}} q^{\frac{r^2}{2}} ,
  & \vartheta_2(\mathfrak{z}) = & \sum_{r \in \mathbb{Z} + \frac{1}{2}} e^{2\pi i r \mathfrak{z}} q^{\frac{r^2}{2}} \ ,\\
  \vartheta_3(\mathfrak{z}) = & \ \sum_{n \in \mathbb{Z}} e^{2\pi i n \mathfrak{z}} q^{\frac{n^2}{2}},
  & \vartheta_4(\mathfrak{z}) = & \sum_{n \in \mathbb{Z}} (-1)^n e^{2\pi i n \mathfrak{z}} q^{\frac{n^2}{2}} \ .
\end{align}

\begin{figure}
  \centering
  \begin{tikzpicture}

  \begin{scope}[scale=1.4]
    \node(00) at (0,0) {$\vartheta_1$};
    \node(10) at (1.2, 0) {$\vartheta_2$};
    \node(20) at (1.2+1.2, 0) {$\vartheta_1$};

    \node(01) at (0.5,1) {$\vartheta_4$};
    \node(11) at (0.5 + 1.2,1) {$\vartheta_3$};
    \node(21) at (0.5 + 1.2 + 1.2,1) {$\vartheta_4$};

    \node(02) at (0.5+0.5,1+1) {$\vartheta_1$};
    \node(12) at (0.5+0.5 + 1.2,1+1) {$\vartheta_2$};
    \node(22) at (0.5+0.5 + 1.2 + 1.2,1+1) {$\vartheta_1$};

    \draw[->=stealth] (00)--node[above]{\scriptsize {$1$}}(10) ;
    \draw[->=stealth] (10)--node[above]{\scriptsize{$-1$}}(20) ;

    \draw[->=stealth] (01)--node[above]{\scriptsize {$1$}}(11) ;
    \draw[->=stealth] (11)--node[above]{\scriptsize {$1$}}(21) ;

    \draw[->=stealth] (02)--node[above]{\scriptsize {$1$}}(12) ;
    \draw[->=stealth] (12)--node[above]{\scriptsize{$-1$}}(22) ;

    \draw[->=stealth] (00) -- node[sloped, above]{\scriptsize {$i\mu$}} (01);
    \draw[->=stealth] (10) -- node[sloped, above]{\scriptsize {$\mu$}} (11);
    \draw[->=stealth] (20) -- node[sloped, above]{\scriptsize {$i\mu$}} (21);

    \draw[->=stealth] (01) -- node[sloped, above]{\scriptsize {$i\mu$}} (02);
    \draw[->=stealth] (11) -- node[sloped, above]{\scriptsize {$\mu$}} (12);
    \draw[->=stealth] (21) -- node[sloped, above]{\scriptsize {$i\mu$}} (22);

  \end{scope}
  \end{tikzpicture}
  \caption{Half shifts of the Jacobi theta functions. \label{half-shifts}}
\end{figure}
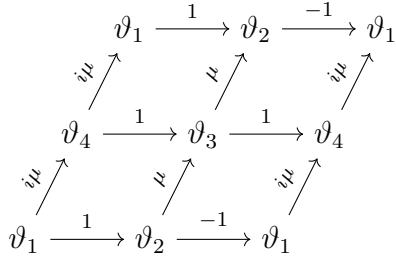

The functions $\vartheta_i(z)$ share similar shift properties under the full period shifts,
\begin{align}
  \vartheta_{1,2}(\mathfrak{z} + 1) = & - \vartheta_{1,2}(\mathfrak{z}) , & 
  \vartheta_{3,4}(\mathfrak{z} + 1) = & + \vartheta_{3,4}(\mathfrak{z}) , & \\
  \vartheta_{1,4}(\mathfrak{z} + \tau) = & - \lambda \vartheta_{1,4}(\mathfrak{z}), & 
  \vartheta_{2,3}(\mathfrak{z} + \tau) = & + \lambda \vartheta_{2,3}(\mathfrak{z}) , & 
\end{align}
where $\lambda \coloneqq e^{-2\pi i \mathfrak{z}}e^{- \pi i \tau}$, while under half-period shifts which can be summarized as in the Figure \ref{half-shifts},
where $\mu = e^{- \pi i \mathfrak{z}} e^{- \frac{\pi i}{4}}$, and $f \xrightarrow{a} g$ means
\begin{align}
  \text{either}\qquad  f\left(\mathfrak{z} + \frac{1}{2}\right) = a g(\mathfrak{z}) \qquad \text{or} \qquad
  f\left(\mathfrak{z} + \frac{\tau}{2}\right) = a g(\mathfrak{z}) \ ,
\end{align}
depending on whether the arrow is horizontal or (slanted) vertical respectively.

% The functions $\vartheta_{i = 2,3,4}(z | \tau)$ transform into each other under the modular $S$ and $T$ transformations, which act, as usual, on the nome and flavor fugacity as $(\frac{\mathfrak{z}}{\tau}, - \frac{1}{\tau})\xleftarrow{~~S~~}(\mathfrak{z}, \tau) \xrightarrow{~~T~~} (\mathfrak{z}, \tau + 1).$ In summary, with $\alpha = \sqrt{-i \tau}e^{\frac{\pi i z^2}{\tau}}$,
% \begin{center}
%   \includegraphics[height=0.2\textheight]{STtheta.pdf}
% \end{center}

Finally, we will frequently encounter residues of the $\vartheta$ functions. In particular,
\begin{align}\label{theta-function-residue}
  \mathop{\operatorname{Res}}\limits_{a \to b^{\frac{1}{n}}q^{\frac{k}{n} + \frac{1}{2n}}e^{2\pi i \frac{\ell}{n}}} \frac{1}{a} \frac{1}{\vartheta_4(n\mathfrak{a} - \mathfrak{b})} = & \ \frac{1}{n} \frac{1}{(q;q)^3} (-1)^k q^{\frac{1}{2} k (k + 1)} \ , \\
  \mathop{\operatorname{Res}}\limits_{a \to b^{\frac{1}{n}}q^{\frac{k}{n}}e^{2\pi i \frac{\ell}{n}}} \frac{1}{a} \frac{1}{\vartheta_1(n\mathfrak{a} - \mathfrak{b})} = & \ \frac{1}{n} \frac{i }{\eta(\tau)^3} (-1)^{k + \ell} q^{\frac{1}{2}k^2}\ .
\end{align}
Note that the $(-1)^\ell$ in the second line is related to the presence of a branch point at $z = 0$ according to \eqref{def:Jacobi-theta}.

\subsection{Eisenstein series}

The twisted Eisenstein series are defined by the following infinite sum,
\begin{align}
  E_{k \ge 1}\left[\begin{matrix}
    \phi \\ \theta
  \end{matrix}\right] \coloneqq & \ - \frac{B_k(\lambda)}{k!} \\
  & \ + \frac{1}{(k-1)!}\sum_{r \ge 0}' \frac{(r + \lambda)^{k - 1}\theta^{-1} q^{r + \lambda}}{1 - \theta^{-1}q^{r + \lambda}}
  + \frac{(-1)^k}{(k-1)!}\sum_{r \ge 1} \frac{(r - \lambda)^{k - 1}\theta q^{r - \lambda}}{1 - \theta q^{r - \lambda}} \ ,
\end{align}
where the parameter $\phi \coloneqq e^{2\pi i \lambda}$ with $\lambda \in [0, 1)$, $B_k(x)$ denotes the $k$-th Bernoulli polynomial, and the prime $^\prime$ in the summation means that the term with $r = 0$ should be omitted whenever $\phi = \theta = 1$. We also define $E_0\big[\substack{\phi \\ \theta}\big] = -1$.

The standard Eisenstein series $E_{2n}$ are the $\theta, \phi \to 1$ limit of the above Eisenstein series. When $k$ is odd, $\theta = \phi = 1$ gives zero except for the special instance with $k = 1$, where there is a simple pole $\mathfrak{z} \to 0$,
\begin{align}
  E_{2n}\left[\begin{matrix}
    +1 \\ +1
  \end{matrix}\right] = E_{2n} \ , \qquad
  E_{2n + 1 \ge 3}\left[\begin{matrix}
    +1 \\ +1
  \end{matrix}\right] = 0 
  \qquad
  E_1\left[\begin{matrix}
    + 1 \\ z
  \end{matrix}\right] = \frac{1}{2\pi i }\frac{\vartheta'_1(\mathfrak{z})}{\vartheta_1(\mathfrak{z})}
  \ .
\end{align}

The Eisenstein series exhibit several useful properties. For example, the symmetry property
\begin{align}\label{Eisenstein-symmetry}
  E_k\left[\begin{matrix}
    \pm 1 \\ z^{-1}
  \end{matrix}\right] = (-1)^k E_k\left[\begin{matrix}
    \pm 1 \\ z
  \end{matrix}\right] \ .
\end{align}
% The twisted Eisenstein series of neighboring weights are related by
% \begin{align}\label{EisensteinDerivative}
%   q \partial_q E_k\left[\begin{matrix}
%     \phi \\ b
%   \end{matrix}
%   \right] = (- k) b \partial_b E_{k + 1}\left[\begin{matrix}
%     \phi \\ b
%   \end{matrix}
%   \right]\ .
% \end{align}

When shifting the argument $\mathfrak{z}$ of the Eisenstein series by half or full periods of $\tau$, or equivalently, shifting $z$ by $q^{\frac{n}{2}}$, one has
\begin{align}\label{Eisenstein-half-shift}
  E_k\left[\begin{matrix}
    \pm 1\\ z q^{\frac{n}{2}}
  \end{matrix}\right]
  =
  \sum_{\ell = 0}^{k} \left(\frac{n}{2}\right)^\ell \frac{1}{\ell !}
  E_{k - \ell}\left[\begin{matrix}
    (-1)^n(\pm 1) \\ z
  \end{matrix}\right] \ , \qquad n \in \mathbb{Z} \ .
\end{align}
A simple consequence is that\footnote{In fact, these equalities remain true even after replacing $1$ by $e^{2\pi i \lambda}$ and $- 1$ by $e^{2\pi i (\lambda + \frac{1}{2})}$.}
\begin{align}\label{Eisenstein-shift-1}
  E_k\left[\begin{matrix}
    \pm 1 \\ zq^{\frac{1}{2}}
  \end{matrix}\right]
  - E_k\left[\begin{matrix}
    \pm 1 \\ zq^{ - \frac{1}{2}}
  \end{matrix}\right]
  = & \ \sum_{m = 0}^{\floor{\frac{k - 1}{2}}} \frac{1}{2^{2m}(2m+1)!}E_{k - 1 - 2m}\left[\begin{matrix}
    \mp 1\\z
  \end{matrix}\right] \ ,
\end{align}
or more generally
\begin{align}
  E_k\left[\begin{matrix}
    \pm 1 \\ zq^{\frac{1}{2} + n}
  \end{matrix}\right]
  - E_k\left[\begin{matrix}
    \pm 1 \\ zq^{ - \frac{1}{2} - n}
  \end{matrix}\right]
  = & \ 2\sum_{m = 0}^{\floor{\frac{k - 1}{2}}} \left(\frac{2n+1}{2}\right)^{2m + 1}\frac{1}{(2m+1)!}E_{k - 1 - 2m}\left[\begin{matrix}
    \mp 1\\z
  \end{matrix}\right] \ .
\end{align}

\subsection{Elliptic function}
In this paper we frequently encounter elliptic functions with respect to a complex structure $\tau$. They are meromorphic functions on $\mathbb{C}$ satisfying the doubly-periodic condition,
\begin{align}
  f(\mathfrak{z}) = f(\mathfrak{z} + \tau) = f(\mathfrak{z} + 1) \ .
\end{align}
Here $\tau \in \mathbb{C}$ with $\operatorname{Im}\tau > 0$. Exploiting the periodicity, one may restrict the domain of $\mathfrak{z}$ to the \emph{fundamental parallelogram} in $\mathbb{C}$ with vertices $0$, $1$, $\tau$, $1 + \tau$. Equivalently, an elliptic function $f$ is a meromorphic function on the torus $T^2_\tau$ with complex structure $\tau$. Using $z = e^{2\pi i \mathfrak{z}}$, $f(\mathfrak{z})$ is sometimes written as $f(z)$.

As a meromorphic function, $f(\mathfrak{z})$ may have poles in the parallelogram. In this paper we mainly focus on elliptic functions $f(\mathfrak{z})$ with only simple poles. We classify the poles $\mathfrak{z}_j$ into two types by the following criteria: we call $\mathfrak{z}_j$ \emph{real} if $\operatorname{Im}\mathfrak{z}_j = 0$, or \emph{imaginary} if $\operatorname{Im}\mathfrak{z}_j > 0$. The residues at the simple poles $\mathfrak{z}_j$ are captured by $R_j$,
\begin{align}
  R_j \coloneqq \mathop{\operatorname{Res}}_{z \to z_j}\frac{1}{z}f(z) \ .
\end{align}
Using the well-known Weierstrass $\zeta$ function and the Eisenstein series, any elliptic function $f(\mathfrak{z})$ with only simple poles\footnote{For functions with higher order poles, one needs to include derivatives of $\zeta$-function or Eisentein series.} can be expanded in various ways,
\begin{align}\label{elliptic-function-decomposition}
  f(\mathfrak{z}) = C_f + \frac{1}{2\pi i}\sum_{j} R_j \zeta(\mathfrak{z} - \mathfrak{z}_j)
  = & \ f(\mathfrak{z}_0) + \sum_j R_j \left(
    E_1 \begin{bmatrix}
      +1 \\ \frac{z_j}{z_0}
    \end{bmatrix}
    - E_1 \begin{bmatrix}
      +1 \\ \frac{z_j}{z}
    \end{bmatrix}
  \right) \nonumber\\
  = & \ f(\mathfrak{z}_0) + \sum_j R_j \left(
    E_1 \begin{bmatrix}
      - 1 \\ \frac{z_j}{z_0} q^{\frac{1}{2}}
    \end{bmatrix}
    - E_1 \begin{bmatrix}
      - 1 \\ \frac{z_j}{z} q^{\frac{1}{2}}
    \end{bmatrix}
  \right) \\
  = & \ f(\mathfrak{z}_0) + \sum_j R_j \left(
    E_1 \begin{bmatrix}
      - 1 \\ \frac{z_j}{z_0} q^{-\frac{1}{2}}
    \end{bmatrix}
    - E_1 \begin{bmatrix}
      - 1 \\ \frac{z_j}{z} q^{-\frac{1}{2}}
    \end{bmatrix}
  \right) \nonumber \\
    = & \ f(\mathfrak{z}_0) + \sum_{\text{real/img} \ \mathfrak{z}_j} R_j \left(
      E_1 \begin{bmatrix}
        - 1 \\ \frac{z_j}{z_0} q^{\pm\frac{1}{2}}
      \end{bmatrix}
      - E_1 \begin{bmatrix}
        - 1 \\ \frac{z_j}{z} q^{\pm\frac{1}{2}}
      \end{bmatrix}
    \right) \nonumber \ .
\end{align}
Here $z_0 = e^{2\pi i \mathfrak{z}_0}$ is an arbitrary and generic reference value. Note that the above expansions are valid for all types of pole combinations, real or imaginary, where the last line incorporates explicitly the realness of the poles to determine the $\pm \frac{1}{2}$. These expansions lead to useful integration formula that we will review later.

\subsection{Useful identities}

The Eisenstein series are related to the Jacobi theta functions,
\begin{align}\label{EisensteinToTheta-2}
  E_k\left[\begin{matrix}
    + 1 \\ z
  \end{matrix}\right] = \sum_{\ell = 0}^{\floor{k/2}}  \frac{(-1)^{k + 1}}{(k - 2\ell)!}\left(\frac{1}{2\pi i}\right)^{k - 2\ell} \mathbb{E}_{2\ell} \frac{\vartheta_1^{(k - 2\ell)}(\mathfrak{z})}{\vartheta_1(\mathfrak{z})} \ ,
\end{align}
where we define
\begin{align}\label{Ebold}
  & \mathbb{E}_{2} \coloneqq E_2, \qquad \mathbb{E}_4 \coloneqq E_4 + \frac{1}{2}(E_2)^2, \qquad
  \mathbb{E}_6 \coloneqq E_6 + \frac{3}{4}E_4E_2 + \frac{1}{8}(E_2)^3 \ , \qquad \ldots\\
  & \mathbb{E}_{2\ell} \coloneqq \sum_{\substack{\{n_p\} \\ \sum_{p \ge 1} (2p)n_p = 2\ell}} \prod_{p\ge 1} \frac{1}{n_p !} \left(\frac{1}{2p}E_{2p}\right)^{n_p}\ .
\end{align}
Similar formula for $E_k\left[\substack{- 1 \\ \pm z}\right]$, $E_k \big[\substack{+1 \\ -z}\big]$ can be obtained by replacing $\vartheta_1$ with $\vartheta_{2,3,4}$. For the reader's convenience we list the first few conversions here.
\begin{align}\label{Ek-thetap}
  E_1\left[\begin{matrix}
    +1 \\ z
  \end{matrix}
  \right] = & \ \frac{1}{2\pi i} \frac{\vartheta'_1(\mathfrak{z})}{\vartheta_1(\mathfrak{z})}\ ,  \\
  E_2\left[\begin{matrix}
    +1 \\ z
  \end{matrix}
  \right] = & \ \frac{1}{8\pi^2}\frac{\vartheta_1''(\mathfrak{z})}{\vartheta_1(\mathfrak{z})} - \frac{1}{2} E_2 \ , \\
  E_3\left[\begin{matrix}
    +1 \\ z
  \end{matrix}
  \right] = & \ \frac{i}{48\pi^3} \frac{\vartheta'''_1(\mathfrak{z})}{\vartheta_1(\mathfrak{z})}
    - \frac{i}{4\pi}\frac{\vartheta'_1(\mathfrak{z})}{\vartheta_1(\mathfrak{z})} E_2,  \\
  E_4\left[\begin{matrix}
    +1 \\ z
  \end{matrix}\right] = & \ - \frac{1}{384\pi^4} \frac{\vartheta''''_1(\mathfrak{z})}{\vartheta_1(\mathfrak{z})} + \frac{1}{16\pi^2}E_2 \frac{\vartheta''_1(\mathfrak{z})}{\vartheta_1(\mathfrak{z})} - \frac{1}{4} \left(E_4 + \frac{1}{2}(E_2)^2\right) \\
  E_5\left[\begin{matrix}
    +1 \\ z
  \end{matrix}\right]
  = & \ - \frac{i}{3840 \pi^5} \frac{\vartheta^{(5)}_1(\mathfrak{z})}{\vartheta_1(\mathfrak{z})} + \frac{i}{96\pi^3}E_2 \frac{\vartheta_1^{(3)}(\mathfrak{z})}{\vartheta_1(\mathfrak{z})} - \frac{i}{8\pi}\left(E_4 + \frac{1}{2}(E_2)^2\right)\frac{\vartheta_1'(\mathfrak{z})}{\vartheta_1(\mathfrak{z})} \\
  E_6\left[\begin{matrix}
    + 1 \\ z
  \end{matrix}\right]
  = & \ \frac{1}{46080\pi^6} \frac{\vartheta^{(6)}_1(\mathfrak{z})}{\vartheta_1(\mathfrak{z})} - \frac{1}{768\pi^4}E_2 \frac{\vartheta_1^{(4)}(\mathfrak{z})}{\vartheta_1(\mathfrak{z})} + \frac{1}{32\pi^2} \left(E_4 + \frac{1}{2}(E_2)^2\right) \frac{\vartheta_1^{(2)}(\mathfrak{z})}{\vartheta_1(\mathfrak{z})} \nonumber\\
  & \ - \frac{1}{6}\left(E_6 + \frac{3}{4}E_4 E_2 + \frac{1}{8}E_2^3\right) \ .
\end{align}

Moreover, the Eisenstein series satisfy the following relations which are generalization of the so-called duplication formula of the Jacobi theta functions,
\begin{align}\label{duplication-Eisenstein}
  \sum_{\pm}E_k\left[\begin{matrix}
    \phi \\ \pm z
  \end{matrix}\right](\tau) = & \ 2 E_k\left[\begin{matrix}
    \phi \\ z^2
  \end{matrix}\right](2\tau) \ , \nonumber \\
  \sum_{\pm} \pm E_k\left[\begin{matrix}
    \phi \\ \pm z
  \end{matrix}\right](\tau)
  = & \ -2 E_k\left[\begin{matrix}
    \phi \\ z^2
  \end{matrix}\right](2\tau)
   + 2 E_k\left[\begin{matrix}
    \phi \\ z
   \end{matrix}\right](\tau)\ , \nonumber
  \\
  E_k\left[\begin{matrix}
    + 1\\z
  \end{matrix}\right](2\tau)
  + E_k\left[\begin{matrix}
    - 1\\z
  \end{matrix}\right](2\tau) = & \ 
  \frac{2}{2^k}E_k\left[\begin{matrix}
    + 1 \\ z
  \end{matrix}\right] \ ,\\
  \sum_{\pm \pm} E_k\left[\begin{matrix}
    \pm 1 \\ \pm z
  \end{matrix}\right](\tau) = & \ \frac{4}{2^k}E_k\left[
  \begin{matrix}
    + 1 \\ z^2
  \end{matrix}\right](\tau)\ . \nonumber\\
  E_1 \begin{bmatrix}
    \phi \\ zq^{-1/2}  
  \end{bmatrix}(2\tau) + E_1 \begin{bmatrix}
    \phi \\ z q^{\frac{1}{2}}  
  \end{bmatrix}(2\tau) = & \ E_1 \begin{bmatrix}
    \phi \\ z  
  \end{bmatrix}(\tau), \qquad \phi = \pm 1 \ .
\end{align}

The $E_1$ function also has some alternative expansions besides its definition, for example,
\begin{align}\label{E1-expansions}
  E_1 \begin{bmatrix}
    -1 \\ b  
  \end{bmatrix}
  =  & \ \frac{b^{-1} q^{\frac{1}{2}}}{1 - b^{-1} q^{\frac{1}{2}}}
  - \frac{b q^{\frac{1}{2}}}{1 - b q^{\frac{1}{2}}}
  + \sum_{k = 1}^{+\infty} q^n \frac{b^n - b^{-n}}{q^{\frac{n}{2}} - q^{-\frac{n}{2}}} \\
  = & \ \frac{1}{2}\left(\frac{b^{-1} q^{\frac{1}{2}}}{1 - b^{-1} q^{\frac{1}{2}}}
  - \frac{b q^{\frac{1}{2}}}{1 - b q^{\frac{1}{2}}}\right)
  + \frac{1-q}{2}\sum_{n = 1}^{+\infty}q^{n/2}\sum_{\substack{m = - n/2 \\ m\ne0 }}^{n/2} \frac{b^{2m} - b^{-2m}}{1 - q^{-2m}} \ .
\end{align}

\section{Integration formula \label{app:integration-formula}}

In this appendix we collect integration formula for contour integrals containing an elliptic function, some products of Eisenstein series and some monomial factors.

\subsection{Integration formula without monomial}

We begin with the simplest formula. Consider an elliptic function $f(\mathfrak{z})$ with only simple poles. Denoting $z = e^{2\pi i \mathfrak{z}}$, then the contour integral of $f$ along the unit circle can be computed analytically,
\begin{align}\label{integration-formula-f}
  \oint_{|z| = 1} \frac{dz}{2\pi i z} f(\mathfrak{z})
  = f(\mathfrak{z}_0)
  + \sum_{\text{real/img} \ \mathfrak{z}_j} R_j E_1 \begin{bmatrix}
    -1 \\ \frac{z_j}{z_0}q^{\pm \frac{1}{2}} 
  \end{bmatrix} \ .
\end{align}
Here, $\mathfrak{z}_0$ (and $z_0 = e^{2\pi i \mathfrak{z}_0}$) denotes an arbitrary and generic reference value, and $\mathfrak{z}_j$ (with $z_j = e^{2\pi i \mathfrak{z}_j}$) are the simple poles of $f$. Recall that $\mathfrak{z}_j$ is real if $\operatorname{Im}\mathfrak{z}_j = 0$, or imaginary if $\operatorname{Im}\mathfrak{z}_j > 0$. This formula follows directly from the decomposition
\begin{align}
  f(\mathfrak{z}) = f(\mathfrak{z}_0) + \sum_{\text{real/img} \ \mathfrak{z}_j} R_j \left(
      E_1 \begin{bmatrix}
        - 1 \\ \frac{z_j}{z_0} q^{\pm\frac{1}{2}}
      \end{bmatrix}
      - E_1 \begin{bmatrix}
        - 1 \\ \frac{z_j}{z} q^{\pm\frac{1}{2}}
      \end{bmatrix}
    \right) \ .
\end{align}
Note that only the last term depends on $z$, and upon integration,
\begin{align}
  \oint \frac{dz}{2\pi i z} E_1 \begin{bmatrix}
    -1 \\ za  
  \end{bmatrix} = 0 \ .
\end{align}
Only the $z$-independent terms survives the contour integration, yielding \eqref{integration-formula-f}.

In computing Schur index, we often encounter more complicated contour integrals involving the product of an elliptic function and several Eisenstein series. For example, for the class-$\mathcal{S}$ $A_1$ index, we need the following integration formula,
\begin{align}
  \label{integration-formula-fE-1}
    & \ \oint_{|z| = 1} \frac{dz}{2\pi i z}f(\mathfrak{z})E_k\left[
    \begin{matrix}
      -1 \\ za
    \end{matrix}\right] \\
    = & \ - \mathcal{S}_{k} \left(f(\mathfrak{z}_0)
      + \sum_{\text{real/img} \mathfrak{z}_j}R_j  E_{1}\left[\begin{matrix}
        -1 \\ \frac{z_j}{z_0}q^{\pm \frac{1}{2}}
      \end{matrix}\right]
    \right)
    - \sum_{\text{real/img} \mathfrak{z}_j} R_j\,  \sum_{\ell = 0}^{k - 1} \mathcal{S}_{\ell}\, E_{k - \ell + 1}\left[
      \begin{matrix}
        1 \\ z_j a q^{\pm \frac{1}{2}}
      \end{matrix}\right] \ .
\end{align}
Here $\mathcal{S}_k$ are rational numbers defined through the series expansion
\begin{align}
  \frac{1}{2} \frac{y}{\sinh(y/2)} = \sum_{\ell \ge 0} \mathcal{S}_\ell y^\ell \ .
\end{align}
Explicitly, we list a few instances of $\mathcal{S}_\ell$ below.
\begin{center}
  \begin{tabular}{c|c|c|c|c|c|c|c|c|c|c|c|c|c}
    $\ell$ & 0 & 1 & 2 & 3 & 4 & 5 & 6 & 7 & 8 & 9 & 10 & 11 & 12 \\
    \hline
    $\mathcal{S}_\ell$ & 1 & 0 & $- \frac{1}{24}$ & 0 & $\frac{7}{5760}$ & 0 & $ - \frac{31}{967680}$ & 0 & $\frac{127}{154828800}$ & 0 & $- \frac{73}{3503554560}$
    & $0$ & 
    $\frac{1414477}{2678117105664000}$
  \end{tabular}
\end{center}

Similarly, we also have
\begin{align}\label{integration-formula-fE-2}
  & \ \oint_{|z| = 1} \frac{dz}{2\pi i z} f(\mathfrak{z}) E_k\left[\begin{matrix}
    + 1 \\ za
  \end{matrix}
  \right] \nonumber\\
  = & \ - \mathcal{A}_k\left(f(\mathfrak{z}_0) + \sum_{\text{real/img } \mathfrak{z}_j}R_j\, E_1\left[\begin{matrix}
      - 1\\ \frac{z_j}{z_0}q^{\pm \frac{1}{2}}
    \end{matrix}\right]\right)  \nonumber\\
    & \ - \sum_{\text{real/img } \mathfrak{z}_j} R_j \left(
      - \mathcal{B}_k\, E_1\left[\begin{matrix}
        -1 \\ z_jaq^{\pm \frac{1}{2}}
      \end{matrix}\right]
    + \sum_{\ell = 0}^{k - 1} \mathcal{S}_{\ell}\, E_{k + 1 - \ell} \left[\begin{matrix}
            -1 \\ z_j a q^{\pm  \frac{1}{2}}
          \end{matrix}\right]\right) \; ,
\end{align}
where
\begin{align}
  \mathcal{A}_{2n} = \frac{B_{2n}}{(2n)!}, \qquad \mathcal{A}_{2n + 1} = \frac{\delta_{n, 0}}{2}, \qquad \mathcal{B}_{2n} = \frac{B_{2n}}{(2n)!} - \mathcal{S}_{2n}, \qquad \mathcal{B}_{2n + 1} = \frac{\delta_{n,0}}{2} \ .
\end{align}

To compute the Schur index of the some class-$\mathcal{S}$ $A_2$ with 2 or 3 maximal punctures and an arbitrary number of minimal punctures, we need the following type of integrals,
\begin{align}\label{EEtype}
  \oint \frac{dz}{2\pi i z} f(\mathfrak{z}) E_1 \begin{bmatrix}
    \pm 1 \\ za  
  \end{bmatrix}
  E_k \begin{bmatrix}
    \pm 1 \\ zb
  \end{bmatrix} \ .
\end{align}

As simplest cases, when $f(\mathfrak{z}) = 1$ we have the following integration formula,
\begin{align}\label{integration-formula-EE}
  & \ \oint \frac{dz}{2\pi i z}E_k \begin{bmatrix}
    -1 \\ za  
  \end{bmatrix}
  E_\ell \begin{bmatrix}
    -1 \\ zb
  \end{bmatrix}\\
  = & \ (-1)^{k + \ell + 1}\Bigg(
  - C_{k + \ell}^{k} \mathcal{S}_{k + \ell} 
  + \sum_{r = 2}^{\ell}\sum_{s = r}^{\ell}
  (-1)^{r + s}C_{k + \ell - s}^{\ell + 1 - r}\mathcal{S}_{k + \ell - s} E_s \begin{bmatrix}
    1 \\ a/b  
  \end{bmatrix}
  \\
  & \ \qquad
  \qquad\qquad\qquad + \sum_{r = \ell + 1}^{k + \ell}(-1)^r C_{k + \ell - 1 - r}^{\ell - 1}
  E_r \begin{bmatrix}
    1 \\ a/b  
  \end{bmatrix}
  \Bigg) \ .
\end{align}
One can also utilize this formula to derive or understand the structure of the other ones we introduce in this appendix. For example, to compute \eqref{integration-formula-fE-1}, one can begin by decomposing $f$ as in (\ref{elliptic-function-decomposition}), and apply (\ref{integration-formula-EE}).

When $f(\mathfrak{z})$ is a nontrivial elliptic function, we have the following series of integration formula,
\begin{align}\label{integration-formula-fEE-1}
  & \ \oint \frac{dz}{2\pi i z} f(\mathfrak{z}) E_1 \begin{bmatrix}
    -1 \\ za
  \end{bmatrix}
  E_{2k} \begin{bmatrix}
    -1 \\ zb
  \end{bmatrix} \nonumber \\
  = & \ 
  \left[\oint \frac{dz}{2\pi i z}f(\mathfrak{z})\right]\sum_{\ell = 0}^{k - 1}
  \mathcal{S}_{2\ell} E_{2k + 1 - 2\ell} \begin{bmatrix}
      1 \\ a/b
    \end{bmatrix} \\
  & \ - \sum_{\text{real/img }\mathfrak{z}_j} R_j E_1 \begin{bmatrix}
      1 \\ a z_j q^{\pm \frac{1}{2}}
    \end{bmatrix}\sum_{\ell = 0}^{k - 1}
  \mathcal{S}_{2\ell} 
    E_{2k + 1 - 2\ell} \begin{bmatrix}
      1 \\ b z_j q^{\pm \frac{1}{2}}
    \end{bmatrix} \nonumber\\
   & \ - \sum_{\text{real/img }\mathfrak{z}_j} R_j  E_1 \begin{bmatrix}
      1 \\ a z_j q^{\pm \frac{1}{2}}
    \end{bmatrix}\sum_{\ell = 0}^{k - 1}
    \mathcal{S}_{2\ell} 
    E_{2k + 1 - 2\ell} \begin{bmatrix}
      1 \\ \frac{a}{b}
    \end{bmatrix}
    \nonumber\\
  & \ - \sum_{\text{real/img }\mathfrak{z}_j} R_j
  \sum_{\ell = 0}^{k}(1-2\ell)\mathcal{S}_{2\ell} E_{2k + 2 - 2\ell}\begin{bmatrix}
    1 \\ b z_j q^{\pm \frac{1}{2}}  
  \end{bmatrix}
  + \mathcal{S}_{2k}\left(E_2\begin{bmatrix}
        1 \\ a z_j q^{\pm \frac{1}{2}}  
      \end{bmatrix}
    + E_2\begin{bmatrix}
        1 \\ b z_j q^{\pm \frac{1}{2}}  
      \end{bmatrix}  \right)\ , \nonumber
\end{align}
and
\begin{align}\label{integration-formula-fEE-2}
  & \ \oint \frac{dz}{2\pi i z} f(\mathfrak{z}) E_1 \begin{bmatrix}
    -1 \\ za  
  \end{bmatrix}
  E_{2k + 1} \begin{bmatrix}
    -1 \\ zb
  \end{bmatrix} \nonumber \\
  = & \ \left[\oint \frac{dz}{2\pi i z}f(\mathfrak{z})\right]\left[- (2k + 1)\mathcal{S}_{2k + 2} + \sum_{\ell = 0}^{k + 1} \mathcal{S}_{2\ell} E_{2k+2- 2\ell}\begin{bmatrix}
        1 \\ a/b  
      \end{bmatrix}\right] \nonumber \\
  & \ - \sum_{\text{real/img.}\ \mathfrak{z}_j} R_j E_1 \begin{bmatrix}
    1 \\ a z_j q^{\pm \frac{1}{2}}
  \end{bmatrix} \sum_{\ell = 0}^{k} \mathcal{S}_{2\ell} E_{2k + 2 - 2\ell}\begin{bmatrix}
    1 \\ bz_j q^{\pm \frac{1}{2}}
  \end{bmatrix} \\
  & \ + \sum_{\text{real/img.}\ \mathfrak{z}_j} R_j E_1 \begin{bmatrix}
    1 \\ a z_j q^{\pm \frac{1}{2}}
  \end{bmatrix} \sum_{\ell = 0}^{k} \mathcal{S}_{2\ell} E_{2k + 2 - 2\ell}\begin{bmatrix}
    + 1 \\ a/b
  \end{bmatrix} \nonumber\\
  & \ - \sum_{\text{real/img.}\ \mathfrak{z}_j} R_j \sum_{\ell = 0}^{k} (1-2\ell)\mathcal{S}_{2\ell} E_{2k + 3 - 2\ell}\begin{bmatrix}
    1 \\ b z_j q^{\pm \frac{1}{2}}
  \end{bmatrix} \nonumber \ .
\end{align}
Variants of these integration formula with $E_1 \big[\substack{+ 1 \\ za}\big]$, $E_1 \big[\substack{+ 1 \\ zb}\big]$ can be obtained by applying (\ref{Eisenstein-half-shift}).

\subsection{Integration formula with monomial}

In the previous discussions, we have encountered integrals involving products of elliptic functions and Eisenstein series. In the following, we further enrich the integration formula by including a monomial of the integration variable,
\begin{align}
  \oint \frac{dz}{2\pi i z} z^n f(\mathfrak{z}), \qquad
  \oint \frac{dz}{2\pi i z} z^n f(\mathfrak{z}) E_k \begin{bmatrix}
    \pm 1 \\
    z a  
  \end{bmatrix}\ , \qquad
  n \in \mathbb{Z}_{\ne 0} \ .
\end{align}
where $f(\mathfrak{z})$ is again an elliptic function in $\mathfrak{z}$. These formula will be important when dealing with loop operator index.

\subsubsection{One Eisenstein and monomial}
In the presence of an Eisenstein series, we have the following integration formula for a generic $a$ independent of $q$,
\begin{align}\label{integration-formula-zE}
  \oint \frac{dz}{2\pi i z} z^n E_k \begin{bmatrix}
    1 \\ za
  \end{bmatrix}
  = & \ \frac{1}{(k-1)!} \frac{q^n}{a^n} \frac{\text{Eu}_{k - 1}(q^n)}{(1 - q^n)^k} \ ,\\
  \oint \frac{dz}{2\pi i z} z^n E_k \begin{bmatrix}
    1 \\ z^{-1}a
  \end{bmatrix}
  = & \ \frac{(-1)^k}{(k-1)!} (aq)^n \frac{\text{Eu}_{k - 1}(q^n)}{(1 - q^n)^k} \ .
\end{align}
Here $\text{Eu}_n(t)$ denotes the Eulerian polynomial defined by the equation
\begin{align}
  \sum_{n = 0}^{+\infty}\text{Eu}_n(t) \frac{x^n}{n!} = \frac{t - 1}{t - e^{(t - 1)x}} \ .
\end{align}
Similarly, we have the following parallel integration formula
\begin{align}
  \oint \frac{dz}{2\pi i z} z^n E_k \begin{bmatrix}
    - 1 \\ za
  \end{bmatrix}
  = & \ \frac{1}{(k-1)!} \frac{q^{n/2}}{a^n} \Phi(q^n, 1 - k, \frac{1}{2}) \ ,\\
  \oint \frac{dz}{2\pi i z} z^n E_k \begin{bmatrix}
    - 1 \\ z^{-1}a
  \end{bmatrix}
  = & \ \frac{(-1)^k}{(k-1)!} a^nq^{n/2} \Phi(q^n, 1 - k, \frac{1}{2}) \ ,
\end{align}
where the $\Phi$ denotes the Lerch transcendent function $\Phi(z, s, a)$ given by
\begin{align}
  \Phi(z, s, a) \coloneqq \sum_{p = 0}^{+\infty} \frac{z^p}{(p + a)^s} \ .
\end{align}
Recall that the Eisenstein series enjoy shift property (\ref{Eisenstein-half-shift}). When inserted into the above integration formula, the shift property translates to
\begin{align}
  \frac{1}{(1 - q^n)^k} \operatorname{Eu}_{k - 1}(q^n)
  = \sum_{\ell = 0}^{k} \left(\frac{1}{2}\right)^\ell \frac{(k-1)!}{\ell!(k - 1 - \ell)!} \Phi(q^n, 1 - k + \ell, \frac{1}{2}) \ .
\end{align}
For readers' convenience, we list a few instances of $\operatorname{Eu}$ and $\Phi$,
\begin{center}
  \begin{tabular}{c|c|c|c|c|c|c}
    $n$ & $1$ & $2$ & $3$ & $4$ & $5$ \\
    \hline
    $\operatorname{Eu}_n(t)$ & $1$ & $1 + q$ & $1 + 4q + q^2$ & $1 + 11q + 11q^2 + q^3$ & $1 + 26 q + 66q^2 + 26q^3 + q^4$ \\
  \end{tabular}
  \begin{tabular}{c|c|c|c|c}
    $n$ & $6$ \\
    \hline
    $\operatorname{Eu}_n(t)$ & $1 + 57q + 302q^2 + 302q^3 + 57q^4 + q^5$
  \end{tabular}
  \begin{tabular}{c|c|c|c|c}
    $n$ & $7$ \\
    \hline
    $\operatorname{Eu}_n(t)$ & $1 + 120q + 1191q^2 + 2416 q^3 + 1191q^4 + 120 q^5 + q^6$
  \end{tabular}
\end{center}

In the presence of certain amount of $q$-shift, the above formula need some modifications. For example, with generic $a$, $0 < \alpha < 1$, $\ell \in \mathbb{N}_{> 0}$
\begin{align}
  \oint \frac{dz}{2\pi i z}z^n E_1 \begin{bmatrix}
    1 \\ z^{-1}aq
  \end{bmatrix}
  = & \ \frac{(a)^n}{1 - q^{-n}} = \oint \frac{dz}{2\pi i z} z^n E_1 \begin{bmatrix}
    1 \\ z^{-1} a  
  \end{bmatrix}, \\
  \oint \frac{dz}{2\pi i z}z^n E_{k} \begin{bmatrix}
    1 \\ z^{-1}aq^\alpha
  \end{bmatrix}
  = & \ \frac{(-1)^k}{(k-1)!} (a q^\alpha)^n q^{n} \frac{\operatorname{Eu}_{k - 1}(q^n)}{(1 - q^n)^k}- \delta_{k = 1}(aq^\alpha)^n \ ,\\
  \oint \frac{dz}{2\pi i z}z^n E_1 \begin{bmatrix}
    -1 \\ z^{-1}aq^\ell
  \end{bmatrix}
  = & \ \frac{-2 + 2q^n}{1 + q^n} \frac{(-1)^k}{(k-1)!} a^n q^{n/2} \Phi(q^n, 1 - k, \frac{1}{2})  \ ,\\
  \oint \frac{dz}{2\pi i z}z^n E_2 \begin{bmatrix}
    -1 \\ z^{-1}aq^\ell
  \end{bmatrix}
  = & \ \frac{-(2\ell - 1) + (2\ell + 1)q^n}{1 + q^n} \frac{(-1)^k}{(k-1)!} a^n q^{n/2} \Phi(q^n, 1 - k, \frac{1}{2})  \ ,\\
  \oint \frac{dz}{2\pi i z}z^n E_3 \begin{bmatrix}
    -1 \\ z^{-1}aq^\ell
  \end{bmatrix}
  = & \ \frac{(2\ell - 1)^2 - 2(-3 + 4\ell^2)q^n + (2\ell + 1)^2 q^{2n}}{1 + q^n} \\
  & \ \qquad\qquad\qquad \times \frac{(-1)^k}{(k-1)!} a^n q^{n/2} \Phi(q^n, 1 - k, \frac{1}{2})  \ .
\end{align}

\subsubsection{Two Eisenstein series}
With two factors of Eisenstein series, the integration formula become much more tedious. For $n \in \mathbb{Z}_{\ne 0}$ and $k_1 \ge k_2$, we have
\begin{align}\label{integration-formula-zEE-1}
  & \ \oint \frac{dz}{2\pi i z} z^n E_{k_1}\begin{bmatrix}
    + 1 \\ z  
  \end{bmatrix}E_{k_2} \begin{bmatrix}
    + 1 \\ za  
  \end{bmatrix} \\
  = & \ \sum_{\ell = 0}^{k_1} \frac{1}{\ell!} \frac{q^n}{a^n}  \frac{\operatorname{Eu}_{k_2 + \ell - 1}(q^n)}{(1 - q^n)^{k_2 + \ell}}\left[
    \frac{(-1)^{k_1 - \ell}}{(k_2 - 1)!} + \frac{\ell!a^n}{(k_1 - 1)! (k_2 - k_1 + \ell)!}
  \right] E_{k_1 - \ell}\begin{bmatrix}
    + 1 \\ a  
  \end{bmatrix} \ , \nonumber
\end{align}
and when $k_1 \le k_2$,
\begin{align}\label{integration-formula-zEE-2}
  & \ \oint \frac{dz}{2\pi i z} z^n E_{k_1}\begin{bmatrix}
    + 1 \\ z  
  \end{bmatrix}E_{k_2} \begin{bmatrix}
    + 1 \\ za  
  \end{bmatrix} \\
  = & \ \sum_{\ell = 0}^{k_2} \frac{1}{\ell!} \frac{q^n}{a^n}  \frac{\operatorname{Eu}_{k_1 + \ell - 1}(q^n)}{(1 - q^n)^{k_1 + \ell}}\left[
    \frac{a^n}{(k_1 - 1)!} + \frac{(-1)^{k_2 - \ell} \ell!}{(k_2 - 1)! (k_1 - k_2 + \ell)!}
  \right] E_{k_2 - \ell}\begin{bmatrix}
    + 1 \\ a
  \end{bmatrix} \ .\nonumber
\end{align}
It may be convenient to merge the two identities into
\begin{align}\label{integration-formula-zEE-3}
  \oint \frac{dz}{2\pi i z} z^n E_{k_1}\begin{bmatrix}
    + 1 \\ z  
  \end{bmatrix}
  E_{k_2}\begin{bmatrix}
    + 1 \\ z  a
  \end{bmatrix}
  = \sum_{\ell = 1}^{\operatorname{max}(k_1, k_2)}
  \frac{1}{\ell!}\frac{q^n}{a^n}\mathcal{E}_{k_1, k_2; \ell}(a^n, q^n) E_{\max(k_1, k_2) - \ell} \begin{bmatrix}
    +1 \\ a  
  \end{bmatrix} \ ,
\end{align}
where $\mathcal{E}$ can be read off from (\ref{integration-formula-zEE-1}) and (\ref{integration-formula-zEE-2}). Note also that
\begin{align}
  \oint \frac{dz}{2\pi i z}z^n E_{k_1} \begin{bmatrix}
    +1 \\ z a  
  \end{bmatrix}
  E_{k_1} \begin{bmatrix}
    +1 \\ z b  
  \end{bmatrix}
  = & \ b^{-n} \oint \frac{dz}{2\pi i z}z^n E_{k_1} \begin{bmatrix}
    +1 \\ z a/b
  \end{bmatrix}
  E_{k_2} \begin{bmatrix}
    +1 \\ z
  \end{bmatrix} \nonumber \\
  = & \ a^{-n} \oint \frac{dz}{2\pi i z}z^n E_{k_1} \begin{bmatrix}
    +1 \\ z
  \end{bmatrix}
  E_{k_2} \begin{bmatrix}
    +1 \\ z b/a
  \end{bmatrix} \ .
\end{align}
The other integration formula variants of (\ref{integration-formula-zEE-1}), (\ref{integration-formula-zEE-2}) of involving $E_k \big[\substack{- 1\\za}\big]$, $E_k \big[\substack{- 1\\zb}\big]$ can be straightforwardly derived using the shift property (\ref{Eisenstein-half-shift}). For example, solving the system of equations
\begin{align}
  q^{- \frac{n}{2}}
  \oint \frac{dz}{2\pi i z}z^n E_{k_1} \begin{bmatrix}
    + 1 \\ z
  \end{bmatrix}
  E_{k_2} \begin{bmatrix}
    + 1 \\ z
  \end{bmatrix}
  = \sum_{\ell_1 = 0}^{k_1}
  \sum_{\ell_2 = 0}^{k_2} \frac{1}{2^{\ell_1 + \ell_2}} \frac{1}{\ell_1! \ell_2!} \oint \frac{dz}{2\pi i z}
  \prod_{i = 1}^2E_{k_i - \ell_i}\begin{bmatrix}
    -1 \\ z  
  \end{bmatrix} \ ,
\end{align}
produces the integration formula for $z^n E_{1 \le \ell_1 \le k_1}\big[\substack{-1\\ z}\big] E_{1 \le \ell_2 \le k_2}\big[\substack{-1\\ z}\big]$ in terms of a linear combination of the known results for $z^n E_{1 \le \ell_1 \le k_i}\big[\substack{-1\\ z}\big]$ and $z^n E_{1 \le \ell_1 \le k_1}\big[\substack{+1\\ z}\big] E_{1 \le \ell_2 \le k_2}\big[\substack{+1\\ z}\big]$.

Combining the above results, one can write down integration formula for $m \in \mathbb{Z}_{\ne 0}$
\begin{align}\label{integration-formula-zfE}
  \oint \frac{dz}{2\pi i z} z^m f(\mathfrak{z}) E_k \begin{bmatrix}
    \pm 1 \\ za
  \end{bmatrix}
  = & \ \left[\oint \frac{dz}{2\pi i z} f(\mathfrak{z})\right]\oint \frac{dz}{2\pi i z}z^m E_k \begin{bmatrix}
    \pm 1 \\ za  
  \end{bmatrix} \nonumber\\
  & \ - \sum_{\operatorname{real/img} \ \mathfrak{z}_j}R_j
    \oint \frac{dz}{2\pi i z}
    z^m
    E_1 \begin{bmatrix}
      - 1 \\ \frac{z_j}{z} q^{\pm\frac{1}{2}} 
    \end{bmatrix}E_k \begin{bmatrix}
      \pm 1 \\ za  
    \end{bmatrix} \ ,
\end{align}
which follows easily from the decomposition
\begin{align}
  f(\mathfrak{z}) = f(\mathfrak{z}_0) + \sum_{\text{real/img} \ \mathfrak{z}_j} R_j \left(
      E_1 \begin{bmatrix}
        - 1 \\ \frac{z_j}{z_0} q^{\pm\frac{1}{2}}
      \end{bmatrix}
      - E_1 \begin{bmatrix}
        - 1 \\ \frac{z_j}{z} q^{\pm\frac{1}{2}}
      \end{bmatrix}
    \right) \ .
\end{align}

\subsubsection{Elliptic functions and monomial}

We proceed with the first integral by recalling that
\begin{align}
  f(\mathfrak{z})
  = C_f(\tau) + \frac{1}{2\pi i}\sum_{j}R_j\zeta (\mathfrak{z} - \mathfrak{z}_j) \ ,
\end{align}
where $\zeta$ can be expanded in Fourier series,
\begin{align}
  \zeta(\mathfrak{z}) = - 4\pi^2 \mathfrak{z} E_2(\tau) - (2m + 1)\pi i
  + \pi \sum_{n}' \frac{1}{\sin n \pi \tau}q^{- \frac{n}{2}} e^{2\pi i n(\mathfrak{z}_0 + \lambda \tau)},
\end{align}
for $\mathfrak{z} = \mathfrak{z}_0 + \lambda \tau + m \tau$, $\lambda \in [0,1)$ and $m \in \mathbb{Z}$. The integral in the presence of $z^n$ with $n \ne 0$ can be carried out easily, which gives
\begin{align}\label{integration-formula-monomial}
  \oint \frac{dz}{2\pi i z} z^n f(\mathfrak{z})
  = & \ - \sum_{\text{real} \ \mathfrak{z}_j} R_j \frac{1}{1 - q^{-n}} z_j^n - \sum_{\text{img}\ \mathfrak{z}_j}R_j \frac{1}{q^n - 1}z_j^n  \nonumber\\
  = & \  - \sum_{\text{real/img} \ \mathfrak{z}_j} R_j \frac{(z_j q^{\pm\frac{1}{2}})^n}{q^{n/2} - q^{-n/2}} \ .
\end{align}
Summing over $n$ with suitable coefficients, we further obtain some useful formulas. For example,
\begin{align}
  \oint \frac{dz}{2\pi i z}\sum_{n \in \mathbb{Z}}' z^nf(\mathfrak{z})
  = - \sum_{\text{real/img } \mathfrak{z}_j} R_j E_1 \begin{bmatrix}
    -1 \\ z_jq^{\pm \frac{1}{2}}  
  \end{bmatrix} \ .
\end{align}
This is simply a special case of (\ref{integration-formula-f}), since
\begin{align}
  f(z = 1) = \oint \frac{dz}{2\pi i z} \delta(z)f(\mathfrak{z}) 
  = \oint \frac{dz}{2\pi i z} (1 + \sum_{n}' z^n)f(\mathfrak{z}) \ .
\end{align}
Also, for $n \in \mathbb{N}$,
\begin{align}
  \oint{\frac{dz}{2\pi iz}\frac{\left( q-1 \right) z}{\left( 1-z \right) \left( 1-qz \right)}f\left( \mathfrak{z} \right)}
  = & \ - \sum_{\text{real } \mathfrak{z}_j}{R_j\frac{qz_j}{1-qz_j}}
    -\sum_{\text{img } \mathfrak{z}_j}{R_j\frac{z_j}{1-z_j}} \\
  = & \ - \sum_{\text{real } \mathfrak{z}_j}{R_j\frac{1}{1-qz_j}}
    -\sum_{\text{img } \mathfrak{z}_j}{R_j\frac{1}{1-z_j}} \ ,\\
  \oint \frac{dz}{2\pi i z} \frac{1}{(1-z^p)(1 - \frac{1}{z^p})} z^n f(\mathfrak{z})
  = & \ \sum_{\text{real/img } \mathfrak{z}_j} R_j (z_jq^{\pm \frac{1}{2}})^n  \sum_{\substack{k \ge 0 \\ k + n \ne 0}} \frac{k(z_j q^{\pm \frac{1}{2}})^{pk}}{q^{\frac{pk + n}{2}} - q^{- \frac{pk + n}{2}}} \\
  & + \frac{n}{p} \delta_{\frac{n}{p} \in \mathbb{Z}_{< 0}}\oint \frac{dz}{2\pi i z} f(\mathfrak{z}) \ .
\end{align}
When $z$ is the $SU(2)$ fugacity, then we have with the insertion of a spin-$J$ character $\chi_J(z) \coloneqq \sum_{m = -J}^{J}z^{2m}$,
\begin{align}\label{integration-formula-χf}
  \oint \frac{dz}{2\pi i z} & \ \chi_{J}(z) f(\mathfrak{z}) \nonumber\\
  = & \ \delta_{J \in \mathbb{Z}} \oint \frac{dz}{2\pi i z}f(\mathfrak{z})
  - \sum_{\substack{m = -J\\ m\ne 0}}^{+J}
  \left(
    \sum_{\text{real} \ \mathfrak{z}_j}R_j  \frac{1}{1 - q^{-2m}}z_j^{2m}
    + \sum_{\text{img} \ \mathfrak{z}_j}R_j \frac{1}{q^{2m} - 1}z_j^{2m}
  \right) \ ,
\end{align}
and
\begin{align}
  \oint \frac{dz}{2\pi i z}\frac{\chi_J(z)}{(1-z^p)(1-1/z^p)}f(\mathfrak{z})
  = & \ \sum_{m = -J}^{J}
        \sum_{\text{real/img } \mathfrak{z}_j}R_j(z_jq^{\pm \frac{1}{2}})^{2m}
        \sum_{\substack{k \ge 0 \\ pk + 2m \ne 0}} \frac{k(z_j q^{\pm \frac{1}{2}})^{pk}}{q^{\frac{pk+2m}{2}} - q^{- \frac{pk+2m}{2}}} \nonumber \\
  & \ + \left[\sum_{m = - J}^{+J} \frac{2m}{p}\delta_{\frac{2m}{p} \in \mathbb{Z}_{< 0}}\right] \oint \frac{dz}{2\pi i z}f(\mathfrak{z}) \ .
\end{align}
Note that for $p = 1$, $J \in\frac{1}{2} \mathbb{N}$, $\sum_{m = -J}^{+J}2m \delta_{2m < 0} = \lceil J \rceil (\lceil J \rceil - 2J - 1) = - \lfloor(J + \frac{1}{2})^2\rfloor$.

\clearpage

%%%%%%%%%%  Bibliography  %%%%%%%%%%%%
{%\small
% \linepenalty=1000
\bibliographystyle{utphys}
\bibliography{ref}
}

\end{document}